\documentclass[a4paper,11pt]{article}
\usepackage{jheppub} 
\usepackage[T1]{fontenc} 
\usepackage{multirow} 
\usepackage{tcolorbox}
\usepackage{graphicx} 
\usepackage{colortbl}
\usepackage{subcaption,xcolor}
\newcommand{\legendbox}[1]{\textcolor{#1}{\rule{10pt}{7pt}}}
\newcommand{\legenddot}[1]{\textcolor{#1}{\Large$\bullet$}}

\usepackage{tikz}

\definecolor{lmax8}{rgb}{0.686,0.780,0.910}  
\definecolor{lmax10}{rgb}{0.73,0.81,0.9}  
\definecolor{lmax12}{rgb}{0.5,0.65,0.78} 
\definecolor{purp}{rgb}{0.5,0,0.4} 
\usepackage{hyperref}
\usepackage{graphicx}
\DeclareMathOperator{\Tr}{Tr}
\DeclareMathOperator{\tr}{tr}

\usepackage{subcaption}
\usepackage{listings}
\usepackage{ytableau}
\usepackage{youngtab} 
\usepackage{booktabs,multirow,siunitx}
\newcommand{\trXX}{\langle \tr\,XX\rangle}

\newcommand{\trXXXX}{\langle \tr\,XXXX\rangle}

\newcommand{\Q}[1]{Q\!\left(#1\right)}

\hypersetup{hidelinks,
	colorlinks=true,
	allcolors=blue!90!black,
	pdfstartview=Fit,
	breaklinks=true}
\title{Bootstrapping Yang-Mills matrix integrals 
}

\author[a]{Wenliang Li}
\author[a,b]{and Xinran Su}
\affiliation[a]{School of Physics, Sun Yat-Sen University,
	 Guangzhou 510275, China}
\affiliation[b]{Perimeter Institute for Theoretical Physics, Waterloo, Ontario N2L 2Y5, Canada}
	 
\emailAdd{liwliang3@mail.sysu.edu.cn}
\emailAdd{xinransu.phys@gmail.com}

\abstract{
	We revisit 	the large $N$ limit of bosonic $D$-matrix Yang-Mills integrals using two complementary bootstrap methods. 
	In the positivity bootstrap, we obtain bounds for $\langle \tr XX \rangle$ and $\langle \tr XXXX \rangle$ at various length cutoffs $L_{\max}$. 
	For $D=3$, we do not find an isolated region until $L_{\max}=12$. 
	For larger $D$, the allowed regions become islands at $L_{\max}=8$ and shrink rapidly as 	$L_{\max}$ increases. 
	The precision of some $L_{\max}=12$ islands is comparable to that of Monte Carlo estimates. 
	For a fixed $L_{\max}$, the allowed region also shrinks with $D$ 
	and converges to the large $D$ expansion results.  
	We further deduce the analytic expressions of various types of trajectories and eigenvalue distributions at large $D$.   
	Based on these explicit formulas, we propose some ansatz for the analytic trajectory bootstrap and obtain accurate results for finite $D$. 
}

\begin{document} 
	\maketitle
	\flushbottom
	
\section{Introduction}

Intriguingly, gravity can be dynamically generated from gauge theory in lower dimensions, 
where space (and time) may emerge from a large number of matrix degrees of freedom \cite{tHooft:1973alw,tHooft:1993dmi,Susskind:1994vu,Maldacena:1997re,Gubser:1998bc,Witten:1998qj}. 
In the holographic framework, large $N$ matrix models provide nonperturbative definitions of quantum gravity that are otherwise difficult to formulate.  
Some representative examples include:
\begin{enumerate}
	\item The  Banks-Fischler-Shenker-Susskind (BFSS) matrix quantum mechanics in $(0+1)$ dimension, which was proposed as a nonperturbative definition of M-theory around flat space~\cite{Banks:1996vh}. This model includes nine  Hermitian $N\times N$ matrices $X^\mu$ and their supersymmetric partners. 
	It is conjectured that this matrix model provides a second-quantized description of M-theory in the infinite momentum frame and captures the nonperturbative scattering amplitudes of massless particles. 
	\item The IIB matrix model or Ishibashi-Kawai-Kitazawa-Tsuchiya (IKKT) matrix model in $(0+0)$ dimensions, conjectured in~\cite{Ishibashi:1996xs} as a nonperturbative formulation of type IIB superstring theory. This model contains 10 Hermitian $N\times N$ matrices $X^\mu$ and their supersymmetric partners. 
	The eigenvalues of $X^\mu$ are interpreted as spacetime coordinates, and their distribution provides a possible mechanism for the dynamical generation of four-dimensional spacetime through spontaneous symmetry breaking. Recently, its holographic aspects have also been discussed in~\cite{Ciceri:2025maa,Ciceri:2025wpb}.
	\item Mass deformations of the above supersymmetric matrix models, such as the Berenstein-Maldacena-Nastase (BMN) matrix quantum mechanics~\cite{Berenstein:2002jq} and the polarized IKKT matrix model~\cite{Bonelli:2002mb,Hartnoll:2024csr}. 
	The BMN model is conjectured to be dual to M-theory on the plane wave background, where the flat directions are lifted by the mass deformation.  
	The holographic duals of the polarized IKKT matrix model were also studied recently in \cite{Hartnoll:2024csr, Komatsu:2024bop,Komatsu:2024ydh,Hartnoll:2025ecj,Chou:2025rwy}. 
\end{enumerate}

One of the minimal holographic settings is perhaps the reduction of $D$-dimensional super-Yang-Mills theory to $(0+0)$ dimensions, namely the supersymmetric Yang-Mills matrix integral. 
For $D=10$, this is precisely the IKKT matrix model mentioned above.  
The reduction to a point may seem to be a drastic simplification, but it is still a challenge to fully solve the IKKT matrix model. 
We would like to examine this problem from the bootstrap perspective.

As the first attempt to bootstrap Yang-Mills matrix integrals, we omit the fermionic part because it usually leads to a sign problem.  
The Pfaffian from a fermionic path integral is generally not real and positive.
In Monte Carlo simulations, it is subtle to interpret a non-positive weight as a probability distribution. 
If the sign problem is mild, one may replace the Pfaffian by its absolute value and obtain reasonable results. 
See \cite{Anagnostopoulos:2022dak} for a review of numerical studies and references therein.  

The main ingredients of modern bootstrap methods are positivity constraints and efficient algorithms~\cite{Rattazzi:2008pe,Poland:2018epd,Rychkov:2023wsd}, 
which can lead to rigorous bounds that systematically shrink with the truncation cutoff. 
As initiated in \cite{Lin:2020mme}, the positivity bootstrap approach has been applied to the studies of bosonic matrix integrals \cite{Lin:2020mme,Kazakov:2021lel,Khalkhali:2020jzr} and matrix quantum mechanics \cite{Han:2020bkb,Lin:2023owt,Lin:2024vvg,Lin:2025srf,Laliberte:2025xvk,Cho:2024kxn}.\footnote{See also \cite{Jevicki:1982jj,Jevicki:1983wu,Koch:2021yeb,Mathaba:2023non},  which are based on the collective representation.}
In the latter case, the positivity condition arises naturally from the positive inner product of the Hilbert space, and this remains true in supersymmetric cases.
However, it is less clear how to formulate the positivity condition for the supersymmetric matrix integrals 
if the path integral measure is not real and positive after integrating out the fermions. 

In this work, we focus on the bosonic $D$-matrix Yang-Mills integral, which consists of only $D$ bosonic Hermitian $N\times N$ matrices. 
 For finite $N$, there are finitely many integrals, so in principle one can carry out the matrix integral explicitly, 
 but this becomes more challenging as $N$ grows. 
For holography, we are more interested in the large $N$ limit. 
For $D\ge 3$, the bosonic Yang-Mills matrix integrals are well-defined at large $N$  
without regularization~\cite{Krauth:1998yu,Austing:2001bd,Austing:2001pk}. 
For $D=10$, this model coincides with the bosonic part of the IIB matrix model, which is also called the bosonic IKKT model. 
Its dynamical aspects such as the extent of spacetime were investigated in~\cite{Hotta:1998en}. 
Such bosonic IKKT matrix integral also arises as the zero-mode sector of the BFSS model, which dominates the path integral at high temperature~\cite{Kawahara:2007ib}. It can further serve as a starting point for analyzing the sign problem of the BFSS model in the high-temperature expansion~\cite{Batra:2026tob}.
In comparison to supersymmetric models, 
an advantage of the bosonic model is that the choice of $D$ is less restricted. 
In particular, it is possible to develop a $1/D$ expansion around the large $D$ saddle \cite{Hotta:1998en,Mandal:2009vz}, 
where $1/D$ furnishes a small expansion parameter. 
The large $D$ expansion provides useful benchmarks for the bootstrap results in this work.  
Our main bootstrap targets are the cases of $D=3, 4, \dots, 10$, 
but we also consider some relatively large $D$, such as $D=10^5$. 

As the positivity condition for the supersymmetric IKKT model is obscured, 
the bootstrap study of the supersymmetric matrix integral cannot rely on positivity. 
Therefore, we also study the bosonic Yang-Mills matrix integral by an alternative bootstrap method 
that does not resort to positivity assumptions. We refer to this method as the analytic trajectory bootstrap. 
In analogy with Regge trajectories, the basic idea of this  bootstrap method is to organize the observables 
by analytic trajectories, which may be associated with some analytic ansatz. 
In the context of matrix models, the matrix moments are connected by various types of trajectories, 
and their analytic structures are closely related to eigenvalue distributions. 
The analytic trajectory bootstrap is still in its infancy and thus far less systematic than the positivity bootstrap. 
Nevertheless, the application to a two-matrix model with quartic potential and Yang-Mills interaction has yielded highly accurate results with low demand on computational resources \cite{Li:2024ggr}.
The multi-matrix Yang-Mills integrals here are far more complicated. 
The explicit formulas from the large $D$ expansion are helpful to the ansatz construction. 

The paper is organized as follows.
 Section~\ref{Preliminaries} introduces the model and sets up the loop equations, 
 together with the $\mathrm{O}(D)$ singlet decomposition and large $D$ expansion. 
 In section~\ref{Positivity}, we formulate the positivity bootstrap and explain how $\mathrm{O}(D)$ representation theory simplifies the problem, allowing us to derive positivity bounds for various $D$ at length cutoffs $L_{\max}=4,6,8,10,12$. 
 For relatively large $D$, we further compare the results with those from the large $D$ expansion. 
 Section~\ref{Analytic} turns to the analytic trajectory bootstrap, where we extract explicit formulas for the moment trajectories and eigenvalue distributions from the large $D$ expansion results. Then we formulate some ansatz for these analytic trajectories and derive nonperturbative results for finite $D$.
In section \ref{sec:discussion}, we summarize our results and discuss some future directions. 
Appendix \ref{Appendix A} explains the procedure for constructing irreducible representations. 
Some technical details can be found in appendix \ref{Appendix:details}. 

\section{Preliminaries}
\label{Preliminaries}
The action of the Yang-Mills matrix model reads
\begin{equation}
	S \;=\; -\,\frac{h\,N}{4}\,\Tr\left([X_\mu,X_\nu][X^\mu,X^\nu]\right),
	\quad \mu,\nu=1,2,\dots,D,
	\label{action}
\end{equation}
where $[X_\mu,X_\nu]= X_\mu X_\nu-X_\nu X_\mu$ is a commutator, 
\(X^\mu\) are \(N\times N\) traceless Hermitian matrices and $h>0$ is a coupling constant. 
Below we set \(h=1\). Alternatively, $h$ can be absorbed by rescaling \(X^\mu\).
The action exhibits an $\mathrm{O}(D)$ symmetry. 
Throughout we write $\mathrm{Tr}$ for the standard trace and 
$\mathrm{tr} \equiv \mathrm{Tr}/N$ for the normalized trace. 
For instance, the traces of the identity matrix $\mathbb{I}_N$ are
\begin{equation}
\mathrm{Tr}\,\mathbf{\mathbb{I}}_N = N,\quad \mathrm{tr}\,\mathbb{I}_N = 1, 
\end{equation} 
where the size of $\mathbb{I}_N$ is $N\times N$. 
As we use the Euclidean signature, 
we do not distinguish upper and lower \(\mathrm{O}(D)\) indices. 
We are interested in the expectation values of various types of \emph{words}, by which we mean
single-trace monomials in the matrices,
\(X^{\mu_1}X^{\mu_2}\!\cdots X^{\mu_L}\). 

\subsection{Loop equations}
\label{Loop Equation}
A basic ingredient for bootstrapping matrix models is the set of \textit{loop equations}, 
i.e., the Schwinger-Dyson equations in the context of gauge theory and matrix model. 
For a real and bounded-from-below action $S$, 
the integral of a total derivative vanishes
\begin{equation}
\int \Big(\prod_{\mu=1}^D dX^\mu\Big) \;
\frac{\partial}{\partial {(X^{\nu})_{ij}}} \bigl(\mathcal{O}_{ij}\, e^{-S}\bigr)
 = 0.
\end{equation}
Since the action \eqref{action} involves a polynomial in matrices, 
we consider the monomial of matrices,  
$\mathcal{O}\equiv X^{\mu_1} X^{\mu_2} \cdots X^{\mu_L}$, 
which is a word of length \(L\). 
In the large $N$ limit, a double-trace moment factorizes into a product of two single-trace moments, 
so the loop equations are quadratic in the single-trace moments
\begin{equation}
	\begin{aligned}
		\big\langle \tr 
		\mathcal{O}\big(
		X^{\nu}X^{\rho}X^{\rho}
		+X^{\rho}X^{\rho}X^{\nu}
		-2\,X^{\rho}X^{\nu}X^{\rho}\big)\big\rangle
		=
		\sum_{p=1}^{L}\,
		\delta^{\mu_p\nu }\;
		\big\langle \tr\mathcal{O}^{(p)}_{\mathrm{l}}\big\rangle\,
		\big\langle \tr\mathcal{O}^{(p)}_{\mathrm{r}}\big\rangle ,
	\end{aligned}
	\label{eq:loop-eq}
\end{equation}
where
$\mathcal{O}^{(p)}_{\mathrm{l}} \equiv X^{\mu_1}\cdots X^{\mu_{p-1}}$
and
$\mathcal{O}^{(p)}_{\mathrm{r}} \equiv X^{\mu_{p+1}}\cdots X^{\mu_L}$. 
The definition of the expectation value $\langle \dots\rangle$ is given by
\begin{equation}
	\langle \tr  \mathcal{O} \rangle = \frac{\int \big(\prod_{\mu=1}^D dX^\mu\big) \;
	(\tr \mathcal{O}\, e^{-S})}
	{\int \big(\prod_{\mu=1}^D dX^\mu\big) \;
		e^{-S}},
\end{equation}
so we have $\langle \tr \mathbb{I} \rangle=1$. 
In principle, we can generate the concrete loop equations by enumerating the explicit superscripts of $\mathcal O$ in \eqref{eq:loop-eq}. 
For $D=3$, some simple examples of the loop equations are
\begin{equation}
	\mathcal{O} = X:\quad
		\langle \tr XXYY \rangle
		-\langle \tr XYXY \rangle
		 \;=\; \frac {1}{4} \,,
\end{equation}

\begin{equation}
	\mathcal{O} = XXX:\quad
		\langle \tr XX  \rangle-2\langle \tr XXXXYY  \rangle+2\langle \tr XXXYXY \rangle
		\;=\; 0 \,,
\end{equation}

\begin{equation}
	\mathcal{O} = YXY:\quad
			\langle \tr X X X Y X Y \rangle
			- \langle \tr X X Y X X Y \rangle 
			+ \langle \tr X X Y Z Y Z \rangle
			- \langle \tr X Y X Z Y Z \rangle=0\,,
\end{equation}

\begin{align}
	\mathcal{O} = XYY:\quad
			& \langle \tr X X \rangle
			-\langle \tr X X X X Y Y \rangle
			+ 2 \langle \tr X X X Y X Y \rangle 
			- \langle \tr X X Y X X Y \rangle
			 \nonumber\\
			& \quad
			- \langle \tr X X Y Y Z Z \rangle
			+ 2 \langle \tr X X Y Z Y Z \rangle
			- \langle \tr X X Y Z Z Y \rangle
			= 0\,,
\end{align}
where $X,Y,Z$ denote $X^\mu$ with $\mu=1,2,3$. 
Note that the last three equations are not independent 
if the $\mathrm{O}(D)$ symmetry of the action \eqref{action} is unbroken. 
In general, the number of concrete equations grows rapidly with $D$ and $L_{\max}$, 
and many of them are redundant due to symmetries. 
Below, we use the $\mathrm{O}(D)$ symmetry to reduce the number of loop equations and unknowns. 

\subsection{$\mathrm{O}(D)$ singlet decomposition}
We assume that the $\mathrm{O}(D)$ symmetry of the action \eqref{action} is unbroken,\footnote{We also assume that the symmetry of the action under the transformation $X_\mu \rightarrow X_\mu^T$ is not broken, which implies that all moments are real.} 
so only $\mathrm{O}(D)$ singlets have non-vanishing expectation values. 
We can express the covariant moments as 
\begin{equation}\label{general-A-decomposition}
\big\langle \tr\, X^{\mu_1}X^{\mu_2}\dots X^{\mu_{L-1}}X^{\mu_L}\big\rangle
= A_{1,2,\dots,L-1,L}\,\delta^{\mu_1\mu_2}\dots\delta^{\mu_{L-1}\mu_L}+(\text{permutations})\,,
\end{equation}
where only inequivalent products of the Kronecker delta are considered. 
Note that the subscripts of the coefficients $A$ encode 
the order of the $\mu$ subscripts. 
The simplest example is at length 2
\begin{equation}
\big\langle \tr\, X^{\mu_1}X^{\mu_2}\big\rangle= A_{1,2}\,\delta^{\mu_1\mu_2}.
\end{equation}
The next example is the singlet decomposition of the length-4 covariant moment 
\begin{equation}\label{singletdecompositionlength4}
\big\langle \tr\, X^{\mu_1}X^{\mu_2}X^{\mu_3}X^{\mu_4}\big\rangle
= A_{1,2,3,4}\,\delta^{\mu_1\mu_2}\delta^{\mu_3\mu_4}
+ A_{1,3,2,4}\,\delta^{\mu_1\mu_3}\delta^{\mu_2\mu_4}
+ A_{1,4,2,3}\,\delta^{\mu_1\mu_4}\delta^{\mu_2\mu_3}.
\end{equation}
After taking into account the cyclic symmetry of the trace, we have 
$
A_{1,4,2,3}=A_{1,2,3,4}
$ and 
\begin{equation}
\big\langle \tr\, X^{\mu_1}X^{\mu_2}X^{\mu_3}X^{\mu_4}\big\rangle
= A_{1,2,3,4}\,(\delta^{\mu_1\mu_2}\delta^{\mu_3\mu_4}+\delta^{\mu_1\mu_4}\delta^{\mu_2\mu_3})
+ A_{1,3,2,4}\,\delta^{\mu_1\mu_3}\delta^{\mu_2\mu_4}.
\label{decompositionL4}
\end{equation}
The singlet decomposition drastically reduces the number of unknowns for relatively small lengths. 
At small lengths, each \(A\) can be identified with a specific word.\footnote{At large lengths, the number of concrete words may be less than that of $A$'s  
because the number of different indices is limited by $D$. 
Accordingly, some coefficients $A$ may be ambiguous, and the concrete moments may be invariant 
under some changes in the coefficients $A$. We can set some coefficients $A$ to zero using this ``gauge symmetry''. }
At length 2, we have $A_{1,2}=\langle \tr\, X X \rangle$. 
In the length-4 example \eqref{decompositionL4}, we have \(A_{1,2,3,4} = \langle \tr\, X X Y Y \rangle\) and \(A_{1,3,2,4} = \langle \tr\, X Y X Y \rangle\). 
Another basic observable of length 4 is
\begin{equation}\label{XXXX-eq-1}
	\big\langle \tr XXXX \big \rangle =2 A_{1,2,3,4}+A_{1,3,2,4}.
\end{equation}
At length 6, the singlet decomposition reads
\begin{equation}
\begin{aligned}
&\big\langle \mathrm{tr}\, X^{\mu_1}X^{\mu_2}X^{\mu_3}X^{\mu_4}X^{\mu_5}X^{\mu_6}\big\rangle
\\[6pt]
={}&  A_{1,2,3,4,5,6}\,\big(\delta^{\mu_1\mu_2}\delta^{\mu_3\mu_4}\delta^{\mu_5\mu_6}
+\delta^{\mu_1\mu_6}\delta^{\mu_2\mu_3}\delta^{\mu_4\mu_5}\big)\\[6pt]
&+ A_{1,2,3,5,4,6}\,\big(\delta^{\mu_1\mu_2}\delta^{\mu_3\mu_5}\delta^{\mu_4\mu_6}
+ \delta^{\mu_1\mu_3}\delta^{\mu_2\mu_4}\delta^{\mu_5\mu_6}
+ \delta^{\mu_1\mu_3}\delta^{\mu_2\mu_6}\delta^{\mu_4\mu_5} \\
&\qquad\qquad\quad + \delta^{\mu_1\mu_5}\delta^{\mu_2\mu_3}\delta^{\mu_4\mu_6}
+\delta^{\mu_1\mu_5}\delta^{\mu_2\mu_6}\delta^{\mu_3\mu_4}
+ \delta^{\mu_1\mu_6}\delta^{\mu_2\mu_4}\delta^{\mu_3\mu_5}
\big)\\
&+ A_{1,2,3,6,4,5}\,\big(\delta^{\mu_1\mu_2}\delta^{\mu_3\mu_6}\delta^{\mu_4\mu_5}
+ \delta^{\mu_1\mu_4}\delta^{\mu_2\mu_3}\delta^{\mu_5\mu_6}
+\delta^{\mu_1\mu_6}\delta^{\mu_2\mu_5}\delta^{\mu_3\mu_4} \big)\\[6pt]
&+ A_{1,3,2,5,4,6}\,\big( \delta^{\mu_1\mu_3}\delta^{\mu_2\mu_5}\delta^{\mu_4\mu_6}
+ \delta^{\mu_1\mu_4}\delta^{\mu_2\mu_6}\delta^{\mu_3\mu_5}
+\delta^{\mu_1\mu_5}\delta^{\mu_2\mu_4}\delta^{\mu_3\mu_6}\big)\\[6pt]
&+A_{1,4,2,5,3,6}\,\delta^{\mu_1\mu_4}\delta^{\mu_2\mu_5}\delta^{\mu_3\mu_6}  ,
\end{aligned}
\end{equation}
where the cyclic symmetry has been implemented. 
We can further express the loop equations in terms of the singlet-decomposition coefficients $A$.  
For example, the explicit loop equations in section \ref{Loop Equation} are encoded in 
\begin{equation}\label{leqA-1}
	2\,(D-1)\,\big(A_{1,2,3,4}-A_{1,3,2,4}\big)=1, 
\end{equation}
\begin{equation}\label{leqA-2}
		D\,A_{1,2,3,5,4,6} -A_{1,2,3,6,4,5}  -(D-2)A_{1,3,2,5,4,6}- A_{1,4,2,5,3,6}= 0,
\end{equation}
\begin{align}\label{leqA-3}
		A_{1,2} +(D-1)(-A_{1,2,3,4,5,6} +A_{1,2,3,5,4,6}-A_{1,2,3,6,4,5}+A_{1,3,2,5,4,6})= 0,
\end{align}
which are independent constraints. 
There is one less equation, and the number of unknowns is reduced from $10$ to $8$. 
In table \ref{Number}, we list the numbers of covariant loop equations 
\footnote{After performing the singlet decomposition, 
we equate the coefficients of each distinct product of Kronecker deltas 
on both sides of \eqref{eq:loop-eq}. 
Then we count the number of these constraints. Note that some of them may still be redundant.   }
and  unknowns \(A\) at some length cutoffs $L_\text{max}$. 
Using the loop equations, we can eliminate some  of the coefficients $A$.
For instance, the first loop equation \eqref{leqA-1} implies 
\begin{equation}\label{A1324-solution}
A_{1,3,2,4}=A_{1,2,3,4}-\frac 1 {2(D-1)}\,.
\end{equation}
Substituting this solution into \eqref{XXXX-eq-1}, 
we can express $\langle \tr XXXX\rangle$ in terms of $A_{1,2,3,4}$
\begin{equation}\label{XXXX-eq-2}
		\big\langle \tr XXXX \big \rangle =3 A_{1,2,3,4}-\frac{1}{2(D-1)}.
\end{equation}

\begin{table}[h]
	\centering
	\begin{tabular}{c|c|c}
		\textbf{Cutoff length} \(L_{\max}\) & \textbf{ \# covariant loop equations} & \textbf{\# unknowns \(A\)} \\
		\hline
		2  & 0   & 1 \\
		4  & 1   & 3 \\
		6  & 3   & 8 \\
		8  & 12  & 25 \\
		10 & 68  & 104 \\
		12 & 553 & 658 \\
	\end{tabular}
	\caption{The total numbers of covariant loop equations and  unknowns \(A\) at some cutoffs for $D\geq 6$.}
	\label{Number}
\end{table}

\subsection{Large $D$ expansion}
\label{Large D expansion}
For the non-supersymmetric Yang-Mills matrix model \eqref{action}, 
we can choose a large $D$ and obtain analytic approximations using the large $D$ expansion.  
They provide consistency checks and concrete, explicit examples of the intricate analytic structures. 
Below we briefly review the large $D$ expansion in \cite{Hotta:1998en}. 
The results for length $6,8,10,12$ are new.

It is useful to expand the Hermitian matrices $X^\mu$ in the adjoint basis 
\begin{equation}
X^\mu = X_a^\mu\, t^a\,,
\end{equation} 
where   $X_a^\mu$ are the expansion coefficients
and  $t^a$ are the $\mathrm{SU}(N)$ generators that satisfy $\tr(t^a t^b)=\delta^{ab}/N$. 
The action \eqref{action} becomes
\begin{equation}
	S[X] \;=\; -\frac{N}{4}\,\lambda^{abcd}\, X_a^\mu X_b^\mu X_c^\nu X_d^\nu\,,
\end{equation}
where the commutator structure is encoded in 
\begin{equation}
	\lambda^{abcd} = \frac{N}{4} \left(
	\tr\big([t^a, t^c][t^b, t^d]\big) + (a \leftrightarrow b) + (c \leftrightarrow d) + 
	(	a \leftrightarrow b,
		c \leftrightarrow d)
	\right)\,.
\end{equation}
By definition, \(\lambda^{abcd}\) is symmetric under \(a \leftrightarrow b\) and \(c \leftrightarrow d\), 
but antisymmetric under \(a \leftrightarrow c\), \(b \leftrightarrow d\).
It is useful to introduce a real symmetric auxiliary field $H_{ab}$. 
By the Hubbard-Stratonovich transformation, the action becomes
\begin{equation}
	S[X,H] \;=\; \frac{N}{4}\,\lambda^{abcd} H_{ab}H_{cd}
	\;+\; \frac{\sqrt{N}}{2}\,K^{ab}\, X_a^\mu X_b^\mu,
	\qquad
	K^{ab} \equiv -\sqrt{N}\,\lambda^{abcd} H_{cd}.
\end{equation}
Since the action is quadratic in $X$, we can integrate out $X$ and derive the effective action
\begin{equation}
	S_{\rm eff}[H] \;=\; \frac{DN}{2}\,\tr\!\ln \sqrt{N}K \;+\; \frac{N}{4}\,\lambda^{abcd} H_{ab}H_{cd}.
\end{equation}
At large \(D\), the path integral is dominated by the saddle point contribution
\(\delta S_{\rm eff}/\delta \tilde H=0\), 
where $H_{ab} = \sqrt{D}\tilde H_{ab}/\sqrt{N}\,$.
Assuming the saddle point solution preserves \(\mathrm{SU}(N)\) symmetry, 
we have \(\tilde H^{(0)}_{ab}=v\,\delta_{ab}\). The saddle point equation implies
\begin{equation}
	v=\frac{1}{\sqrt{2N}}\,,\qquad
	K^{(0)}_{ab}=\sqrt{2ND}\,\delta_{ab}\,,\qquad
	\big(K^{(0)}\big)^{-1}_{ab}=\frac{1}{\sqrt{2ND}}\,\delta_{ab}\,.
\end{equation}
To study the $1/D$ corrections, we parameterize the small fluctuations as
\begin{equation}
	\tilde H_{ab}=\frac{1}{\sqrt{2N}}\,\delta_{ab}
	\;+\;2\sqrt{\frac{N}{D}}\,\phi_{ab}\,,
	\qquad \phi_{ab}=\phi_{ba}\in\mathbb{R}\,.
\end{equation}
We then obtain
\begin{equation}
	K_{ab}
	= -\,\lambda^{abcd}\,\tilde H_{cd}\,\sqrt{D}
	=\sqrt{2ND}\,(\delta_{ab}-\varepsilon\,\Theta_{ab})\,,
\end{equation}
with
\begin{equation}
	\varepsilon=\sqrt{\frac{2}{D}}\,,\qquad
	\Theta_{ab}=\lambda_{ab}{}^{cd}\,\phi_{cd}\,.
\end{equation}
To subleading order in \(1/D\),  the inverse of the kernel is given by
\begin{equation}
	(K^{-1})_{ab}
	=\frac{1}{\sqrt{2ND}}\Big(\delta_{ab}
	+\varepsilon\,\Theta_{ab}
	+\varepsilon^{2}\,\Theta_{ac}\Theta_{cb}
	+O(\phi^{3})\Big).
\end{equation}

To compute singlet moments, 
we express them in terms of the traces of the $\mathrm{SU}(N)$ generators
\begin{equation}
\langle \tr(X^{\mu_1}X^{\mu_2}\dots) \rangle=
\langle X_{a_1}^{\mu_1}X_{a_2}^{\mu_2}\dots  \rangle \tr(t^{a_1}t^{a_2}\dots)\,,
\end{equation}
where $\{X_{a_i}^{\mu_i}\}$ are numbers and commute with each other. 
The first part $\langle X_{a_1}^{\mu_1}X_{a_2}^{\mu_2}\dots  \rangle$ can be decomposed into products of $K^{-1}$ by Wick's theorem.  
Then we use the identity
\begin{equation}
\sum_a (t^a)_{ij}(t^a)_{kl}=\delta_{il}\delta_{jk}-\frac 1 N\delta_{ij}\delta_{kl}
\end{equation}
to derive the final expression. 
For example, the quadratic singlet reads
\begin{equation}
	\big\langle \tr(X^{\mu}X^{\mu})\big\rangle
	=\frac{D}{\sqrt N }\big\langle{{({K^{ - 1}})}_{aa}}\big\rangle\,.
\end{equation}
By rewriting $K^{-1}$ and $S_{\mathrm{eff}}$ in terms of $\phi$ and evaluating the  2-point function $\langle \phi\phi\rangle$, we obtain the $1/D$ series to subleading order
\begin{align}
	\big\langle \tr (X^{\mu}X^{\mu})\big\rangle 
	&= \frac{\sqrt{D}}{\sqrt{2}}
	\left(\left(1 - \frac{1}{N^{2}}\right)
	+ \left(\frac{7}{6} - \frac{1}{6N^{2}}\right)\frac{1}{D}
	+{O}\!\left(\frac{1}{D^{2}}\right)\right)
	\nonumber\\
	&\xrightarrow{N\rightarrow \infty} \frac{\sqrt{D}}{\sqrt{2}}\left(1+\frac {7}{6D}+O(D^{-2})\right).
\end{align}
We further assume that the $D \to \infty $ limit commutes with the $N\to \infty$ limit, and 
take the large $N$ limit before the large $D$ limit. 
To compute longer singlet moments, 
we need to evaluate higher moments of $K^{-1}$.  
For example, the quartic singlets can be derived from
\begin{equation}
	\big\langle X_a^\mu X_b^\mu X_c^\nu X_d^\nu \big\rangle
	= \frac{1}{N} \Big\langle
	D^{2} ({K^{-1}})_{ab} ({K^{-1}})_{cd}
	+ D ({K^{-1}})_{ac} ({K^{-1}})_{bd}
	+ D({K^{-1}})_{ad} ({K^{-1}})_{bc} 
	\Big\rangle,
\end{equation}
whose contraction with $\tr(t^{a}t^{b}t^{c}t^{d})$ or $\tr(t^{a}t^{c}t^{b}t^{d})$ gives
\begin{equation}
	\begin{gathered}
		\big\langle \operatorname{tr}  X^\mu X^\mu X^\nu X^\nu \big\rangle
		= \frac{D}{2} + \frac{3}{2}+O(D^{-1})\,, \quad
		\big\langle \operatorname{tr} X^\mu X^\nu X^\mu X^\nu \big\rangle
		= \frac{3}{2}+O(D^{-1})\, .
	\end{gathered}
\end{equation}
In principle, this procedure applies to arbitrarily long singlets,  
but the computation becomes more involved at large lengths.
 For example, the derivation of sextic singlets is shown in appendix \ref{Appendix:details}.
We verify the literature results for length $2,4$ in \cite{Hotta:1998en} and 
derive new results for length $6,8,10$ and $12$. 

We can also translate the singlet results into the singlet decomposition coefficients $A$. 
For example, the contraction of $\mathrm{O}(D)$ indices in \eqref{decompositionL4} gives
\begin{align}
		\big\langle \tr\, X^{\mu}X^{\mu}X^{\nu}X^{\nu}\big\rangle
		&= D(D+1) A_{1,2,3,4}
		+D A_{1,3,2,4}\,,
	\\
		\big\langle \tr\, X^{\mu}X^{\nu}X^{\mu}X^{\nu}\big\rangle
		&= 2D A_{1,2,3,4}
		+D^2  A_{1,3,2,4}\,,
\end{align}
so the singlet decomposition coefficients can be determined by the singlets
\begin{align}
A_{1,2,3,4}&=\frac{D\big\langle \tr\, X^{\mu}X^{\mu}X^{\nu}X^{\nu}\big\rangle-\big\langle \tr\, X^{\mu}X^{\nu}X^{\mu}X^{\nu}\big\rangle}{D(D-1)(D+2)}\,,
\\
A_{1,3,2,4}&=\frac{-2\big\langle \tr\, X^{\mu}X^{\mu}X^{\nu}X^{\nu}\big\rangle+(D+1)\big\langle \tr\, X^{\mu}X^{\nu}X^{\mu}X^{\nu}\big\rangle}{D(D-1)(D+2)}\,.
\end{align}
The explicit results for $L\leq 6$ are
\begin{align}
		A_{1,2} &=\frac{1}{\sqrt{2}}D^{-1/2}+ \frac{7}{6\sqrt{2}}D^{-3/2} +O(D^{-5/2}),\\
		A_{1,2,3,4} &=  \frac{1}{2}D^{-1}+D^{-2} +O(D^{-3}), \quad
		A_{1,3,2,4} = \frac{1}{2}D^{-2}+O(D^{-3}),\\
		A_{1,2,3,4,5,6} &= \frac{1}{2\sqrt{2}}D^{-3/2}+ \frac{5}{4\sqrt{2}}D^{-5/2} +O(D^{-7/2}), \\
		A_{1,2,3,6,4,5} &=  \frac{1}{2\sqrt{2}}D^{-3/2}+\frac{17}{12\sqrt{2}}D^{-5/2} +O(D^{-7/2}), \\
		A_{1,2,3,5,4,6} &= \frac{1}{2\sqrt{2}}D^{-5/2}+O(D^{-7/2}), \\
		A_{1,3,2,5,4,6} &= O(D^{-7/2}), \quad
		A_{1,4,2,5,3,6} = O(D^{-7/2}).
\end{align}
For reference, the length-8 coefficients are
\begin{align}
	A_{1,2,3,4,5,6,7,8} &= \frac{1}{4D^{2}} + \frac{2}{3D^{3}} + O(D^{-4}), \quad
	A_{1,2,3,4,5,7,6,8} = \frac{1}{4D^{3}} + O(D^{-4}), \\
	A_{1,2,3,4,5,8,6,7} &= \frac{1}{4D^{2}} + \frac{5}{6D^{3}} + O(D^{-4}), \quad
	A_{1,2,3,5,4,8,6,7} = \frac{1}{4D^{3}} + O(D^{-4}), \\
	A_{1,2,3,7,4,8,5,6} &= \frac{1}{4D^{3}} + O(D^{-4}), \quad
	A_{1,2,3,8,4,6,5,7} = \frac{1}{4D^{3}} + O(D^{-4}), \\
	A_{1,2,3,8,4,7,5,6} &= \frac{1}{4D^{2}} + \frac{11}{12D^{3}} + O(D^{-4}), 
\end{align}
where the other length-8 coefficients are  $O(D^{-4})$. 
Then we further derive the large $D$ expansion of the concrete moments. 
Below are some explicit results for $L\leq 8$: 
\begin{itemize}
\item
$\langle\tr X^{2n}\rangle$ type
\begin{align}
		\langle \tr X^2\rangle &= \frac{1}{\sqrt{2}}D^{-1/2} + \frac{7}{6\sqrt{2}}D^{-3/2} + O(D^{-5/2}),\label{large-D-XX}\\
		\langle \tr X^4\rangle &= D^{-1} + \frac{5}{2}D^{-2} + O(D^{-3}),
		\label{large-D-XXXX}\\
		\langle \tr X^6\rangle &= \frac{5}{2\sqrt{2}}D^{-3/2} + \frac{39}{4\sqrt{2}}D^{-5/2} + O(D^{-7/2}),\\
		\langle \tr X^8\rangle &= \frac{7}{2}D^{-2} + \frac{56}{3}D^{-3} + O(D^{-4}).
\end{align}
\item $\langle\tr X^{2n} Y^{2n'}\rangle$ type
\begin{align}
		\langle \tr X^2 Y^2\rangle &= \frac{1}{2}D^{-1} + D^{-2} + O(D^{-3}),
		\\
		\langle \tr X^4 Y^2\rangle &=\frac{1}{\sqrt{2}}D^{-3/2} + \frac{19}{6\sqrt{2}}D^{-5/2} + O(D^{-7/2}),
		\\
		\langle \tr X^6Y^2\rangle &=\frac{5}{4}D^{-2} + \frac{67}{12}D^{-3} + O(D^{-4}),
		\\
		\langle \tr X^4Y^4\rangle &= D^{-2} + \frac{17}{4}D^{-3} + O(D^{-4}).
\end{align}
\item $\langle\tr X^{2n} (XY)^2\rangle$ type
\begin{align}
		\langle \tr X^0 (XY)^2\rangle &= \frac{1}{2}D^{-2} + O(D^{-3}),
		\\
		\langle \tr X^2 (XY)^2\rangle &= \frac{1}{\sqrt{2}}D^{-5/2} + O(D^{-7/2}),
		\\
		\langle \tr X^4 (XY)^2\rangle &= \frac{5}{4}D^{-3} + O(D^{-4}).
\end{align}
\item $\langle\tr X^{2n}(YZ)^2\rangle$ type
\begin{align}
		\langle \tr X^2(YZ)^2\rangle &= \frac{1}{2\sqrt{2}}D^{-5/2} + O(D^{-7/2}),
		\\
		\langle \tr X^4(YZ)^2\rangle &= \frac{1}{2}D^{-3} + O(D^{-4}).
\end{align}
\item $\langle\tr X^{2n}Y^2Z^2\rangle$ type
\begin{align}
		\langle \tr X^2Y^2Z^2\rangle &= \frac{1}{2\sqrt{2}}D^{-3/2} + \frac{5}{4\sqrt{2}}D^{-5/2} + O(D^{-7/2}),
		\\
		\langle \tr X^4Y^2Z^2\rangle &= \frac{1}{2}D^{-2} + \frac{7}{4}D^{-3} + O(D^{-4}).
\end{align}
\item $\langle\tr X^{2n} YZ^2Y\rangle$ type
\begin{align}
		\langle \tr X^2YZ^2Y\rangle &= \frac{1}{2\sqrt{2}}D^{-3/2} + \frac{17}{12\sqrt{2}}D^{-5/2} + O(D^{-7/2}),
		\\
		\langle \tr X^4 YZ^2Y\rangle &= \frac{1}{2}D^{-2} + 2D^{-3} + O(D^{-4}).
\end{align}
\end{itemize}
We use $X,Y,Z,W$ to denote $X^\mu$ with $\mu=1,2,3,4$.

An obvious question is how to organize these explicit results for the concrete words. 
In fact, the above results are already presented in accordance with some one-length trajectories. 
As noted in \cite{Li:2024ggr}, 
the trajectory interpretation of a concrete word may not be unique,  
so the one-length trajectories can intersect. 
Furthermore, they can be embedded in a higher-dimensional space 
through the more intricate multi-length trajectories. 
We postpone the detailed discussion to  section \ref{Analytic}, 
where we revisit these large $D$ results 
from the perspective of analytic trajectories and eigenvalue distributions. 

\section{Positivity bounds}
\label{Positivity}
In this section, we examine the implications of the positivity condition.  
Our main results are the positivity bounds on $\left\langle\tr XX\right\rangle$ and $\left\langle\tr XXXX\right\rangle$ at length cutoffs $L_{\max}=4,6,8,10,12$. 
We briefly discuss the direct procedure of the positivity bootstrap. 
Then we explain how to reduce the computational complexity of the semi-definite programming 
by using the $\mathrm{O}(D)$ irreducible representations (irreps). 
The use of $\mathrm{O}(D)$ irreps also enables us to study the cases of relatively large $D$. 
As a result, we further examine the positivity bounds for relatively large $D$ and compare them with the predictions of the $1/D$ expansion. 
\subsection{Direct procedure}
\label{Direct procedure}
The positivity condition stems from the positivity of the path integral measure \cite{Lin:2020mme,Kazakov:2021lel}. 
More specifically, we consider a linear combination of monomial words with a length cutoff
\begin{equation}\mathcal{O} = \sum_{i} c_i \,\mathcal{O}_i,
\qquad
\mathcal{O}_i \in \{1, X^{\mu_1}, X^{\mu_1}X^{\mu_2},\dots,X^{\mu_1}X^{\mu_2}\cdots X^{\mu_{L_{\max}/2}}\},
\end{equation}
where the coefficients $c_i$ are arbitrary complex numbers. The positivity condition
\begin{equation}
\langle \mathrm{tr}\, \mathcal{O}^\dagger \mathcal{O} \rangle \geq 0 
\end{equation}
implies the positive semi-definiteness of the matrix \(\mathcal{M} \)
\begin{equation}
\mathcal{M}_{i,j} = \langle \mathrm{tr}\,\mathcal{O}_i^\dagger \mathcal{O}_j \rangle,\quad \mathcal {M}  \succeq 0,
\label{positivity}
\end{equation}
where $\mathcal{M}_{i,j}$ denotes the entry of $\mathcal{M}$ in the $i$-th row and $j$-th column.
It is straightforward to enumerate the possible $\mathcal O_i$ 
and deduce the explicit expression for the matrix $\mathcal{M}$. 
After substituting the solutions of the loop equations, 
we determine the region compatible with the positive semi-definiteness condition. 
However, for multi-matrix models, 
the dimension of the matrix $\mathcal{M}$, 
\begin{equation}
\text{dim}(D,L_\text{max})
=\sum_{l=0}^{L_{\text{max}}/2}D^l=\frac{D^{L_\text{max}/2+1}-1}{D-1},
\label{dimension_direct}
\end{equation}
grows rapidly with $L_\text{max}$, 
so this direct approach is doable only for relatively small $D$ and low $L_\text{max}$. 
We manage to derive the positivity bounds using the concrete matrix $\mathcal M$ 
for $D=3,4$ with $L_\text{max}\leq 10$,  
for $D=5,6$ with $L_\text{max}\leq 8$, 
and for $D=7, 8, 9, 10$ with $L_\text{max}\leq 6$. 
The explicit dimensions of $\mathcal M$ at the maximum length cutoffs are
\begin{align}
\text{dim}(3,10)=364,\, \text{dim}(4,10)=1365,\,
\text{dim}(5,8)=781, \,\text{dim}(6,8)=1555,\\
\text{dim}(7,6)=400, \,\text{dim}(8,6)=585, \,
\text{dim}(9,6)=820, \,\text{dim}(10,6)=1111,
\end{align}
which are of order $10^3$ or less. 
The results of the direct procedure also provide independent checks for 
the bounds from the representation-theory-based approach below. 

\subsection{$\mathrm{O}(D)$ basis and factorization}
As suggested in \cite{Lin:2024vvg,Lin:2025srf} (see also \cite{Bachoc:2012,Kazakov:2022xuh}), 
one can use representation theory to significantly reduce the sizes of positive semi-definite matrices. 
In this way, we can derive the positivity bounds more systematically, 
and study their large $D$ behavior. 
To be more specific, 
we use the $\mathrm{O}(D)$ symmetry to block-diagonalize the matrix $M$, which is a standard technique. 
A more subtle aspect is that these smaller matrices take a tensor-product form, 
so the positive semi-definite conditions can be imposed on significantly smaller matrices.  
Furthermore, the sizes of these smaller matrices are associated with 
the multiplicities of the representations, 
so the computational complexity does not grow with $D$ for large enough $D$. 
Below we explain the general procedure step by step:
\begin{enumerate}
	\item \emph{$\mathrm{O}(D)$ irreducible representations as a basis.}
	
	For a fixed length cutoff \(L_{\max}\), the space of  operators
\begin{equation}\label{Vspan}
	V=\mathrm{span}\{1, X^{\mu_1},X^{\mu_1}X^{\mu_2},\dots,X^{\mu_1}X^{\mu_2}\cdots X^{\mu_{L_{\max}/2}}\}
\end{equation}
	carries a representation of \(\mathrm{O}(D)\). 
	We introduce the notation:
\begin{itemize}
	\item $k$ indicates the tensor rank.\footnote{Here  “tensor rank” \(k\) counts the number of free \(\mathrm{O}(D)\) indices of a word
	, which can be smaller than the word length $L$. A word of length \(L\) can contribute to several ranks \(k=L,L-2,L-4,\dots\), depending on the number of pairs of contracted indices. 
	For example, the length \(L=4\) word 
		\(X^{I_1}X^{I_2}X^{I_3}X^{I_4}\) has rank-4, rank-2 (after one contraction), and rank-0 (after two contractions) components. 
		}
	\item $r$ labels the inequivalent irreducible representations (irreps) for a given rank.  
	\item \(m_{k,r}\) denotes the number of equivalent representations, i.e., multiplicity. 
	\item $n_k$ counts the number of possible ways to contract the indices of the basis vectors in \eqref{Vspan}, 
	i.e., $X^{\mu_1}X^{\mu_2}\cdots X^{\mu_{l}}$ with $k\leq l\leq L_\text{max}/2$,   such that $k$ free indices remain.\footnote{The case of no contraction is also taken into account. }
\end{itemize}
	Accordingly, $V$ can be decomposed as
	\(V=\bigoplus_{k,r} V^{(k,r)}\), 	where each  \( V^{(k,r)} \) is an invariant subspace transforming under the $r$th inequivalent irrep at tensor rank \(k\). 
	Any operator \( \mathcal{O} \in V \) can be written as a sum of basis vectors in different irrep sectors 
	\begin{equation} 
	\mathcal{O} = \sum_{k,r,a,\{\mu_i\}} c_{k,r,a}^{\mu_1,\dots,\mu_k}\, \mathcal{O}^{\mu_1,\dots,\mu_k}_{k,r,a} ,
	\end{equation} 
	where \(\mathcal{O}_{k,r,a} \in V^{(k,r)}\) and $a=1,...,n_km_{k,r}$. 
	Note that some $c_{k,r,a}^{\mu_1,\dots,\mu_k}$ are redundant. 
	We then use $\{\mathcal{O}^{\mu_1,\dots,\mu_k}_{k,r,a}\}$ to build the positive semi-definite matrix $M$. 
	More details about the construction of $\mathrm{O}(D)$ irreducible representations can be found in appendix \ref{Appendix A}.

\begin{table}[h]
	\centering
	\begin{tabular}{c|c|c}
		\hline
		\multirow{2}{*}{rank $k$} & \multirow{2}{*}{$r_{\max}(k)$} & \multirow{2}{*}{multiplicity $m_{k,r}$} \\
		&  &   \\
		\hline
		0 & 1  & $\{1\}$ \\
		1 & 1  & $\{1\}$  \\
		2 & 2  & $\{1,1\}$  \\
		3 & 3  & $\{1,2,1\}$  \\
		4 & 5  & $\{1,3,2,3,1\}$ \\
		5 & 7  & $\{1,4,5,6,5,4,1\}$ \\
		6 & 11 & $\{1,5,5,9,10,16,10,9,5,5,1\}$ \\
		\hline
	\end{tabular}
	\caption{The numbers of inequivalent irreducible representations and their multiplicities at rank $k=0,1,\dots,6$. 
	}	\label{tab:rmax-mkr}
\end{table}

\begin{table}[h]
	\centering
	\begin{tabular}{c|c|c|c|c|c}
		\hline
		\multirow{2}{*}{rank $k$}  & \multicolumn{5}{c}{$n_k$} \\
		\cline{2-6}
		& ${L_\text{max}}=2\times 2$ & ${L_\text{max}}=2\times 3$ & ${L_\text{max}}=2\times 4$ & ${L_\text{max}}=2\times 5$ & ${L_\text{max}}=2\times 6$ \\
		\hline
		0 & 2& 2& 5 & 5 & 20 \\
		1 & 1 & 4 & 4 & 19 & 19 \\
		2 & 1 & 1 & 7 & 7 & 52 \\
		3 & 0 & 1 & 1 & 11 & 11 \\
		4 & 0 &  0& 1 & 1 & 16 \\
		5 & 0 & 0 & 0 & 1 & 1 \\
		6 & 0 & 0 &0& 0& 1 \\
		\hline
	\end{tabular}
	\caption{The number of possible ways to contract the indices of the basis operators in \eqref{Vspan} 
	such that $k$ indices remain free. 
	}	\label{tab:nk}
\end{table}

	\item \emph{Block diagonalization.}
	
	Since only operators in the same irrep \((k,r)\) can  form an \(\mathrm{O}(D)\) singlet, 
	the positive semi-definite matrix $M$ takes a block-diagonalized form
\begin{equation}
	M=\bigoplus_{k,r} M^{(k,r)}.
\end{equation}
    As a result, the positivity requirement reduces to the positive semi-definite condition on $M^{(k,r)}$.  
    In each block $M^{(k,r)}$, the basis vectors are given by \(\mathcal{O}^{\mu_1,\dots,\mu_k}_{k,r,a} \) with $a=1,...,n_k m_{k,r}$.  
    The matrix elements are defined as
    \begin{equation}\label{Mkr-matrix-element}
   (M^{(k,r)})_{\{a, \{\mu_1,\dots,\mu_k\}\},\{b,\{\mu_1',\dots,\mu_k'\}\}} 
   =\left\langle \tr\; (\mathcal O_{k,r,a}^{\mu_1,\dots,\mu_k})^\dagger \mathcal O_{k,r,b}^{\mu_1',\dots,\mu_k'}\right\rangle. 
    \end{equation}
    Some matrix elements are redundant because the number of independent components of an irrep is less than $D^k$, but this definition simplifies the discussion below. 
	
	\item \emph{Tensor product factorization.}
	
	A further reduction comes from the factorization structure. 
	As indicated by their Young diagrams, 
	the equivalent representations share similar index-permutation properties. 
	For some concrete rank $k$, we notice that the block \(M^{(k,r)}\)  further factorizes into a tensor product 
\begin{equation}
	M^{(k,r)} =\widehat{ M}^{(k,r)}  \otimes U^{(k,r)},
\end{equation}
	where \(\widehat{ M}^{(k,r)} \) is a \( n_{k}m_{k,r} \times n_{k}m_{k,r} \) matrix 
	and $U^{(k,r)}$ carries the $\mathrm{O}(D)$ indices. 
	In terms of matrix elements, the factorization takes the form
\begin{equation}
	(M^{(k,r)})_{\{a, \{\mu_1,\dots,\mu_k\}\},\{b,\{\mu_1',\dots,\mu_k'\}\}} =(\widehat{ M}^{(k,r)})_{a,b}\,  ( U^{(k,r)})_{\{\mu_1,\dots,\mu_k\},\{\mu_1',\dots,\mu_k'\}},
\end{equation}
where $a,b=1,\dots, n_{k}m_{k,r} $. 
If the multiplicity of the \((k,r)\) irrep is greater than 1, i.e., \(m_{k,r}>1\), 
	we may need to rearrange the order of some $\mathrm{O}(D)$ indices in accordance with 
	their Young tableaux $T_{k,r,a}$.

	The factorization form 
	is not unique. 
	It turns out that we can choose a basis such that \( U^{(k,r)} \) is  positive semi-definite.\footnote{If \( U^{(k,r)} \) is neither positive semi-definite nor negative semi-definite, 
	then all the eigenvalues of $\widehat{M}^{(k,r)}$ should be zero. } 
	Then we fix the normalization of $U^{(k,r)}$ by the trace condition $\Tr U^{(k,r)}=1$. 
	According to the definition in \eqref{Mkr-matrix-element}, 
	the contraction of the matrix elements of $M^{(k,r)}$ with $\delta_{\mu_1,\mu_1'}\dots\delta_{\mu_k,\mu_k'}$ implies 
	\begin{equation}
	\label{Mhat-definition}
	(\widehat{M}^{(k,r)})_{a,b}=\langle \tr\; (\mathcal O_{k,r,a}^{\mu_1\dots\mu_k})^\dagger \mathcal O_{k,r,b}^{\mu_1\dots\mu_k}\rangle, 
	\end{equation}
	where repeated $\mathrm{O}(D)$ indices indicate summation over $\mu_i=1,2,\dots D$. 
	Note that $k,r$ are fixed labels. 
	In the end, the positivity condition on the full matrix blocks  further reduces to 
	the positivity requirement on the much smaller matrices
	\footnote{In \cite{Kazakov:2022xuh,Lin:2024vvg,Lin:2025srf,Bachoc:2012}, the authors did not explicitly discuss the tensor-product factorization, but our formulation is strongly inspired by these works. 
	}
	\begin{equation}\label{positivity-Mhat}
	M^{(k,r)}\succeq 0 \quad \Longrightarrow \quad  \widehat{M}^{(k,r)}\succeq  0.
	\end{equation}
	
\end{enumerate}

The matrix dimensions can be compared with those from the direct procedure.
For a given cutoff length $L_{\max}$, the total number of positive semi-definite matrices is 
$\sum_{k=0}^{L_{\max}/2} r_{\max}(k)$, 
where $r_{\max}(k)$ is the number of inequivalent irreducible representations at rank $k$. 
For the $r$-th irrep of rank $k$, 
the dimension of the positive semi-definite matrix \(\widehat{M}^{(k,r)}\) is given by $m_{k,r} n_k$. 
At our maximum length cutoff $L_{\max}=12$, there are $1+1+2+3+5+7+11=30$ matrices. 
Their individual dimensions are 
\begin{align}
&\{20\},\{19\}, \{52, 52\}, \{11, 22, 11\}, \{16, 48, 32, 48, 16\},
\\ 
 &\{1, 4, 5, 6, 5, 4, 1\},\{1, 5, 5, 9, 10, 16, 10, 9, 5, 5, 1\},
\end{align}
whose total sum is $449$. 
According to \eqref{dimension_direct}, the dimensions of the matrix $\mathcal M$ in the direct procedure are
\begin{equation}
\{1\,093, \;5\,461,\;19\,531, \;55\,987, \;137\,257, \;299\,593,\; 597\,871,\; 1\,111\,111\}
\end{equation}
for $D=3,4,\dots,10$. 
The use of $\mathrm{O}(D)$ irreducible representations significantly reduces the sizes of the positive semi-definite matrices, 
making the cases of larger $L_{\max}$ and larger $D$ more computationally tractable.

Below we use some simple examples to illustrate the general procedure, 
and present our main results,  
the positivity bounds for $\langle \tr XX \rangle$ and $\langle \tr XXXX \rangle$. 
For the concrete cases considered in the direct procedure, 
	we verify that the positivity bounds are the same as those from $ \widehat{M}^{(k,r)}\succeq  0$ up to numerical errors. 
	Using the above $\mathrm{O}(D)$-based procedure, we further derive many positivity bounds that 
	are beyond the scope of the direct procedure. 

\subsection{Bounds at different length cutoffs}
We begin with the simple cases of $L_{\max}=4,6$, 
where the procedure is carried out explicitly and analytic bounds are obtained. 
We then present the numerical bounds associated with the length cutoffs $L_{\max}=8,10,12$.
\subsubsection{Analytic bounds at $L_{\max}=4$}
Our first example is the case of $L_{\max}=4$. 
In the direct procedure, the positive semi-definite matrix is associated with the basis vectors 
$\{1, X^{\mu_1}, X^{\mu_1}X^{\mu_2}\}$
\begin{equation}
\mathcal{M}=
	\begin{pmatrix}
		1 & \textbf{0} &\mathcal{M}_{1,3} \\
		\textbf{0} & \mathcal{M}_{2,2} & \textbf{0} \\
		\mathcal{M}_{3,1}& \textbf{0} & \mathcal{M}_{3,3}
	\end{pmatrix}
	\label{explicitmatrixL4}.
\end{equation} 
The nonvanishing matrix elements are given by
\begin{equation}
	\begin{aligned}
		(\mathcal{M}_{1,3})_{1,\,\{\mu_1', \mu_2'\}}=\left\langle \operatorname{tr} X^{\mu_1'} X^{\mu_2'}\right\rangle, \quad
	(\mathcal{M}_{3,1})_{\{\mu_1,\mu_2\},\,1}&=\left\langle \operatorname{tr} X^{\mu_1} X^{\mu_2} \right\rangle ,\\
	(\mathcal{M}_{2,2})_{\mu_1,\,\mu_1'}=\left\langle \operatorname{tr} X^{\mu_1} X^{\mu_1'} \right\rangle,  \quad
	(\mathcal{M}_{3,3})_{\{\mu_1,\mu_2\},\,\{\mu_1',\mu_2'\}}&= \left\langle \operatorname{tr} X^{\mu_1} X^{\mu_2} X^{\mu_1'} X^{\mu_2'} \right\rangle,
	\end{aligned}
\end{equation}
where 
$\mathcal{M}_{1,3}\in \mathbb{R}^{1\times D^2},
		\mathcal{M}_{2,2}\in \mathbb{R}^{D\times D},
		\mathcal{M}_{3,1}\in \mathbb{R}^{D^2\times 1},
		\mathcal{M}_{3,3}\in \mathbb{R}^{D^2\times D^2}$. 
Using the explicit matrix $\mathcal M$, 
we derive the positivity bounds for $D=3,4,\dots,10$.

To further investigate the general $D$ behavior, 
we use the $\mathrm{O}(D)$ irrep basis to construct the positive semi-definite matrix. 
The length-$0,1$ words are irreducible, 
but the length-$2$ covariant word admits the decomposition
\begin{equation}
	X^{\mu_1}X^{\mu_2}
=\underbrace{\tfrac{1}{D}X^{\mu_3}X^{\mu_3}}_{\text{S, rank-0: } \mathcal{O}_{0,1,2}}\,\delta^{\mu_1\mu_2}
\;+\;
\underbrace{\tfrac12\{X^{\mu_1},X^{\mu_2}\}-\tfrac{1}{D}\delta^{\mu_1\mu_2}X^{\mu_3}X^{\mu_3}}_{\text{T, rank-2: }\mathcal{O}_{2,1,1}^{\mu_1\mu_2}}
\;+\;
\underbrace{\tfrac12[X^{\mu_1},X^{\mu_2}]}_{\text{A, rank-2: }\mathcal{O}_{2,2,1}^{\mu_1\mu_2}}
,
\label{rank-2-decomposition}
\end{equation}
where $\{X,Y\}=XY+YX$. Accordingly, the positive semi-definite matrix $M$ reads
\begin{equation}
	M = M^{(0,1)} \oplus M^{(1,1)} \oplus M^{(2,1)}\oplus M^{(2,2)}.
	\label{diagL4}
\end{equation}
According to the basis vectors $\{1, X^\mu X^\mu/D\}$, the rank-0 block $M^{(0,1)}$ is given by a 2 by 2 matrix. 
The rank-1 block $M^{(1,1)}$ is the same as $\mathcal{M}_{2,2}$ in \eqref{explicitmatrixL4}. 
The rank-2 part is further block diagonalized into 
the matrix $ M^{(2,1)}$ for the traceless symmetric (T) sector 
and the matrix $M^{(2,2)}$ for the antisymmetric (A) sector. 
The positivity of \eqref{explicitmatrixL4} reduces to the positivity of each block in \eqref{diagL4}. 
Below we examine these blocks one by one: 
\paragraph{Rank-0 sector.}
For the singlet representations ($k=0$, $r_\text{max}=1$), we have $m_{k=0,r=1}=1$ and $n_{k=0}=2$. 
The positive semi-definite matrix reads
\begin{equation}
	\begin{aligned}
		M^{(0,1)} &= 
	\begin{pmatrix}
		1 & \frac{1}{D}\left\langle \operatorname{tr} X^{\mu_1'} X^{\mu_1'} \right\rangle \\
		\frac{1}{D}	\left\langle \operatorname{tr} X^{\mu_1} X^{\mu_1} \right\rangle &  \frac{1}{D^2}\left\langle \operatorname{tr} X^{\mu_1} X^{\mu_1} X^{\mu_1'} X^{\mu_1'} \right\rangle
	\end{pmatrix}\\
	&=
	\begin{pmatrix}
		1 & A_{1,2} \\
		A_{1,2} & \dfrac{(1+D) A_{1,2,3,4} + A_{1,3,2,4}}{D}
	\end{pmatrix}\,.
	\end{aligned}
\end{equation}
Substituting the solution for $A_{1,3,2,4}$ in \eqref{A1324-solution}, 
the positive semi-definite condition for $D\geq 3$ implies
\begin{equation}\label{lmax4rank0}
	A_{1,2,3,4} \ge \frac{1 + 2D(D-1)\,A_{1,2}^{2}}{2(D-1)(D+2)},
\end{equation}
so the corresponding boundary of the allowed region takes a quadratic form. 

\paragraph{Rank-1 sector.}
For the vector representation ($k=1$, $r_\text{max}=1$), we have 
$m_{k=1,r=1}=1$ and $n_{k=1}=1$, so $a_\text{max}=1$. 
The matrix elements of the corresponding block $M^{(1,1)} $ are
\begin{equation}
	(M^{(1,1)} )_{\mu_1,\mu_1'}= \left\langle \operatorname{tr} X^{\mu_1} X^{\mu_1'} \right\rangle
	=A_{1,2}\,\delta^{\mu_1 \mu_1'}. 
\end{equation}
It is clear that the matrix takes a factorized form
\begin{equation}
	M^{(1,1)}= A_{1,2}\,\mathbb{I}_{D},
\end{equation}
where \(\mathbb{I}_{D}\) is the $D\times D$ identity matrix. 
We obtain
\begin{equation}\label{lmax4rank1}
	A_{1,2}\ge0,
\end{equation}
which is also the simplest positivity bound associated with $L_\text{max}=2$.

\begin{figure}[tbp]
	\centering 
	\includegraphics[width=1\textwidth]{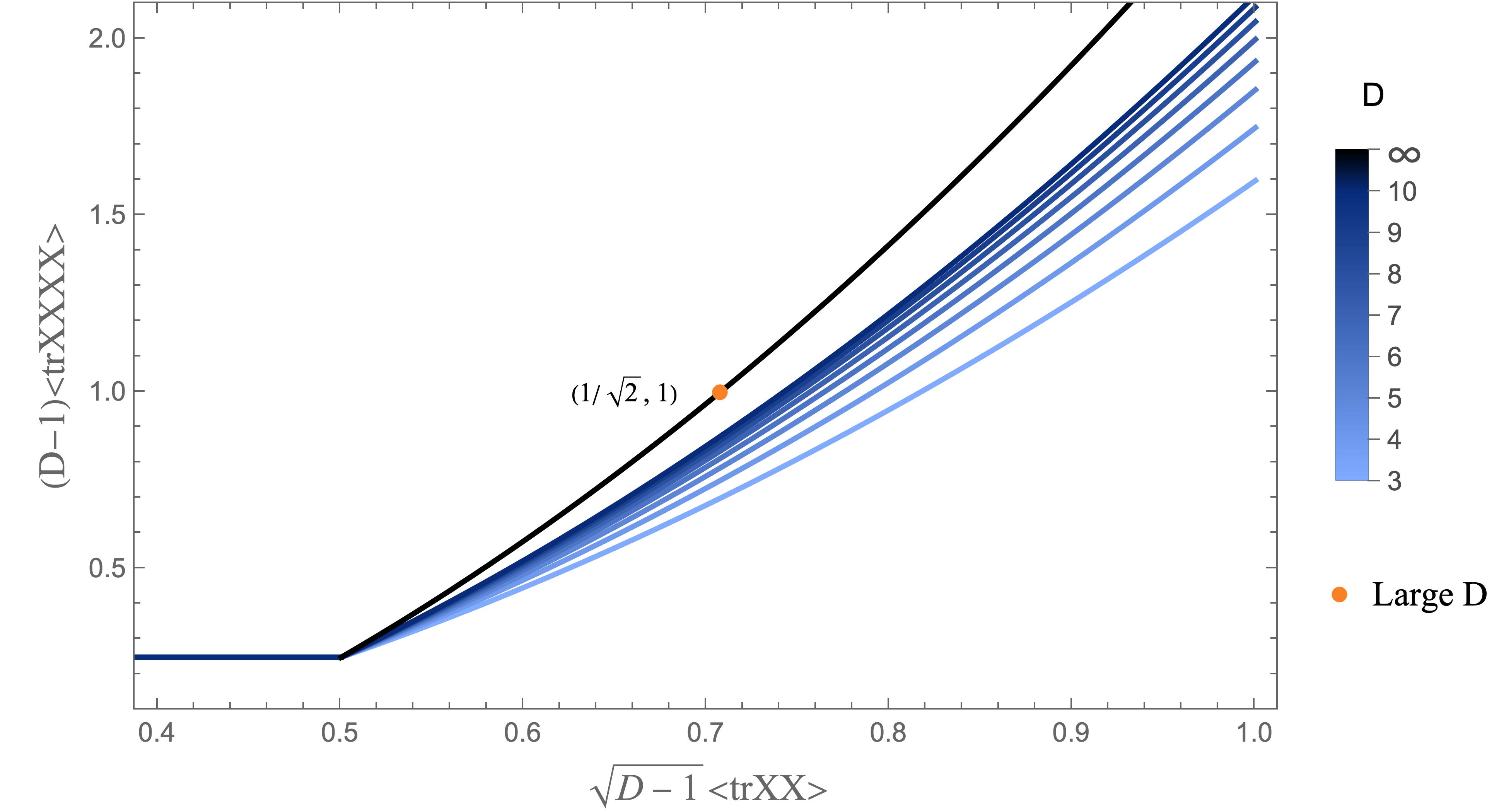}
	\caption{
		The positivity bounds at $L_{\max}=4$. From light to dark, the  blue curves correspond to the lower bounds of $\left\langle\tr XXXX\right\rangle$ for $\left\langle\tr XX\right\rangle\geq 0$ with $D=3,4,\dots,10,\infty$. 
		The black solid curve indicates the large $D$ limit of the lower bound.
		The leading prediction of the large $D$ expansion (orange point) is on the black curve. 
		We insert some $D$-dependent factors so that the large $D$ results are associated with finite coordinates.  
	}
			\label{fig:Lmax4}
\end{figure}

\paragraph{Rank-2 sector.} 
As mentioned above, there are two different irreps at rank $k=2$, i.e., $r_\text{max}=2$. 
We also have $m_{k=2,r=1}=m_{k=2,r=2}=1$ and $n_{k=2}=1$, so $a_\text{max}=1$ for both $r=1,2$. 
The two inequivalent rank-$2$ irreps are of the traceless-symmetric type \(\ytableausetup{boxsize=10pt}\scalebox{1}{$\begin{ytableau}1&2\end{ytableau}$}\)
and the antisymmetric type \(\scalebox{1}{$\begin{ytableau}1\\2\end{ytableau}$}\). 
The corresponding matrix elements are
\begin{equation}
\text{T}:\quad
	(M^{(2,1)})_{\{\mu_1,\mu_2\},\{\mu_1',\mu_2'\}} =
	\left\langle \operatorname{tr}\; (\mathcal{O}_{2,1,1}^{\mu_1 \mu_2})^\dagger \mathcal{O}_{2,1,1}^{\mu'_1 \mu'_2} \right\rangle,
\end{equation}
\begin{equation}
\text{A}:\quad
(M^{(2,2)})_{\{\mu_1,\mu_2\},\{\mu_1',\mu_2'\}}  =
	\left\langle \operatorname{tr}\;(\mathcal{O}_{2,2,1}^{\mu_1 \mu_2})^\dagger \mathcal{O}_{2,2,1}^{\mu'_1 \mu'_2} \right\rangle.
\end{equation}
The two matrices can be factorized as
\begin{equation}
M^{(2,1)}  = \widehat{M}^{(\mathrm{2,1})} \otimes U^{(\mathrm{2,1})}\,,\quad
M^{(2,2)}  = \widehat{M}^{(\mathrm{2,2})} \otimes U^{(\mathrm{2,2})}\,.
\end{equation}
As $a_\text{max}=1$, the smaller matrices have only one element
\begin{align}
(\widehat{M}^{(\mathrm{2,1})})_{1,1} &\equiv
		\left\langle \operatorname{tr}\;(\mathcal{O}_{2,1,1}^{\mu_1 \mu_2})^\dagger  \mathcal{O}_{2,1,1}^{\mu_1 \mu_2} \right\rangle 
		=\dfrac{1}{2}(D+2)(D-1) 
	\,\bigl(A_{1,2,3,4}+A_{1,3,2,4}\bigr),
		\\
		(\widehat{M}^{(\mathrm{2,2})})_{1,1} &\equiv
\left\langle \operatorname{tr}\; (\mathcal{O}_{2,2,1}^{\mu_1 \mu_2})^\dagger  \mathcal{O}_{2,2,1}^{\mu_1 \mu_2} \right\rangle 
	=\dfrac{1}{2}(D-1)\,D \,
		\bigl(A_{1,2,3,4}-A_{1,3,2,4}\bigr)	\,.
\end{align}
If we use the index ordering \(\{\mu_1,\mu_2\}=
\{1,1\},\dots,\{D,D\},\{1,2\},\{2,1\},\{1,3\},\dots,\{D-1,D\},\{D,D-1\}\), the $U$ matrices are
\begin{equation}
U^{(\mathrm{2,1})}
=\frac{1}{(D-1)(D+2)}\begin{pmatrix}
U^{(\mathrm{2,1})}_{1,1}  & \textbf{0}\\[3pt]
\textbf{0} &U^{(\mathrm{2,1})}_{2,2}  \,
\end{pmatrix},
\quad
U^{(\mathrm{2,2})}
=\frac{1}{D(D-1)}\begin{pmatrix}
\textbf{0}  & \textbf{0}\\[3pt]
\textbf{0} &\; U^{(\mathrm{2,2})}_{2,2}
\end{pmatrix},
\end{equation}
where the nonzero matrix elements are given by
\begin{equation}
(U^{(\mathrm{2,1})}_{1,1} )_{\mu_1,\mu_2} =2\,\delta_{\mu_1\mu_2} - 2/{D},
\quad \mu_1,\mu_2=1,\dots,D,
\end{equation}
\begin{equation}
U^{(\mathrm{2,1})}_{2,2} =\mathbb{I}_{D(D-1)/2} \otimes 
\begin{pmatrix}
	1&\;1\\[3pt]
	1&\;1
\end{pmatrix}, \; U^{(\mathrm{2,2})}_{2,2}=\mathbb{I}_{D(D-1)/2} \otimes 
\begin{pmatrix}
	1&-1\\[3pt]
	-1&1
\end{pmatrix}.
\end{equation}
For $D\ge 3$, the constant matrices $U^{(2,1)}$, $U^{(2,2)}$ are positive semi-definite, 
so the positivity requirement on $M^{(2,1)} $ and $M^{(2,2)} $ reduces to that on $\widehat{M}^{(2,1)}$ and $\widehat{M}^{(2,2)}$
\begin{equation}
A_{1,2,3,4}+A_{1,3,2,4}\geq 0\,,\quad
A_{1,2,3,4}-A_{1,3,2,4}\geq 0. 
\end{equation}
Substituting \eqref{A1324-solution} into these constraints, the second one is automatically satisfied, but the first one gives an analytic bound
\begin{equation}\label{lmax4rank2}
	A_{1,2,3,4} \;\ge\; \frac{1}{4(D-1)}\,.
\end{equation}

\paragraph{Analytic bounds.}
The above analyses imply the analytic bounds: 
\begin{equation}\label{eq:L4-bound-length-2}
\langle \tr XX\rangle\geq 0\,,
\end{equation}
and
\begin{equation}
	\langle \tr XXXX\rangle\ \ge\
	\begin{cases}
		\dfrac{1}{4\,(D-1)}, 
		&\frac{1} {2\sqrt{D-1}}\ge \langle \tr XX\rangle\ge 0\,,\\
		\dfrac{6D\,\langle \tr XX\rangle^2 - 1\,}{ 2(D+2)}\,,
		& \langle \tr XX\rangle>\frac{1} {2\sqrt{D-1}}\;,
	\end{cases}
	\label{eq:L4-bound}
\end{equation}
where we have used \eqref{singletdecompositionlength4}, \eqref{XXXX-eq-2}, \eqref{lmax4rank0}, \eqref{lmax4rank1}, and \eqref{lmax4rank2}. 
The  first constraint comes from the positivity of the rank-1 sector. 
The two cases below are associated with the rank-2 traceless-symmetric sector and the rank-0 sector. 
In figure \ref{fig:Lmax4}, we present the lower bounds for $\langle \tr XXXX\rangle$
with $D=3,...,10$.
At large $D$, the lower bound of $\langle \tr XXXX\rangle$ converges to 
\begin{equation}
	(D-1)\langle \tr XXXX\rangle\ \ge\
	\begin{cases}
		1/4, 
		&1/2\ge \sqrt{D-1}\,\langle \tr XX\rangle\ge 0\,,\\
		3(D-1)\,\langle \tr XX\rangle^2 - 1/2\,,
		& \sqrt{D-1}\,\langle \tr XX\rangle>1/2\;.
	\end{cases}
\end{equation}
The leading prediction of the large $D$ expansion is on the boundary associated with the second case. 

It is also interesting to consider the positivity bounds without using the constraints from the loop equations. 
We obtain these universal bounds 
\begin{equation}
\langle \tr XX\rangle\geq0,\quad
\langle \tr XXXX\rangle\ge  \langle \tr XX\rangle^2. 
\end{equation}
The former is identical to \eqref{eq:L4-bound-length-2}, 
but the latter is weaker than \eqref{eq:L4-bound} except at $\langle \tr XX\rangle=\frac{1} {2\sqrt{D-1}}$.

\subsubsection{Analytic bounds at $L_{\max}=6$}
In the second example, we consider the case of $L_{\max}=6$, where inequivalent irreps occur at rank 3.
As the new basis operators are of length $3$, 
they do not appear in the rank-0 and rank-2 blocks, 
so these two cases are the same as those at $L_{\max}=4$.
On the other hand, the rank-$1$ block is enlarged and leads to stronger positivity constraints.  
The cases of rank-$1$ and rank-$3$ are discussed below. 
\paragraph{Rank-1 sector.}
The basis operators in the rank-$1$ sector are
\begin{equation}
\{\mathcal{O}_{1,1,1}^{\mu},...,\mathcal{O}_{1,1,4}^{\mu}\}=
\left\{\,X^{\mu},\ \frac{1}{D} X^{\mu} X^{\nu} X^{\nu},\ \frac{1}{D} X^{\nu} X^{\mu} X^{\nu},\ \frac{1}{D} X^{\nu} X^{\nu} X^{\mu}\,\right\},
\label{rank-1-decomposition}
\end{equation}
and we have $n_{k=1}=4$.  
The multiplicity remains $m_{k=1,r=1}=1$, so $a_\text{max}=4$. 
The $1/D$ factors are included for proper normalization. 
The corresponding matrix can be factorized as
\begin{equation}
M^{(1,1)} = \widehat{M}^{(1,1)} \otimes(D^{-1} \mathbb{I}_{D}),
\end{equation}
where $ \widehat{M}^{(1,1)}$ is a $4\times 4$ matrix. 
As the matrix elements are more involved, we do not write their explicit expressions. 
The corresponding positivity bound is 
\begin{equation}\label{Lmax6-A12}
A_{1,2} \geq \dfrac{1}{\sqrt{2(D-1)}}.
\end{equation}
When the lower bound of $A_{1,2}$ is saturated, 
the value of $A_{1,2,3,4} $ is fixed by the positive semi-definite condition to be
\begin{equation}\label{0-eigenvalue}
A_{1,2,3,4} =\frac{1}{2(D-1)}\,,\quad\text{if}\quad A_{1,2}=\dfrac{1}{\sqrt{2(D-1)}}.
\end{equation}
If $A_{1,2,3,4}$ takes other values, the positive semi-definite condition is always violated.  
Furthermore, we notice that $\widehat{M}^{(1,1)}$ has at least one vanishing eigenvalue at the special point\footnote{According to the characteristic polynomial in $ \widehat{M}^{(1,1)}$, 
the product of three eigenvalues is proportional to 
\begin{equation}
\left(A_{1,2,3,4}-\frac 1{2(D-1)}\right)^2
\end{equation}
with a negative coefficient. 
As the eigenvalues of $ \widehat{M}^{(1,1)}$ are non-negative, their product is also non-negative. 
Then the positive semi-definite condition implies
\begin{equation}
A_{1,2,3,4}=\frac 1{2(D-1)},
\end{equation}
and at least one eigenvalue of $ \widehat{M}^{(1,1)}$ is zero. 
}
\begin{equation}\label{0-eigenvalue-sol}
(A_{1,2},\,A_{1,2,3,4})=\left(\dfrac{1}{\sqrt{2(D-1)}},\,\frac{1}{2(D-1)}\right).
\end{equation}
Curiously, the special solution \eqref{0-eigenvalue-sol} is consistent with the leading terms from the large $D$ expansion and 
furthermore coincides with the leading ansatz solution \eqref{leading-ansatz-sol} of the analytic trajectory bootstrap in section \ref{sec:Leading ansatz}. 
There are intimate connections among special points that saturate the positivity bound, the existence of null eigenvectors~\cite{Lin:2020mme, Li:2022prn,Li:2023nip,Guo:2023gfi}, and good analytic properties~\cite{Li:2023ewe,Li:2024rod}.

\paragraph{Rank-3 sector.} We have three inequivalent irreps at rank $k=3$, so $r_{\max}=3$. 
Their multiplicities $m_{k,r}$ are $(m_{3,1}, m_{3,2},m_{3,3})=(1,2,1)$. 
Since $k=L_\text{max}/2$, we have $n_{k=3}=1$. 
They correspond to the Young tableaux $T_{k,r,a}$
\[
\begin{aligned}
	T_{3,1,1} &= \scalebox{1}{\begin{ytableau}
			1 & 2 & 3
	\end{ytableau}}\,,
	\quad
	T_{3,2,1} &= \scalebox{1}{\begin{ytableau}
			1 & 2 \\
			3
	\end{ytableau}}\,,
	\quad
	T_{3,2,2} &= \scalebox{1}{\begin{ytableau}
			1 & 3 \\
			2
	\end{ytableau}}\,,
	\quad
	T_{3,3,1} &= \scalebox{1}{\begin{ytableau}
			1 \\
			2 \\
			3
	\end{ytableau}}\,.
\end{aligned}
\]
The middle two Young tableaux $T_{3,2,1}$ and $T_{3,2,2}$  correspond to equivalent irreps, 
but they exhibit different permutation properties with respect to the $\mathrm{O}(D)$ indices $(\mu_1, \mu_2, \mu_3)$.\footnote{There are two common definitions of the action of a Young symmetrizer associated with a Young tableau on tensors. Consider a permutation $P$ of the form
	\begin{equation}
		P =
		\begin{pmatrix}
			1 & 2 & \cdots & n \\
			q_1 & q_2 & \cdots & q_n
		\end{pmatrix}.
	\end{equation}
	(1) $P$ means that the index originally in position $i$ moves to position $q_i$. 
	(2) $P$ means that the index $\mu_i$ turns into $\mu_{q_i}$. 
	The two definitions should lead to the same positivity bounds.
	We have checked this equivalence explicitly for $D=3,\ldots,10$ at 
	$L_{\max}=4,6,8,10$, and for $D=3$ at $L_{\max}=12$. 
		In the numerical implementation at $L_{\max}=12$, we use the second convention.
The first convention is adopted in the main text.
	\label{fn-action}} 
	For the Young tableau $T_{3,2,1}$, the Young symmetrizer first symmetrizes the indices originally in the first and second positions, and then antisymmetrizes	the indices originally in the first and third positions.
	In contrast, the Young tableau $T_{3,2,2}$ first symmetrizes the indices originally in the first and third positions, and then antisymmetrizes the 	indices originally in the first and second positions.
To preserve the factorization structure, we need to adjust the order of some $\mathrm{O}(D)$ indices. 
A simple way is to exchange the second and third indices of the mixed-symmetry operator 
$\tilde{\mathcal{O}}_{3,2,2}^{\mu_1,\mu_2,\mu_3} $ defined by $T_{3,2,2}$,  
i.e., the second basis operator is 
$\mathcal{O}_{3,2,2}^{\mu_1,\mu_2,\mu_3}=\tilde{\mathcal{O}}_{3,2,2}^{\mu_1,\mu_3,\mu_2}$. 
The corresponding two by two matrix is 	
\begin{equation}
		\widehat{M}^{(3,2)}=\frac{2D(D+2)(D-2)}{9}\begin{pmatrix}
			2\,\tilde A &
			\tilde A \\[6pt]
			\tilde A&
			-A_{1,2,3,5,4,6}+2A_{1,2,3,6,4,5}-2A_{1,3,2,5,4,6}+A_{1,4,2,5,3,6}
		\end{pmatrix},
\end{equation} 
where the combination $\tilde A$ is defined as
\begin{equation}
	\tilde A=A_{1,2,3,5,4,6}+A_{1,2,3,6,4,5}-A_{1,3,2,5,4,6}-A_{1,4,2,5,3,6}\,.
\end{equation}
On the other hand,  $\widehat{M}^{(3,1)}$ for the totally symmetric sector and $\widehat{M}^{(3,3)}$ for the totally antisymmetric sector are $1\times 1$ matrices
\begin{align}
(\widehat{M}^{(3,1)})_{1,1}&=\dfrac{D(D+4)(D-1)}{6}\,
(2A_{1,2,3,5,4,6}+A_{1,2,3,6,4,5}+2A_{1,3,2,5,4,6}+A_{1,4,2,5,3,6})\,	,
			\\
(\widehat{M}^{(3,3)})_{1,1}&=
			\dfrac{D(D-1)(D-2)}{6}\,
			(-2A_{1,2,3,5,4,6} + A_{1,2,3,6,4,5} + 2A_{1,3,2,5,4,6}- A_{1,4,2,5,3,6})\,.
\end{align}
These matrix elements can be directly derived from \eqref{Mhat-definition}. 
For $D>4$, the positive semi-definite condition implies
\begin{equation}
\frac{ A_{1,2}}{D-1}\geq
A_{1,2,3,4,5,6} +\frac{D}{2} A_{1,2,3,5,4,6}-\frac{D+2}{2} A_{1,3,2,5,4,6}\,,
\end{equation}
and
\begin{equation}
A_{1,3,2,5,4,6} \le
	\begin{cases}
		\frac{D+2}{D-4}A_{1,2,3,5,4,6} , 
		&A_{1,2,3,5,4,6} \leq 0\,,\\
		A_{1,2,3,5,4,6} \,,
		&A_{1,2,3,5,4,6} > 0\;.
	\end{cases}
\end{equation}

Above we use the rank-3 mixed-symmetry sector to illustrate how to preserve the factorization structure by adjusting the order of $\mathrm{O}(D)$ indices. 
Since the dimensions of these matrices are relatively low, 
the positivity condition on the rank-3 sectors does not 
lead to a bound for $\langle \tr XX\rangle$ or $\langle \tr XXXX\rangle$.\footnote{If we combine the rank-3 positivity condition with those of lower ranks, 
we obtain some useful constraints for the length-6 moments, 
such as some lower bounds for $A_{1,2,3,4,5,6}$ and $\langle \tr XXXXXX\rangle$. }
As $L_{\max}$ increases,  the rank-3 sectors also provide crucial constraints. 

\begin{figure}[tbp]
	\centering 
	\includegraphics[width=1\textwidth]{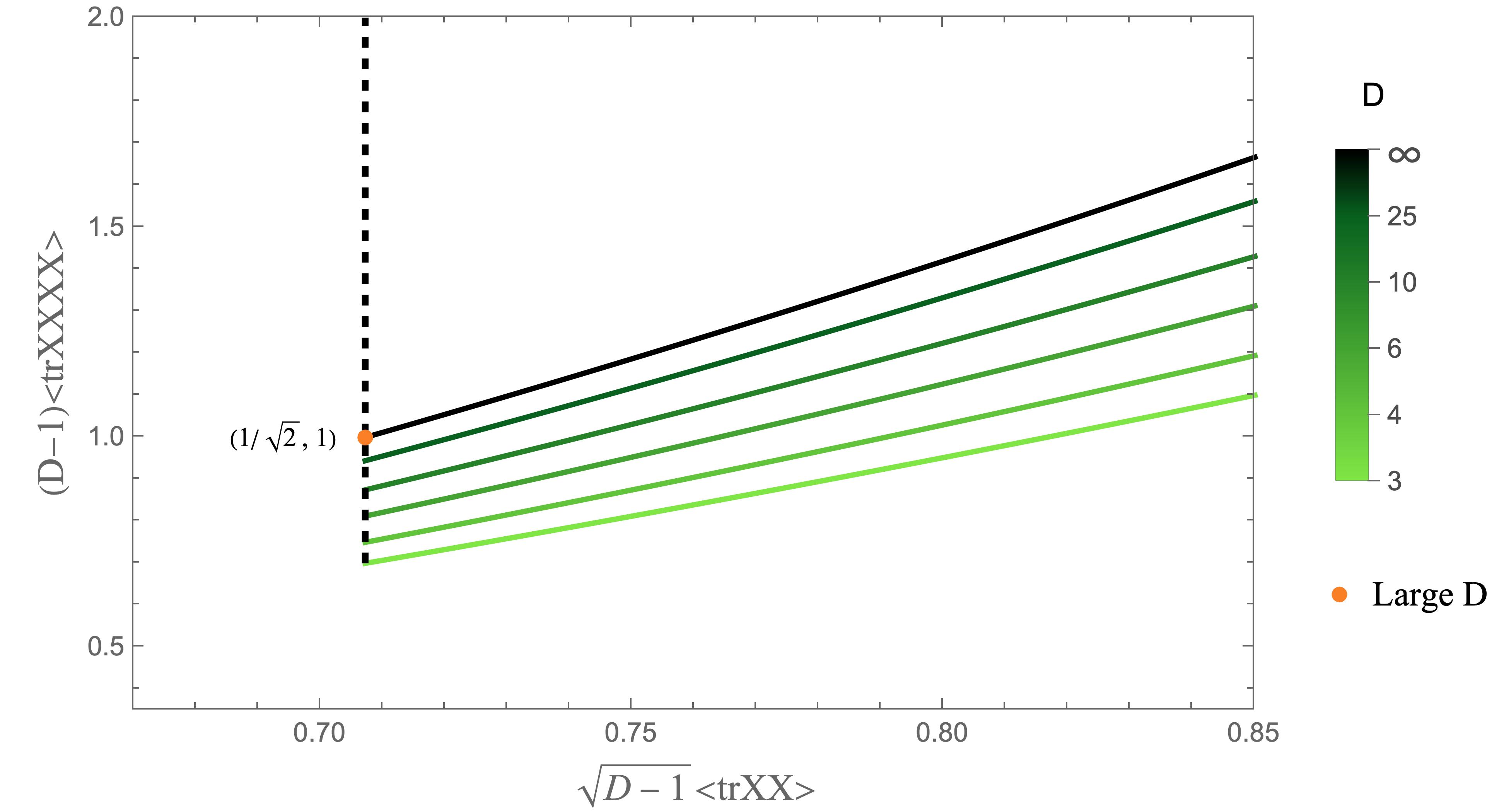}
	\caption{
		 The positivity bounds at $L_{\max}=6$. From light to dark, the green curves correspond to the lower bounds of $\left\langle\tr XXXX\right\rangle$ for $D=3,4, 6,10,25, \infty$.  The orange dot represents the leading prediction of the large $D$ expansion. 
		 The dashed line indicates that the left boundary violates the positive semi-definite condition except for the orange point. 
		 As in figure \ref{fig:Lmax4}, we insert some $D$-dependent factors for the clarity of  the $D\rightarrow\infty$ results. 
	}
			\label{fig:Lmax6}
\end{figure}

\paragraph{Analytic bounds.}
After the $\mathrm{O}(D)$ irrep reduction, 
we substitute the solutions of the $L_\mathrm{max}=6$ loop equations into the positivity matrices. 
For $\langle \tr XX\rangle$, the constraint \eqref{Lmax6-A12} from the rank-1 sector 
implies the analytic bound 
\begin{equation}
{\langle \tr XX\rangle} \geq  {\frac{1}{\sqrt{2(D - 1)}}}. 
\end{equation}
For $\langle \tr XXXX\rangle$, we use \eqref{XXXX-eq-2}, \eqref{lmax4rank0}, \eqref{Lmax6-A12} and \eqref{0-eigenvalue} to obtain
\begin{equation}{\langle \tr XXXX\rangle} \left\{ {\begin{array}{*{20}{l}}
		={\dfrac{1}{{D - 1}} ,}&\quad{\langle \tr XX\rangle} =  {\frac{1}{\sqrt{2(D - 1)}}}  ,\\ 
		 \ge\dfrac{6D\,\langle \tr XX\rangle^2 - 1\,}{ 2(D+2)},&\quad{\langle \tr XX\rangle} > {\frac{1}{ \sqrt{2(D - 1)}}}  .
\end{array}} \right.\end{equation}
The first case comes from the rank-1 sector, 
while the second case is associated with the rank-0 sector. 
In the large $D$ limit, the boundary of the allowed region converges to 
\begin{equation}
\langle \tr XXXX\rangle= 3 \langle \tr XX\rangle^2-\frac {1}{2D}+O(D^{-2})
\quad \text{with}\quad
\langle \tr XX\rangle\geq \frac {1}{\sqrt{2D}}+O(D^{-\frac 3 2}).
\end{equation}
The left end point reproduces the leading terms of the large $D$ expansion 
in \eqref{large-D-XX}, \eqref{large-D-XXXX}
\begin{equation}
\langle \tr XX\rangle= \frac {1}{\sqrt{2D}}+O(D^{-3/2})\,,\quad
\langle \tr XXXX\rangle= \frac 1 D+O(D^{-2})\,.
\end{equation}
In figure \ref{fig:Lmax6}, we present the $L_{\max}=6$ bounds for $\langle \tr XX\rangle$ and $\langle \tr XXXX\rangle$ at various $D$ and the leading prediction of the large $D$ expansion.

\begin{figure}[tbp]
	\centering 
	\includegraphics[width=.48\textwidth]{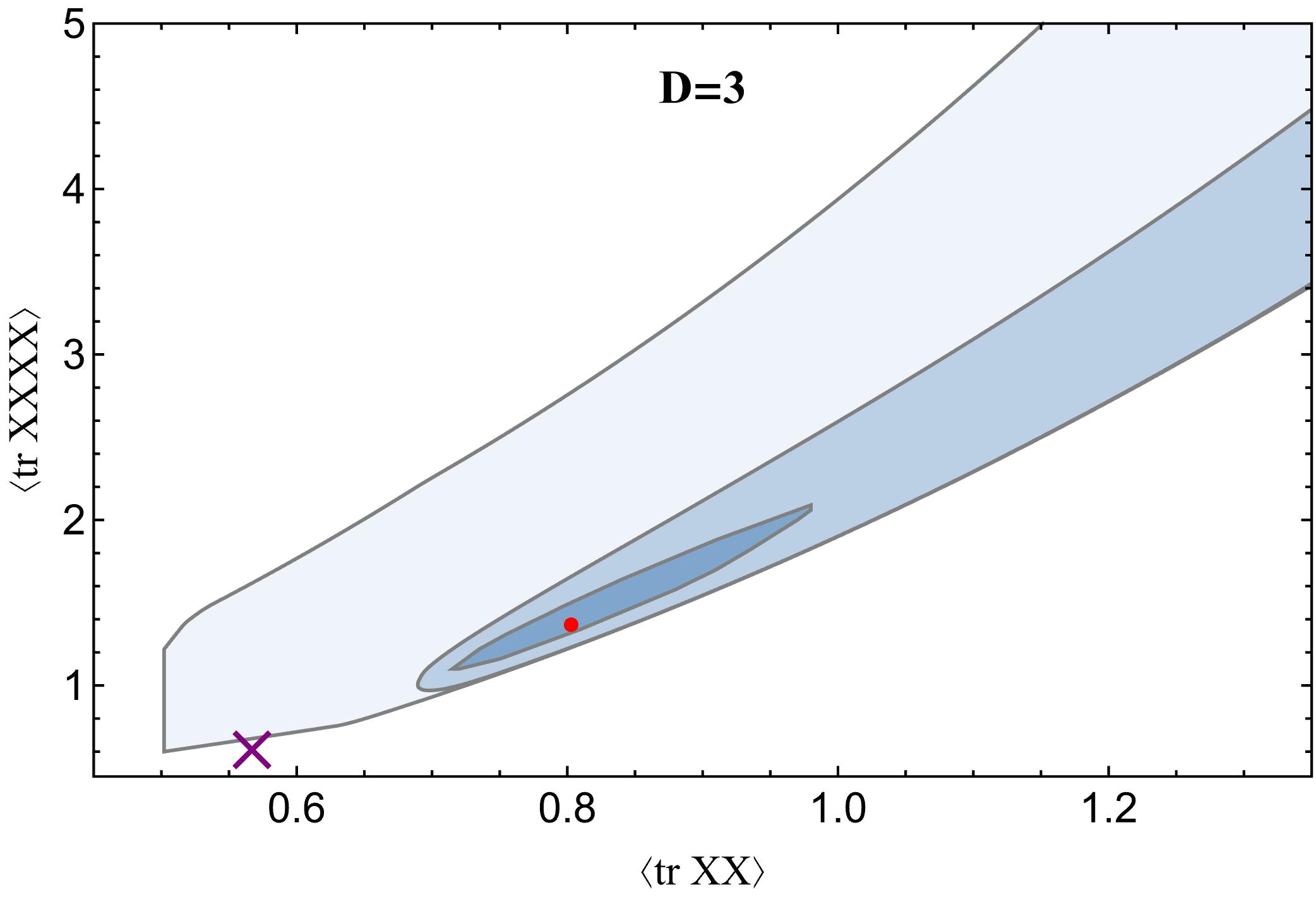}
	\hfill
	\includegraphics[width=.48\textwidth]{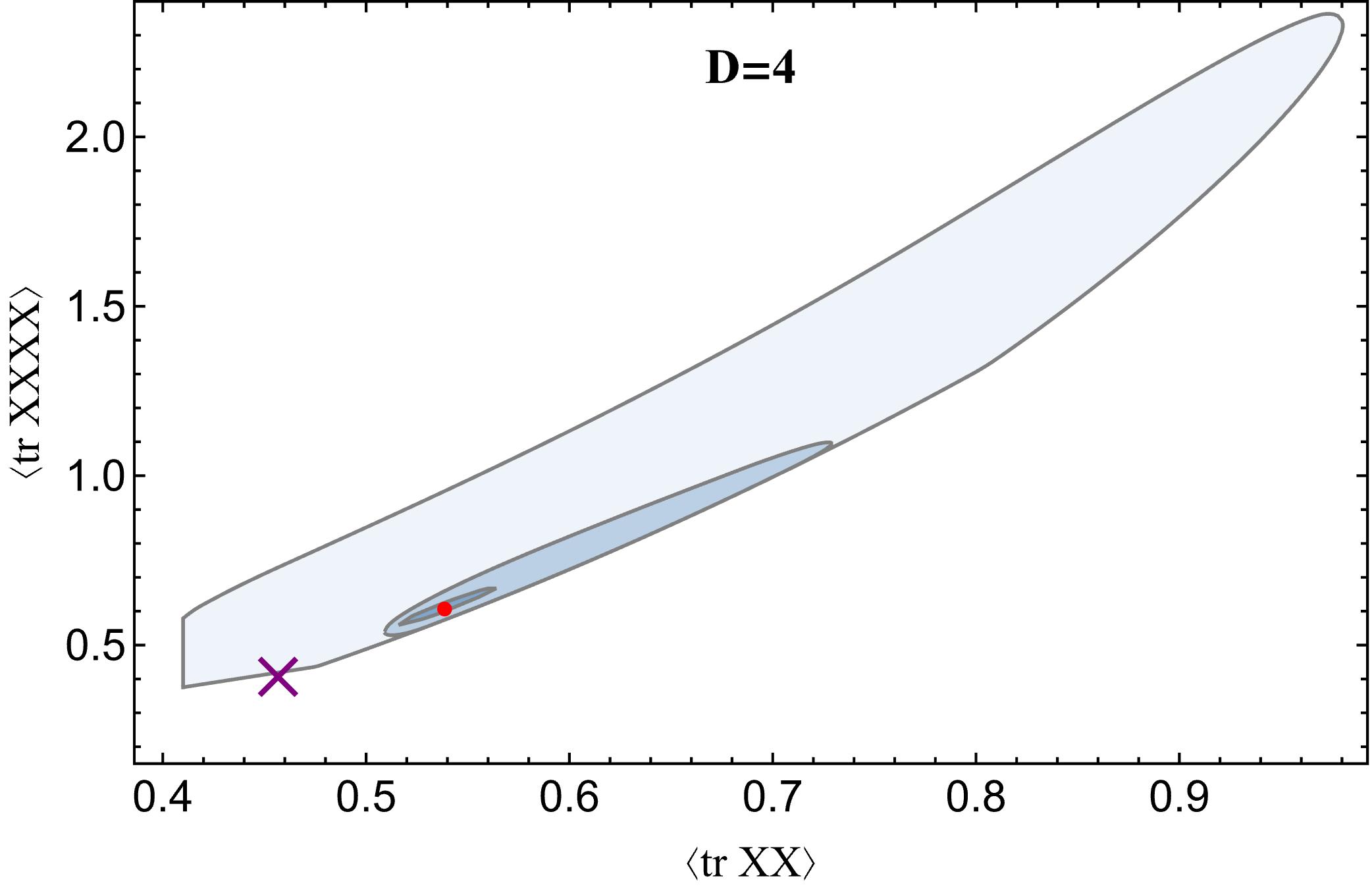}
	
	\vspace{8pt}
	\includegraphics[width=.48\textwidth]{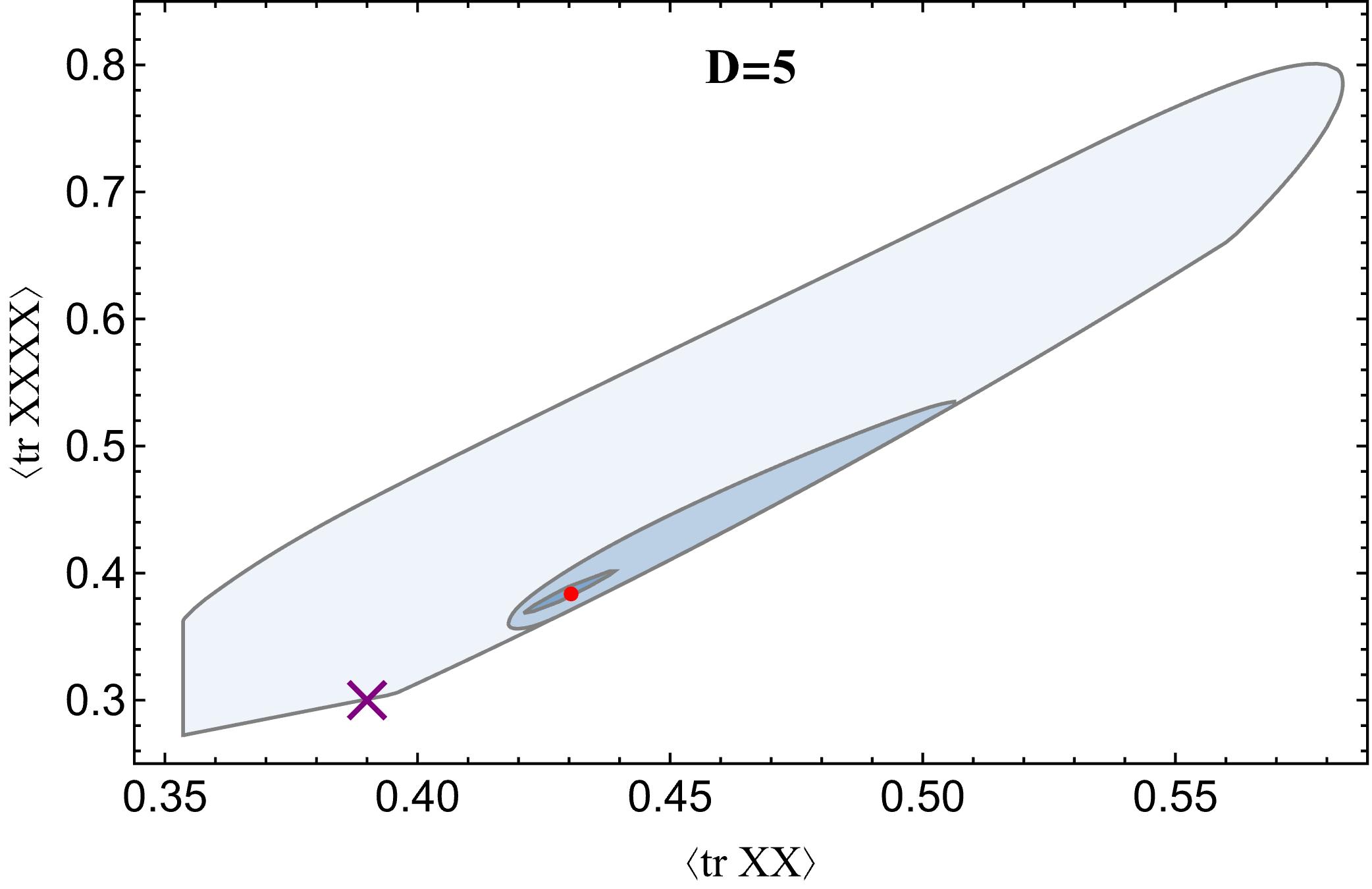}
	\hfill
	\includegraphics[width=.48\textwidth]{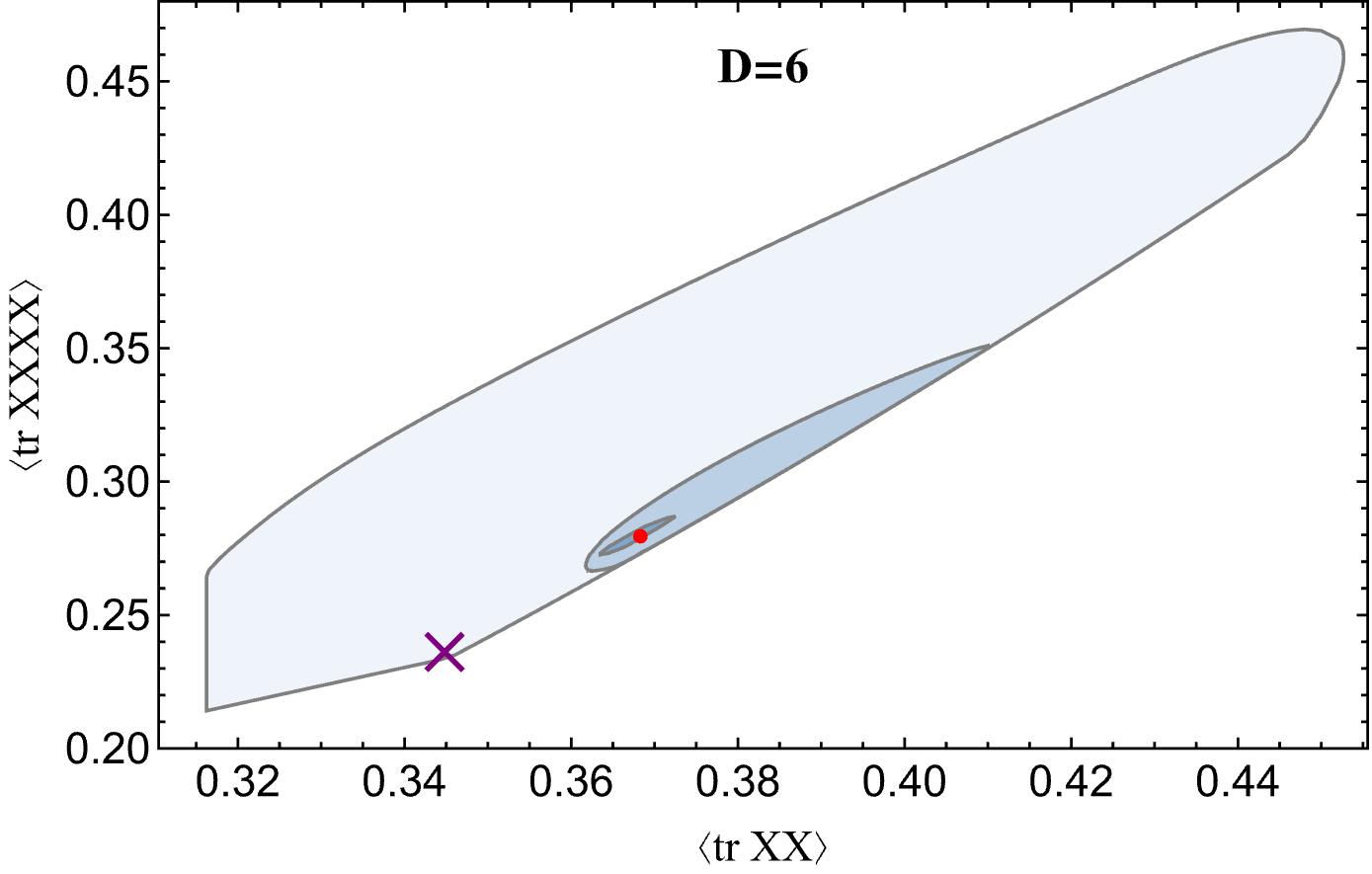}
	
	\vspace{8pt}
	\includegraphics[width=.48\textwidth]{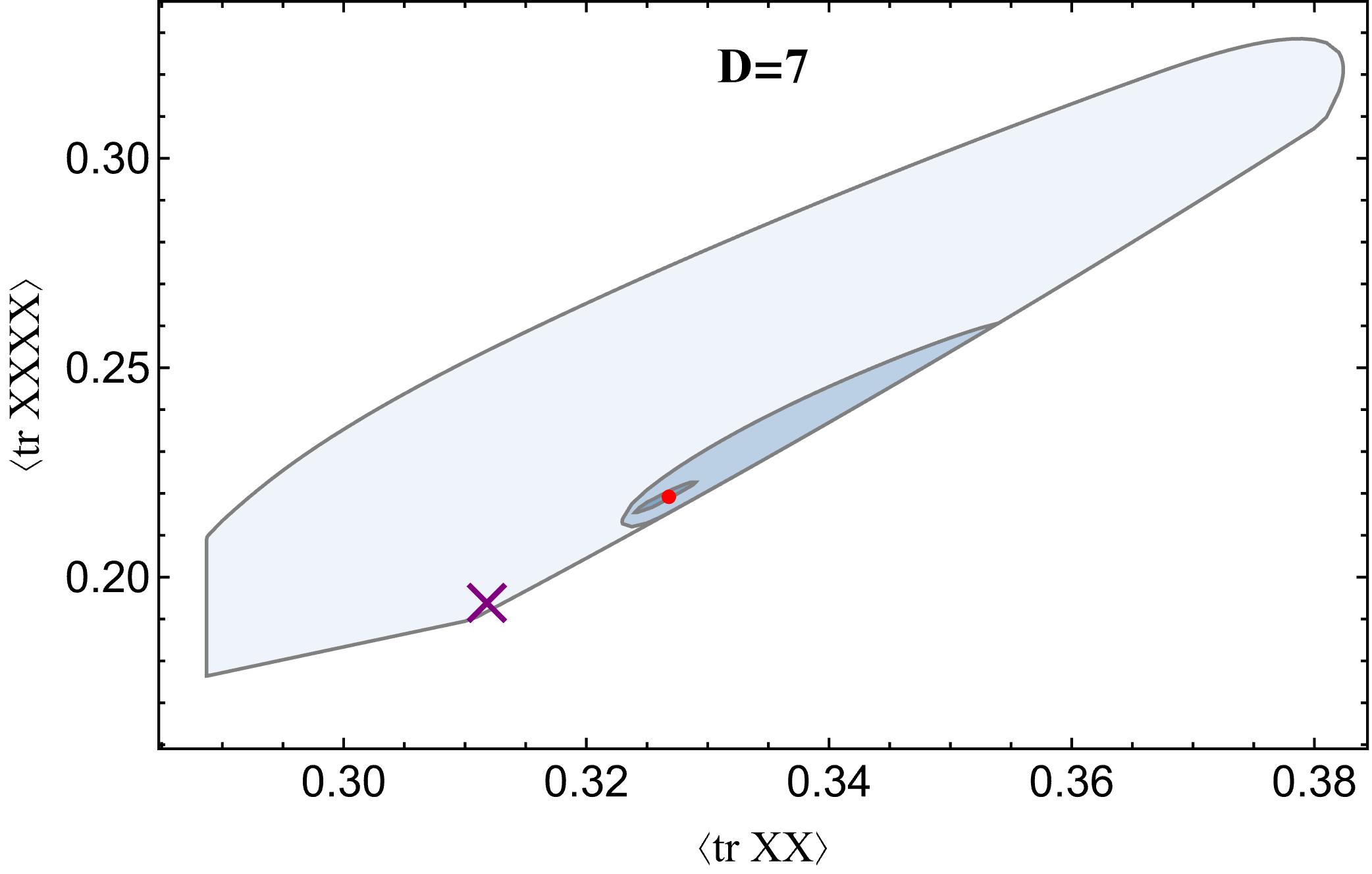}
	\hfill
	\includegraphics[width=.48\textwidth]{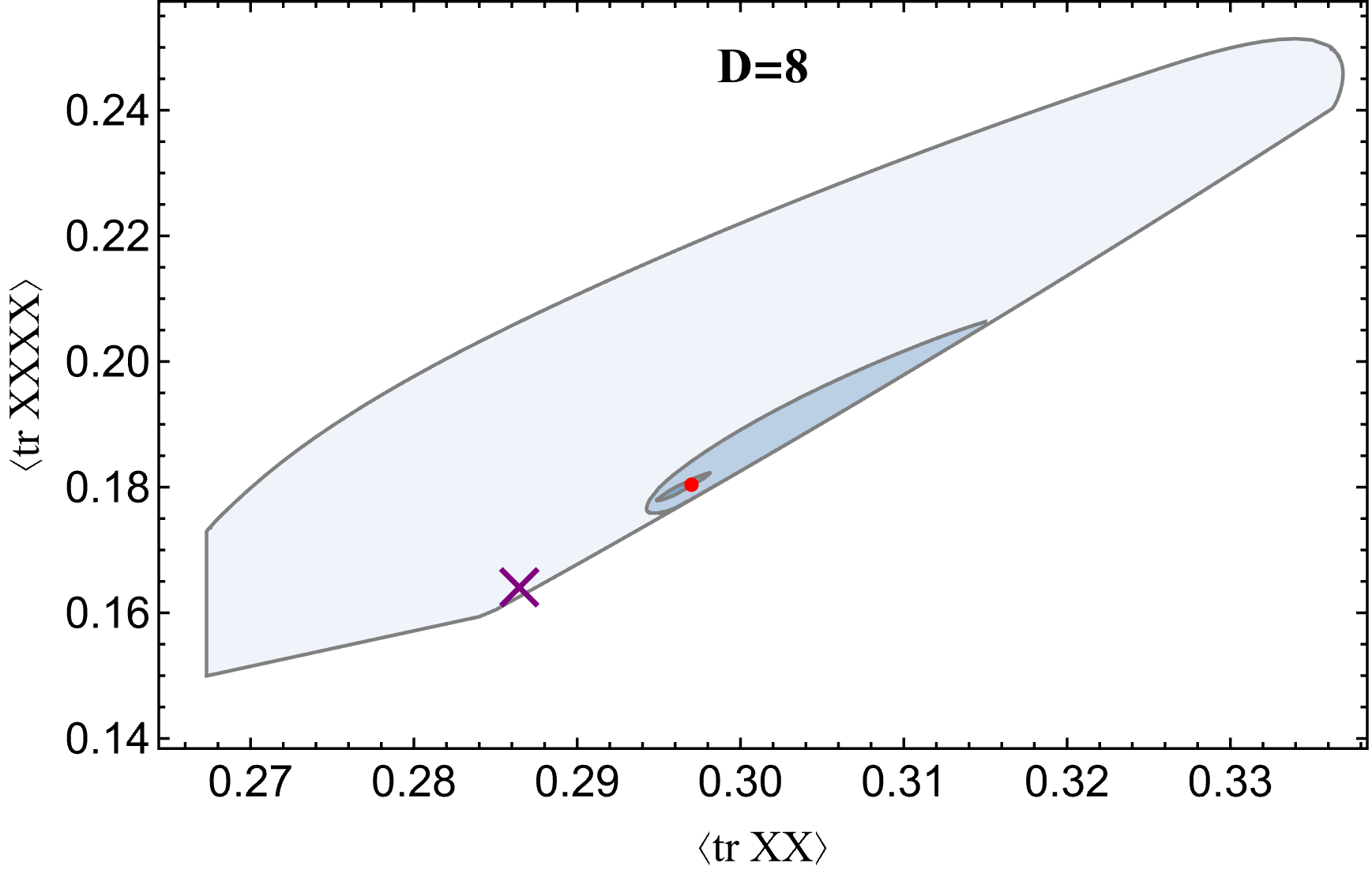}
	
	\vspace{8pt}
	\includegraphics[width=.48\textwidth]{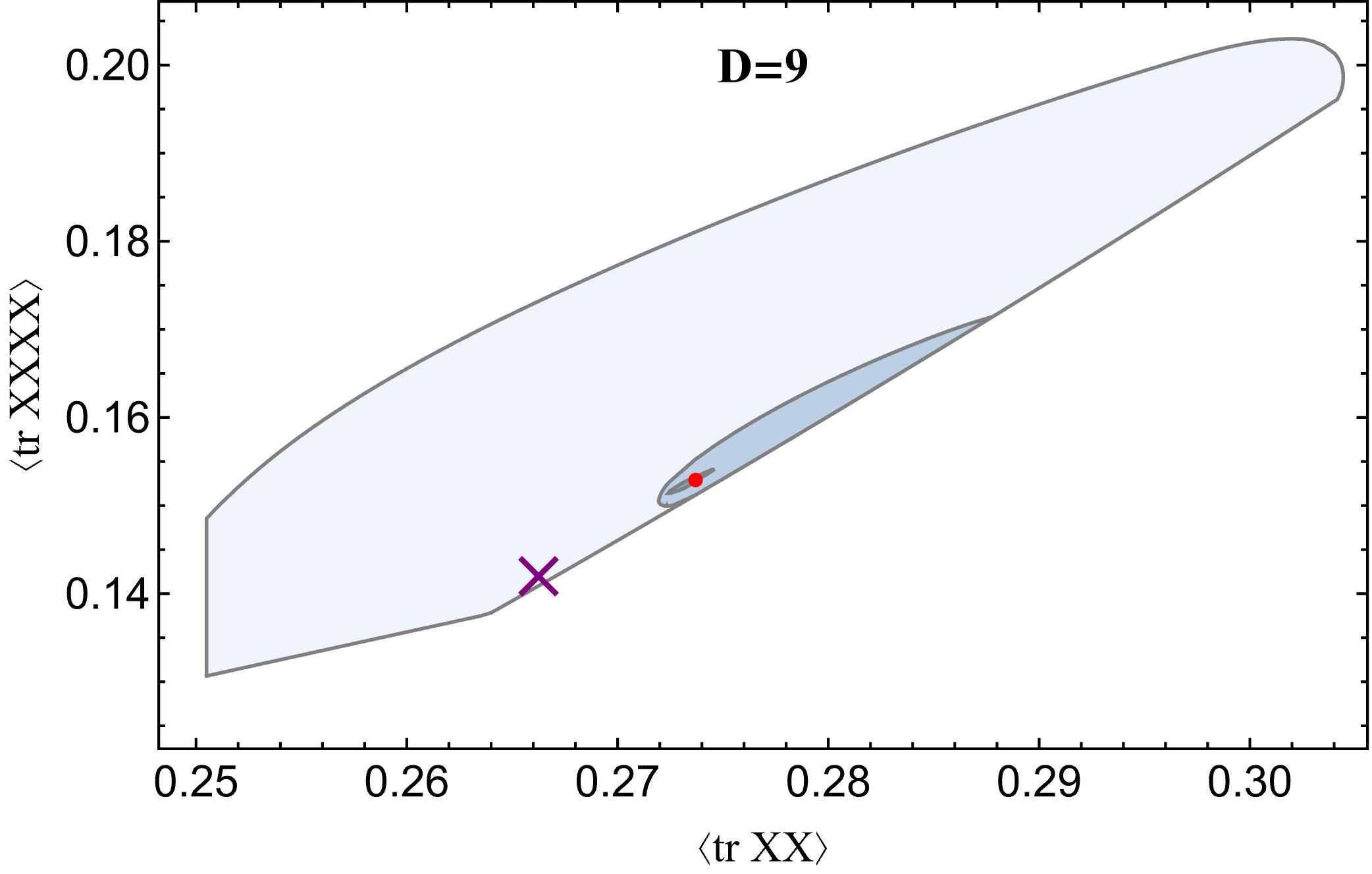}
	\hfill
	\includegraphics[width=.48\textwidth]{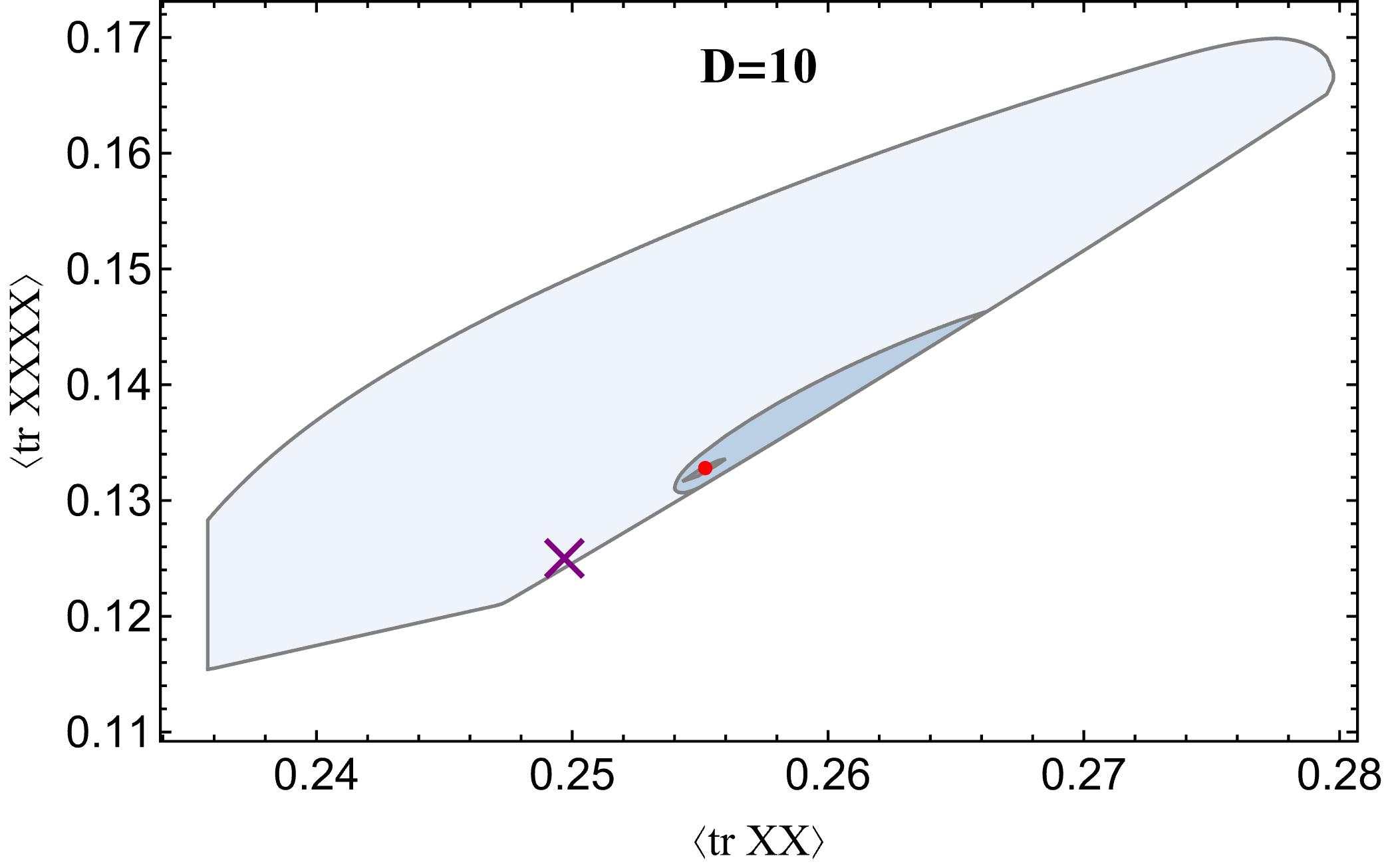}
	\vspace{8pt}
	\begin{minipage}{0.9\textwidth}\centering
		\legendbox{lmax8!60}\; $L_{\max}=8$ \quad
		\legendbox{lmax10}\; $L_{\max}=10$ \quad
		\legendbox{lmax12}\; $L_{\max}=12$\quad
		\legenddot{red}\; Monte Carlo\quad
		{\tikz[baseline=-0.72ex]\draw[purp, line width=1.3pt] 
			(-2.5pt,-2.5pt)--(2.5pt,2.5pt) 
			(-2.5pt,2.5pt)--(2.5pt,-2.5pt);}\; $1/D$
	\end{minipage}
	\caption{\label{fig:i} Positivity bounds on $\left\langle\mathrm{tr}\,XX\right\rangle\,,\left\langle\mathrm{tr}\,XXXX\right\rangle$ for $D=3,...,10$. 
		From light to dark, the shaded regions correspond to cutoffs $L_{\max}=8,10,12$. 
		The allowed regions shrink rapidly as $L_{\max}$ increases, and become islands  at $L_{\max}\ge8$ for $D\geq 4$, but at $L_{\max}=12$ for $D=3$. The red points indicate the Monte Carlo results, while the purple cross represents the $1/D$ expansion series to subleading order, i.e., $\left(\frac 1 {\sqrt{2D}}+\frac {7}{6\sqrt{2}D^{3/2}},\,\frac 1 D+\frac{5}{2D^2}\right)$.
	}
	\label{Lmax81012}
\end{figure}

\subsubsection{Numerical bounds at $L_{\max}=8,10,12$}
We now turn to the cases of larger length cutoffs $L_{\max}=8,10,12$. 
As explained above, we can use \eqref{Mhat-definition} to compute the matrix elements of  
\(\widehat{ M }^{(k,r)}\) and impose the positive semi-definite condition on these smaller matrices. 
Although these matrices are considerably smaller than those from the direct procedure, 
their sizes still grow rapidly with $L_\text{max}$. 
Therefore, we can only compute the numerical bounds.

At $L_{\max}=8,10$, the squared terms in the loop equations are associated with $A_{1,2}$. 
If we scan $A_{1,2}$, then the positivity constraints can be implemented by semi-definite programming. 
We determine the maximum and minimum of $A_{1,2,3,4}$ as functions of $A_{1,2}$.  For numerical stability, the basis operators are normalized according to the large $D$ expansion results. Here the solver we use is \texttt{MOSEK}  in \texttt{Mathematica}\cite{MOSEK}.

At $L_{\max}=12$, the computation of the positivity bounds requires more care. 
First, the squared terms in the loop equations involve both $A_{1,2}$ and $A_{1,2,3,4}$. 
To avoid nonlinear dependence, we need to scan both $A_{1,2}$ and $A_{1,2,3,4}$.   
Second, the feasibility checks become less reliable due to numerical errors. 
In the usual bootstrap implementation, 
one works in \emph{feasibility mode}. 
For each trial point in moment space, the semi-definite program (SDP) is solved 
to decide whether all positivity constraints admit a feasible solution. 
The output is therefore binary (``allowed'' or ``excluded''). 
However, this approach may rule out some positive region due to numerical errors.
We instead introduce a matrix bootstrap version of the \emph{navigator function} \cite{Reehorst:2021ykw}, which is more stable than the binary mode, and use \texttt{SDPA-GMP}~\cite{SDPA-GMP}, which is an arbitrary-precision SDP solver.
To be more specific, we define the navigator function as\footnote{We thank Zechuan Zheng for suggesting this.}
\begin{equation}
	\label{eq:navigator}
\mathcal N\;=\;\max\Bigl\{\,\lambda\;\Big|\;\widehat{M}^{(k,r)}-\lambda\,\mathbb{I}\ \succeq\ 0\,,\quad \forall\,(k,r)\Bigr\},
\end{equation}
where $\mathbb{I}$ indicates the identity matrix of the corresponding dimension. 
The  criterion for the positive region is $\mathcal N>0$.\footnote{For $L_{\max}=12$, the computation was first performed with \texttt{MOSEK} in \texttt{Mathematica}, which is significantly faster than \texttt{SDPA-GMP}. 
However, due to the precision effects, we find that the navigator function for $D=3,4$ at $L_{\max}=12$ is always negative. Thus, the strict criterion would rule out the physical solution as well. 	Therefore, we need to relax the criterion and introduce a tolerance.  
	A negative point is also accepted if
	\begin{equation}
		\label{eq:tolerance}
		\mathcal{N} \;>\; -\varepsilon \,,
	\end{equation}
	where $\varepsilon$ is a positive number. 
	Accordingly, the negative region satisfying $ 0> \mathcal{N}>-\varepsilon$ is also included. 
	In practice, we choose the tolerance $\varepsilon$ to be $10^{-8}$ for $D=3$ and $10^{-10}$ for $D=4$. 
	The resulting positivity bounds are more reasonable than those from the approach without using the navigator function. For example, the boundaries of the allowed regions become less jagged and more convex. When using \texttt{SDPA-GMP}, the precision can instead be increased systematically, so that no manually chosen tolerance is needed. 
The bounds presented in the main text agree with those obtained with \texttt{MOSEK} based on the above choices of tolerance.}

\begin{table}[t]
	\centering
	\caption{The predictions for \(\trXX\) and \(\trXXXX\) from the positivity bounds at length cutoffs $L_\text{max}=8,10,12$ with \(D=3,4,\dots,10\). For comparison, we also list the rough estimates of Monte Carlo simulations. }
	\label{tab:boot-vs-mc}
	\setlength{\tabcolsep}{6.5pt}
	\renewcommand{\arraystretch}{1.15}
	\begin{tabular}{c c c c c c}
		\toprule
		\textbf{moments} & \(\mathbf{D}\) &
		{$L_{\max}=8$} & {$L_{\max}=10$} & {$L_{\max}=12$} &\textbf{Monte Carlo} \\
		\midrule
		\multirow{8}{*}{$\trXX$}
		& 3  & $>0.500 $ & $>0.689$ & 0.85(14) & $0.8028^{+0.0010}_{-0.0006}$ \\
		& 4  & 0.69(29) &0.62(11)  & 0.540(26) & $0.5386^{+0.0007}_{-0.0005}$ \\
		& 5  &  0.47(12) & 0.463(45)& 0.43(1)& $0.4304^{+0.0012}_{-0.0007}$ \\
		&6&0.384(68)&0.386(25)& 0.368(5)&$0.3683^{+0.0006}_{-0.0006}$\\
		&7&0.336(47)&0.339(16)&0.3265(27)&$0.3268^{+0.0009}_{-0.0004}$\\
		&8&0.302(35)&0.305(11)&0.2965(18)&$0.2970^{+0.0006}_{-0.0006}$\\
		&9&0.277(27)&0.280(8)&0.2735(12)&$0.2737^{+0.0016}_{-0.0007}$\\
		&10&0.258(22)&0.2602(62)&0.25515(95)&$0.2552^{+0.0018}_{-0.0006}$\\
		\midrule
		\multirow{8}{*}{$\langle \mathrm{tr}\,XXXX\rangle$}
		& 3  & $>0.596$ & $>0.965$ & 1.61(54)   & $1.367^{+0.0048}_{-0.0020}$ \\
		& 4  & 1.37(99) & 0.81(29)  & 0.613(60)  & $0.6065^{+0.0025}_{-0.0015}$ \\
		& 5  & 0.54(27) & 0.448(90) & 0.385(18)  & $0.3836^{+0.0032}_{-0.0006}$\\
		& 6  & 0.34(13) & 0.310(43) & 0.2798(78) & $0.2795^{+0.0005}_{-0.0015}$ \\
		& 7  & 0.253(78) & 0.237(25) & 0.2191(42) & $0.2192^{+0.0005}_{-0.0007}$ \\
		& 8  & 0.201(51) & 0.192(16) & 0.1801(23) & $0.1804^{+0.0011}_{-0.0009}$ \\
		& 9  & 0.166(37) & 0.161(11) & 0.1528(17) & $0.1529^{+0.0024}_{-0.0009}$\\
		& 10 & 0.143(28)  & 0.1385(81) & 0.1326(11) & $0.1328^{+0.0022}_{-0.0008}$ \\
		\bottomrule
	\end{tabular}
\end{table}

For $L_{\max} =8, 10$, 
the positivity condition on the sectors of ranks  $k=L_\text{max}/2, L_\text{max}/2-1$ has negligible effects on the positivity bounds of $\langle \tr XX\rangle$ and $\langle \tr XXXX\rangle$. 
In fact, we have already seen this phenomenon in the explicit example for $L_{\max}=6$, 
where the rank-$2$ and rank-$3$ sectors fail to strengthen the bounds of $\langle \tr XX\rangle$ and $\langle \tr XXXX\rangle$. 
The reason is that these high-rank sectors involve a large number of free parameters $A$, but their positivity matrices are of relatively low dimensions. 
Accordingly, for $L_{\max}=12$, we do not impose positivity conditions on the rank-5 and rank-6 sectors, which further reduces the computational effort,
as the derivation of their matrix elements involves more complicated traceless projectors and contractions.\footnote{The rank-5 sector is the same as that of $L_\text{max}=10$, so its matrix elements are already derived. We also remove them to reduce the potential numerical instability. } 
Moreover, there are some subtleties in the traceless projector for $k\ge D+2$, where 
the general $D$ expressions become divergent, such as the cases of rank-5 for $D=3$ and rank-6 for $D=3,4$. 
This issue is avoided by omitting the rank-$5$ and rank-$6$ sectors in the $L_\text{max}=12$ computation.\footnote{
At $L_\text{max}=10$, we also omit the positivity condition on the rank-$5$ sector when $D=3$. 
The positivity bounds for $\langle \tr XX\rangle$ and $\langle \tr XXXX\rangle$ are consistent with those from the direct procedure in section \ref{Direct procedure}.}

\begin{figure}[tbp]
	\centering 
	\includegraphics[width=\textwidth]{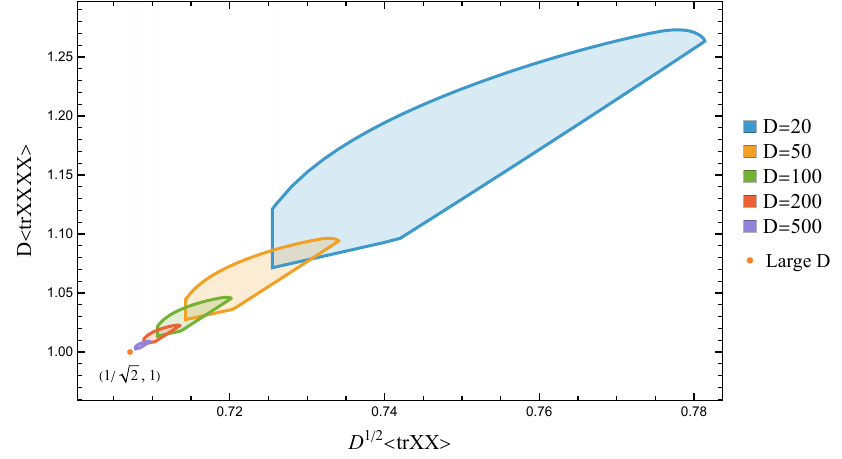}
	\vspace{8pt}
	
	\includegraphics[width=\textwidth]{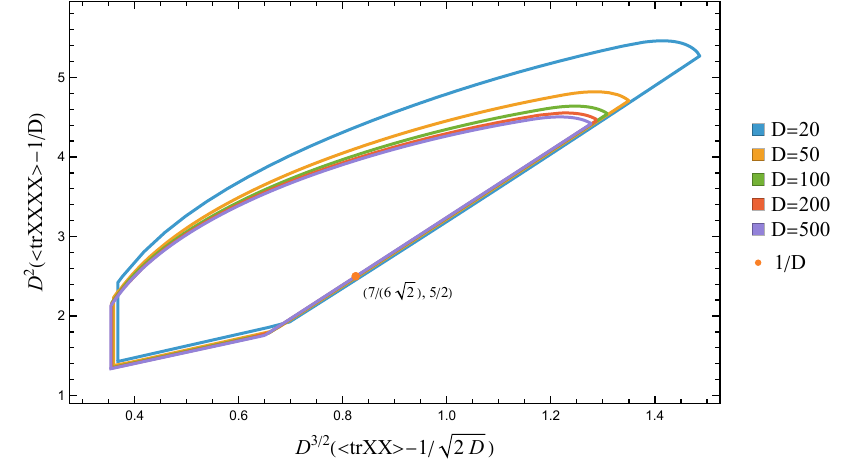}
	\caption{Positivity bounds at $L_{\max}=8$ with $D=20, 50,100, 200, 500$. 
		In the first plot, the islands shrink to the point associated with the leading terms of the large $D$ limit. 
		In the second plot, the islands converge as $D$ grows. The point associated with the subleading terms of the $1/D$ expansion is located on the boundary of the allowed region.
	}
	\label{Lmax8largeD}
\end{figure}
\begin{figure}[tbp]
	\centering 
	\includegraphics[width=1\textwidth]{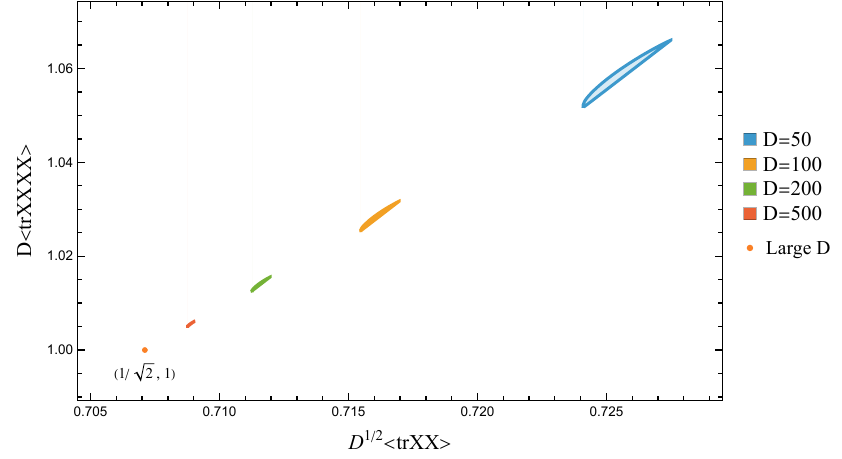}
		\vspace{8pt}
		
	\includegraphics[width=1\textwidth]{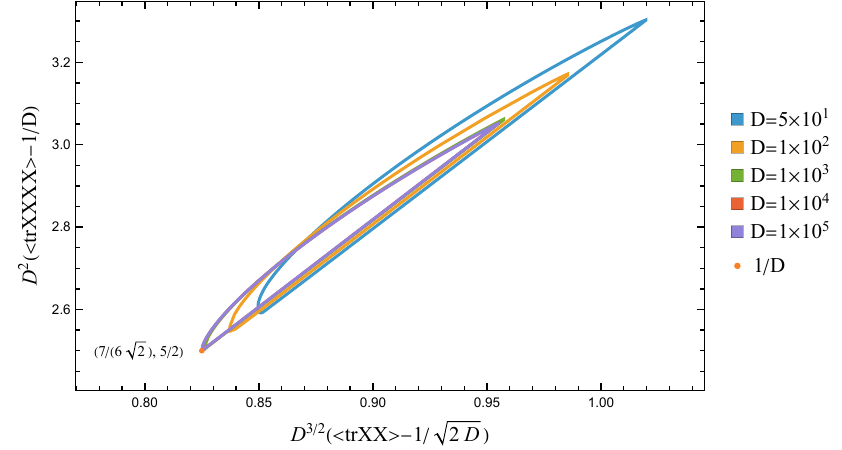}
	\caption{Positivity bounds at $L_{\max}=10$ with large values of $D$. 
		In the first plot, as $D$ increases, the islands shrink to the leading large $D$ results more rapidly than the $L_\text{max}=8$ islands. 
		In the second plot, the left lower tip of the islands converges to the subleading results of the $1/D$ expansion.
	}
	\label{Lmax10largeD}
\end{figure}

In figure~\ref{Lmax81012}, we present the positivity bounds for $L_{\max}=8,10,12$ 
with $D=3,4,\dots,10$.  For $D\ge 4$, an island appears at $L_{\max}=8$ and shrinks rapidly with $L_{\max}$.  
However, the $D=3$ bound becomes an island only at $L_{\max}=12$, which is much greater than the higher $D$ cases.\footnote{It is also possible that the $D=3$ bound becomes an island at lower $L_\text{max}$, but the upper bounds for $\langle \tr XX\rangle$ and $\langle \tr XXXX\rangle$ are excessively large and 
beyond the numerically reliable range of the SDP solver. }
In table \ref{tab:boot-vs-mc}, we summarize the predictions for $\langle \tr XX\rangle$ and $\langle \tr XXXX\rangle$ from the positivity bounds, which improve with $L_\text{max}$ and $D$. 
They are in good agreement with the Monte Carlo estimates.\footnote{We used and adapted the code from \cite{Jha:2021exo} to run Monte Carlo simulations. }  
For relatively large $D$, the precision of some $L_{\max}=12$ islands is of the same order as that of the Monte Carlo results.

For $L_{\max}=8,10$, the use of the $\mathrm{O}(D)$ irrep basis  and the arbitrary-precision SDP solver \texttt{SDPA-GMP} enables us to 
study the positivity bounds for significantly larger $D$. 
In this way, we can further study the asymptotic behavior of the numerical bounds 
and compare the results with the predictions of the large $D$ expansion. 
As shown in figure \ref{Lmax8largeD} and \ref{Lmax10largeD}, the islands at $L_{\max}\ge 8$ shrink to the point
\begin{equation}
D^{1/2}\langle \tr XX\rangle\Big|_{D\rightarrow \infty}=\frac 1 {\sqrt{2}}, 
\quad D\langle \tr XXXX\rangle\Big|_{D\rightarrow \infty}=1,
\end{equation}
which is associated with the leading asymptotic behavior at large $D$ in \eqref{large-D-XX}, \eqref{large-D-XXXX}. 
In this sense, the leading terms of the $1/D$ expansion are captured by the positivity bounds at $L_\text{max}\geq 8$. 

To examine the $1/D$ corrections, we subtract $(\langle \tr XX\rangle, \langle \tr XXXX\rangle)$ by the leading terms and then multiply the differences by $(D^{3/2}, D^2)$. 
We also present these subtracted bounds in figure \ref{Lmax8largeD} and \ref{Lmax10largeD}. 
In both cases, the results seem to converge to  islands of finite size, which gives a range of the subleading term.
According to \eqref{large-D-XX}, \eqref{large-D-XXXX}, 
the subleading asymptotic behavior at large $D$ is associated with the point
\begin{equation}
D^{3/2}\left(\langle \tr XX\rangle-\frac 1{\sqrt{2D}}\right)\Big|_{D\rightarrow \infty}
=\frac 7 {6\sqrt{2}},
\quad 
D^2\left(\langle \tr XXXX\rangle-\frac 1 D\right)\Big|_{D\rightarrow \infty}=\frac 5 2.
\end{equation}
For $L_\text{max}=8$, this point is located on the boundary of the allowed region.
For $L_\text{max}=10$, interestingly, this point happens to lie at the left lower tip of the island.
We expect that the higher order terms of the $1/D$ expansion can be extracted from the positivity bounds as the length cutoff $L_\text{max}$ is increased.

\section{Analytic trajectories and eigenvalue distributions}
\label{Analytic}
Above, the positivity bootstrap works well for the Yang-Mills matrix integrals. 
However, as mentioned in the introduction,
the supersymmetric Yang-Mills matrix integrals usually do not lead to a manifestly positive measure after integrating out the fermions, due to the Pfaffians from fermionic path integrals.
This calls for alternative bootstrap methods that do not rely on positivity assumptions. 

In this section, we switch to a complementary bootstrap approach, which is based on analytic continuation of matrix moments in length. 
To illustrate these moment trajectories, we use the large $D$ expansion results to derive the explicit expressions for various types of moment trajectories and the associated eigenvalue densities. 
Inspired by these large $D$ results, we then construct some ansatz for the analytic trajectories at finite $D$. 
The analytic trajectories must satisfy certain consistency conditions arising from trajectory intersections, which are automatically enforced by the multi-length trajectory ansatz. In the end, we determine the matrix moments using symmetry constraints, contraction limits and loop equations, without resorting to positivity constraints.

\subsection{Explicit examples from the large $D$ expansion}\label{largeD}
Based on the large $D$ expansion results in section \ref{Large D expansion}, 
we can extract the explicit formulas for the analytic trajectories in terms of analytic functions of lengths. 
Below, we start from the one-length trajectories and then extend the results to the cases of multi-length trajectories. 
We also discuss some explicit examples of the resolvents and eigenvalue densities 
associated with analytic trajectories.

\subsubsection{One-length trajectories}\label{1-length-large-D}
In section \ref{Large D expansion}, we presented the concrete expressions of moments 
in various types of one-length trajectories. 
It is not hard to guess their general expressions, such as
\begin{align}
		\big\langle \tr X^{2n} \big\rangle
		&= f_{n}\left(1 + \frac{n(3n+4)}{2(n+2)}\,\frac{1}{D}+O(D^{-2})\right),\label{X2n}
		\\[6pt]
		\big\langle \tr X^{2n} Y^2 \big\rangle
		&= f_{n}\,f_{1} \left(1+ \frac{9n^{2}+13n+14}{6(n+2)}\,\frac{1}{D}+O(D^{-2})\right),\label{X2nY2}
		\\[6pt]
		\big\langle \tr X^{2n} Y^4 \big\rangle
		&= f_{n}\,f_{2} \left(1+ \frac{3n^{2}+6n+10}{2(n+2)}\,\frac{1}{D}+O(D^{-2})\right),\label{X2nY4}
	\\[6pt]
		\big\langle \tr X^{2n} Y^2Z^2 \big\rangle
		&= f_{n}\,f_{1}^2 \left(1+ \frac{3n^{2}+4n+8}{2(n+2)}\,\frac{1}{D}+O(D^{-2})\right),\label{X2nY2Z2}
		\\[6pt]
		\big\langle \tr X^{2n} YZ^2Y \big\rangle
		&= f_{n}\,f_{1}^2  \left(1+ \frac{3n^{2}+6n+8}{2(n+2)}\,\frac{1}{D}+O(D^{-2})\right),
\end{align}
and
\begin{align}
		\big\langle \tr X^{2n} (XY)^2 \big\rangle
		&= f_{n+1}\,f_{1}\,f_{2}\left(1+O(D^{-1})\right),
		\\[6pt]
		\big\langle \tr X^{2n} (YZ)^2 \big\rangle
		&= f_{n}\,f_{1}^2 \,f_{2}\left(1+O(D^{-1})\right).
\end{align}
The building block function $f_n$ is defined as
\begin{equation}
	f_n=\frac{C_n}{(2D)^{\,n/2}},
	\qquad
	C_n = \frac{1}{n+1}\binom{2n}{n},
\end{equation}
where $C_n$ denotes the $n$-th Catalan number. 
We verify that these general length formulas are consistent with the concrete results of the large $D$ expansion. 

\subsubsection{Multi-length trajectories}\label{Multi-length trajectories}

The above one-length trajectories provide beautiful unification of many concrete words,  
as well as new predictions that are not accessible by direct computation. 
As noted in \cite{Li:2024ggr}, there exists higher unification through multi-length trajectories.  
For example, natural unification of $\langle \tr X^{2n} \rangle$, $\langle \tr X^{2n} Y^2\rangle$, $\langle \tr X^{2n} Y^4\rangle$ in  \eqref{X2n}, \eqref{X2nY2}, \eqref{X2nY4}  is given by
the two-length trajectory
\begin{align}\label{2-length-trajectory}
	\big\langle \tr  X^{n_1}Y^{n_2}    \big\rangle
	=
	\left(
	1+\frac{3(n_1+n_2)}{4D}
	-\frac{
		G_{n_1}+G_{n_2}
		+G_{n_1}G_{n_2}}{3D}
	\right)g_{n_1}\,g_{n_2}\,
	+O(D^{-2-\frac{n_1+n_2}{4}})\;.
\end{align}
The building block functions are
\begin{equation}
	G_n=\frac{g_{n+2}}{g_{n}\,g_2}-1\,,
	\quad
	g_n=\frac{1+(-1)^n}{2}\,
	\frac{(1/2)_{\,n/2}}{(2)_{\,n/2}}\left(\frac{2\sqrt 2}{\sqrt{D}}\right)^{n/2}
	=
	\begin{cases}
		\displaystyle f_{n/2} &  \text{even}\ n\\
		0 & \text{odd} \ n\,,
	\end{cases}
\end{equation}
where \((a)_m\) is the Pochhammer symbol. 
By taking into account $\langle \tr X^{2n} Y^2Z^2 \rangle$ in \eqref{X2nY2Z2}, 
we further determine the three-length trajectory $\langle \tr (X^1)^{n_1}(X^2)^{n_2} (X^3)^{n_3}  \rangle$. 
It turns out that a further extension leads to the general expression
\begin{align}
		&\qquad\big\langle \tr\; \bigl(X^{1}\bigr)^{n_1}\bigl(X^{2}\bigr)^{n_2}\dots \bigl(X^{j_\text{max}}\bigr)^{n_{j_\text{max}}}\big\rangle
		\nonumber\\
		&=
		\left(
		1+\frac{3\sum_{j=1}^{j_\mathrm{max}} n_j
		}{4D}
		-\frac{
		\sum_{j=1}^{j_\mathrm{max}}G_{n_j}
		+
		\sum_{1\le j<j'\le j_\mathrm{max}}G_{n_{ j}}\,G_{n_{j'}}}{3D}
		\right)\prod_{j=1}^{j_\mathrm{max}}g_{n_j}\,
		 +O(D^{-2-\frac{\sum_{j=1}^{j_\mathrm{max}} n_j}{4}})\,,
\end{align}
where we use $n_j$ to denote the power of $X^j$. 
By construction, we have $G_0=0$ and $g_0=1$, 
so one can reduce the number of lengths by setting an $n_j$ to zero. 

We also find other types of multi-length trajectories, i.e., those with duplicate matrices.
It is convenient to introduce
\begin{equation}
	Q(x_1,x_2,\dots,x_{j_\mathrm{max}}) \;\equiv\; \sum_{j=1}^{j_\mathrm{max}} G_{x_j}\,.
\end{equation}
The explicit expressions for four-length trajectories are
\begin{align}\label{1213-trajectory}
	&\quad\Big\langle \tr X^{n_1}Y^{n_2}X^{n_3}Z^{n_4}\Big\rangle \nonumber\\
	&= \left( 1+\frac{3\sum_{j=1}^{4} n_j}{4D}
	-\frac{\Q{n_1+n_3,n_2,n_4}
		+\Q{n_1+n_3}\,\Q{n_2,n_4}}{3D} \right)
	g_{n_1+n_3}\,g_{n_2}\,g_{n_4}\nonumber\\
	&\quad -\frac{\Q{n_2}\,\Q{n_4}}{3D}\,g_{n_1}\,g_{n_2}\,g_{n_3}\,g_{n_4} +O({D^{-2-\sum_{j=1}^{4} n_j/4}})\,,
\end{align}

\begin{align}
	&\quad\Big\langle \tr X^{n_1}Y^{n_2}X^{n_3}Y^{n_4}\Big\rangle \nonumber\\
	&= - \left( 1+\frac{3\sum_{j=1}^{4} n_j}{4D}
	-\frac{Q(n_1,n_2,n_3,n_4)
		+Q(n_1,n_3)\,Q(n_2,n_4)}{3D} \right)
	g_{n_1}\,g_{n_2}\,g_{n_3}\,g_{n_4}\nonumber\\
	&\quad + \left( 1+\frac{3\sum_{j=1}^{4} n_j}{4D}
	-\frac{Q(n_1+n_3,n_2,n_4)
		+Q(n_1+n_3)\,Q(n_2,n_4)}{3D} \right)
	g_{n_1+n_3}\,g_{n_2}\,g_{n_4} \nonumber\\
	&\quad + \left( 1+\frac{3\sum_{j=1}^{4} n_j}{4D}
	-\frac{Q(n_2+n_4,n_1,n_3)
		+\Q{n_2+n_4} Q(n_1,n_3)}{3D} \right)
	g_{n_2+n_4}\,g_{n_1}\,g_{n_3}\nonumber\\
	&\quad +2\,g_{n_1+1}\,g_{n_2+1}\,g_{n_3+1}\,g_{n_4+1}
	+O(D^{-2-\sum_{j=1}^{4} n_j/4})\,,
\end{align}
which belong to the $(1,2,1,3)$, $(1,2,1,2)$ types. 
They can be extended to five-length trajectories of the $(1,2,1,3,4), (1,2,1,2,3)$ types 
\begin{align}
	&\quad\Big\langle \tr X^{n_1}Y^{n_2}X^{n_3}Z^{n_4}W^{n_5}\Big\rangle \nonumber\\
	&= \Bigl( 1+\tfrac{3\sum_{j=1}^{5} n_j}{4D}
	-\tfrac{\Q{n_1+n_3,n_2,n_4,n_5}
		+\Q{n_1+n_3}\,\Q{n_2,n_4,n_5}
		+\Q{n_4}\,\Q{n_5}}{3D} \Bigr)\,
	g_{n_1+n_3}\,g_{n_2}\,g_{n_4}\,g_{n_5} \nonumber\\
	&\quad - \tfrac{\Q{n_2}\,\Q{n_4,n_5}}{3D}\;
	g_{n_1}\,g_{n_2}\,g_{n_3}\,g_{n_4}\,g_{n_5}+O(D^{-2-\sum_{j=1}^{5} n_j/4})\,,
\end{align}

\begin{align}
	&\quad\Big\langle \tr X^{n_1}Y^{n_2}X^{n_3}Y^{n_4}Z^{n_5}\Big\rangle \nonumber\\
	&= - \Bigl( 1+\tfrac{3\sum_{j=1}^{5} n_j}{4D}
	-\tfrac{\Q{n_1,n_2,n_3,n_4,n_5}
		+\Q{n_1,n_3}\,\Q{n_2,n_4}
		+\Q{n_1,n_4}\,\Q{n_5}}{3D} \Bigr)\,
	g_{n_1}\,g_{n_2}\,g_{n_3}\,g_{n_4}\,g_{n_5} \nonumber\\
	&\quad + \Bigl( 1+\tfrac{3\sum_{j=1}^{5} n_j}{4D}
	-\tfrac{\Q{n_1+n_3,n_2,n_4,n_5}
		+\Q{n_1+n_3}\,\Q{n_2,n_4,n_5}
		+\Q{n_4}\,\Q{n_5}}{3D} \Bigr)\,
	g_{n_1+n_3}\,g_{n_2}\,g_{n_4}\,g_{n_5} \nonumber\\
	&\quad + \Bigl( 1+\tfrac{3\sum_{j=1}^{5} n_j}{4D}
	-\tfrac{\Q{n_2+n_4,n_1,n_3,n_5}
		+\Q{n_2+n_4}\,\Q{n_1,n_3,n_5}
		+\Q{n_1}\,\Q{n_5}}{3D} \Bigr)\,
	g_{n_2+n_4}\,g_{n_1}\,g_{n_3}\,g_{n_5},\nonumber\\
	&\quad +2\,g_{n_1+1}\,g_{n_2+1}\,g_{n_3+1}\,g_{n_4+1}\,g_{n_5} 
	+O(D^{-2-\sum_{j=1}^{5} n_j/4})\,.
\end{align}
We verify that these multi-length formulas are consistent with the concrete results from the large $D$ expansion.

\subsubsection{Eigenvalue distributions}
In one-matrix models, there are close connections between the resolvents, eigenvalue distributions and analytic trajectories of the matrix moments. 
As in \cite{Li:2024ggr}, we introduce the generalized resolvent for the $D$-matrix model
\begin{equation}
	R_{M,\mathcal{O}}(z) 
	= \Big\langle \tr \frac{\mathcal O}{z - M} \Big\rangle
	= \sum_{n=0}^\infty \langle \tr M^n \mathcal O \rangle \, z^{-n-1}.
\end{equation}
By contour deformation, the mixed moments in the same one-length trajectory are associated with a generalized eigenvalue density~$\rho_{M,\mathcal{O}}(z)$
\begin{equation}
	\left\langle \tr M^n \mathcal O \right\rangle
	=\frac{1}{2\pi i}\oint_{C} dz\,z^{n} R_{M,\mathcal{O}}(z) 
	=\int_{z_{\text{min}}(M)}^{z_{\text{max}}(M)} dz \; z^{n} \,\rho_{M,\mathcal{O}}(z),
\end{equation}
where the original contour $C$ encircles  all the branch points and poles of $R_{M,\mathcal{O}}(z)$ anticlockwise,  
and the eigenvalue density is related to the discontinuity of the generalized resolvent $R_{M,\mathcal{O}}(z)$. 
For $\mathcal{O}=\mathbb{I}$, they reduce to the standard resolvent and eigenvalue distribution associated with $M$. 
In some sense, the generalized density 
$\rho_{M,\mathcal{O}}(z)$ can be understood as the standard eigenvalue density dressed 
by the $\mathcal O$ insertion. 
Accordingly, the endpoints of the generalized distribution, i.e., $z_{\text{min}}(M)$ and $z_{\text{max}}(M)$, are expected to be identical to the minimum and maximum eigenvalues of $M$.\footnote{It is also possible that the endpoints are inside the range of the eigenvalue distribution of $M$. } 
Below are some explicit expressions of the resolvents associated 
with the one-length trajectories in section \ref{1-length-large-D}:
\begin{align}
	R_{X,\mathbb{I}}(z) &= \tilde R\left(-\tfrac 7 6,-\tfrac{\sqrt{2}}{3};z\right), \\[6pt]
	R_{X,YY}(z) & = f_1\,\tilde R\left(\tfrac 1 3,-\tfrac {2\sqrt{2}}{3};z\right),\quad
	R_{X,YYYY}(z)  =f_2\, \tilde R\left(\tfrac {11}{6},-\tfrac {5}{3\sqrt{2}};z\right), \\[6pt]
	R_{X,YYZZ}(z)  &= f_1^2\, \tilde R\left(\tfrac 3 2,-\sqrt{2};z\right), \quad
	R_{X,YZZY}(z)  = f_1^2\,\tilde R\left(\tfrac 7 6,-\tfrac {2\sqrt{2}}{3};z\right),\\[6pt]
	R_{X,YXXY}(z)  &=f_1^2\, \tilde R\left(\tfrac {7} {6},-\tfrac {2\sqrt{2}}3;z\right), \\[6pt]
	R_{X,XYXY}(z) &= \frac{1}{2D}
	\left( z^3-\tfrac{\sqrt{2}}{\sqrt{D}}z -z^2 \sqrt{z^2-\tfrac{2\sqrt{2}}{\sqrt{D}}}\right)\,,\\
	R_{X,YZYZ}(z)  &= \tfrac{1}{2\sqrt{2}\,D^{3/2}}
	\left( z - \sqrt{z^2-\tfrac{2\sqrt{2}}{\sqrt{D}}}\right),
\end{align}
where higher-order terms in $1/D$ are omitted  and we have used the ansatz
\begin{align}
 	\tilde R(a_1,a_2; z)  &= 
	\frac{\sqrt{D}}{\sqrt{2}}
 	\left(
 	 1+\frac{a_1 }{D} +\frac{a_2}{\sqrt{D}}\, z^2
	 -\frac{\sqrt{2}\,a_2}{D}\Big(1+\frac {3}{2D}\Big) \right)z\nonumber\\
	&\quad- \frac{\sqrt{D}}{\sqrt{2}}\,\Big(1+\frac{a_1}{D}+\frac{a_2}{\sqrt{D}}z^2\Big)\,\sqrt{z^2-\frac{2\sqrt{2}}{\sqrt{D}}\left(1+\frac{3}{2D}\right)}.
 	\label{eq:ansatz2}
 \end{align}
The regularity of the large $z$ limit 
indicates that the polynomial part is determined by the large $z$ expansion of the square root part.  
We can further extract the eigenvalue densities from the discontinuity of the square root part
\begin{align}
	\rho_{X,\mathbb{I}}(z) &= \tilde \rho\left(-\tfrac 7 6,-\tfrac{\sqrt{2}}{3};z\right), \label{rho-X-I}\\[6pt]
	\rho_{X,YY}(z) & = f_1\,\tilde \rho\left(\tfrac 1 3,-\tfrac {2\sqrt{2}}{3};z\right),\quad
	\rho_{X,YYYY}(z)  =f_2\, \tilde \rho\left(\tfrac {11}{6},-\tfrac {5}{3\sqrt{2}};z\right), \\[6pt]
	\rho_{X,YYZZ}(z)  &= f_1^2\, \tilde \rho\left(\tfrac 3 2,-\sqrt{2};z\right), \quad
	\rho_{X,YZZY}(z)  = f_1^2\,\tilde \rho\left(\tfrac 7 6,-\tfrac {2\sqrt{2}}{3};z\right),\\[6pt]
	\rho_{X,YXXY}(z)  &=f_1^2\, \tilde \rho\left(\tfrac {7} {6},-\tfrac {2\sqrt{2}}3;z\right),
\end{align}
\begin{align}
	\rho_{X,XYXY}(z) &= \frac{z^2 }{2\pi D}\,\sqrt{\tfrac{2\sqrt{2}}{\sqrt{D}}-z^2},\quad
	\rho_{X,YZYZ}(z)  = \frac{1}{2\sqrt{2}\,\pi D^{3/2}}\sqrt{\tfrac{2\sqrt{2}}{\sqrt{D}}-z^2},
\end{align}
where the ansatz for the eigenvalue density is
\begin{align}
	\tilde \rho(a_1,a_2;z)  &=  \frac{\sqrt{D}}{\sqrt{2}\pi}
	\Big(1+\frac{a_1}{D}+\frac{a_2}{\sqrt{D}}z^2\Big)
	\sqrt{\frac{2\sqrt{2}}{\sqrt{D}}\left(1+\frac{3}{2D}\right)-z^2}.
\end{align}
As expected, these eigenvalue densities are invariant under $z\rightarrow -z$ due to the $X\rightarrow -X$ symmetry.  Furthermore, the maximal eigenvalue
\begin{equation}\label{lambdamax}
z_{\text{max}}(X)=\sqrt{\frac{2\sqrt{2}}{\sqrt{D}}\left(1+\frac{3}{2D}\right)}+O(D^{-9/4})
\end{equation} 
does not depend on the $\mathcal O$ insertion. 
We can see that the $1/D$ corrections take a polynomial form.
Note that $z$ is of order $D^{-1/4}$ around the endpoints of the eigenvalue distributions. 

We can also extract the multi-eigenvalue distributions from the multi-length trajectories. 
In fact, the universal term $\frac{3}{4D}\sum_{j=1}^{j_\text{max}} n_j$ in section \ref{Multi-length trajectories} 
is due to the $1/D$ correction to the maximal eigenvalue in \eqref{lambdamax}. 
The other $1/D$ corrections also take a polynomial form. 
For instance, the simplest two-length trajectory 
\begin{equation}
\big\langle \tr  X^{n_1}Y^{n_2}    \big\rangle=\int_{-z_{\text{max}}}^{z_{\text{max}}}
dz_1dz_2\,z_1^{n_1}z_2^{n_2}\,\rho^{(2)}(z_1,z_2)
\end{equation}
in \eqref{2-length-trajectory} is associated with the two-eigenvalue density
\begin{equation}
\rho^{(2)}(z_1,z_2)= \frac{D}{2\pi^2}
	\left(1-\frac 8 {3D}-\frac{2z_1^2z_2^2}{3}\right)
	\sqrt{(z_{\text{max}}^2-z_1^2)(z_{\text{max}}^2-z_2^2)}\,,
\end{equation}
where $z_\text{max}=z_\text{max}(X) $ is defined in \eqref{lambdamax}. 
As a consistency check, 
we verify that the contraction limit $n_2\rightarrow 0$ is associated with integrating out $z_2$
\begin{equation}
\int_{-z_{\text{max}}}^{z_{\text{max}}}dz_2\,\rho^{(2)}(z_1,z_2)
=\tilde \rho\left(-\frac 7 6,-\frac{\sqrt{2}}{3};z_1\right)
=\rho_{X,\mathbb{I}}(z_1)\,,
\end{equation}
which reduces to the standard one-eigenvalue density \eqref{rho-X-I} to subleading order in $1/D$. 
The cases with duplicate matrices are more subtle. 
The simplest example is the four-length trajectory 
\begin{equation}
\big\langle \tr  X^{n_1}Y^{n_2}X^{n_3}Z^{n_4}    \big\rangle=\int_{-z_{\text{max}}}^{z_{\text{max}}}
dz_1dz_2dz_3dz_4\,z_1^{n_1}z_2^{n_2}z_3^{n_3}z_4^{n_4}\,\rho^{(4)}(z_1,z_2,z_3,z_4)\,,
\end{equation}
where the four-eigenvalue density reads
\begin{align}
&\rho^{(4)}(z_1,z_2,z_3,z_4)
\nonumber\\
=\,&\delta(z_1-z_3)\frac{D^{3/2}}{(2\pi^2)^{3/2}}
\left(1-\frac{25}{6D}+\frac{\sqrt 2 z_1^2}{3\sqrt D}-\frac{2z_1^2(z_2^2+z_4^2)}{3}\right)
\sqrt{\prod_{i=1,2,4}(z_{\text{max}}^2-z_i^2)}
\nonumber\\&
-\frac{D}{12\pi^4}
\left(1-\frac 6 D+2(D-9)z_2^2 z_4^2-\frac{(2D-15)(z_2^2+z_4^2)}{\sqrt{2D}}\right)
\sqrt{\prod_{i=1,2,3,4}(z_{\text{max}}^2-z_i^2)}\,.
\end{align}
Note that the delta function $\delta(z_1-z_3)$ in the first part leads to the sum of two length variables, $(n_1+n_3)$, in \eqref{1213-trajectory}. 
In general, we observe that duplicate matrices lead to delta functions in the multi-eigenvalue density and thus the associated lengths are added together.

\subsubsection{Singlet trajectories}
\label{sec:singlet-trajectories}
It is also interesting to use the quadratic $\mathrm{O}(D)$ singlet as the building block
\begin{equation}
M=X^\mu X_\mu,
\end{equation} 
where a summation over $\mu=1,2,\dots,D$ is implied by the repeated indices. 
The general expression for the moments of $X^\mu X_\mu$ is
\begin{equation}
\langle\tr (X^\mu X_\mu)^n\rangle=(D/2)^{\tfrac{n}{2}}
\left(1 + \frac{n(2n+5)}{6}D^{-1}+O(D^{-2})\right)\,.
\end{equation}
To subleading order in the large $D$ expansion, the standard resolvent of $X^\mu X_\mu$ can be written as
\begin{equation}\label{XX-singlet-resolvent}
	R_{X^\mu X_\mu,\mathbb{I}}(z)=\frac{2}
	{\left(\sqrt{z_{\max}} - \sqrt{z_{\min}}\right)^{2}}
	\left(1-\sqrt{\left(1- \frac{z_{\max}}{z} \right)\left(1 - \frac{z_{\min}}{z}\right)}
	-
	\frac{\sqrt{z_{\max}\, z_{\min}}}{z}\right),
\end{equation}
where the definitions of $z_\text{max}$ and $z_\text{min}$ are given below. 
As this resolvent has a pole at $z=0$, the contour deformation requires more care at small length. 
The corresponding eigenvalue density of $M=X^\mu X_\mu$ is
\begin{equation}\label{rho_XmuXmu}
	\rho_{{X^\mu X_\mu}, \mathbb{I}}(z)=\frac{2}{\pi} \, \frac{\sqrt{\,\bigl(z_{\max} -z\bigr)\,\bigl(z - z_{\min}\bigr)}}{(\sqrt{z_{\max}} -\sqrt{z_{\min}})^2\,z}\,,
\end{equation}
where the maximum and minimum eigenvalues are
\begin{equation}
	z_{\max}(X^\mu X_\mu) = \frac{\sqrt{D}}{\sqrt{2}}\left(1 + \frac{11}{6D}\right) +  \frac{2}{\sqrt{3}}\, ,
	\quad
	z_{\min}(X^\mu X_\mu) = \frac{\sqrt{D}}{\sqrt{2}}\left(1 + \frac{11}{6D}\right)  -  \frac{2}{\sqrt{3}}\, .
\end{equation}
The eigenvalues of $X^\mu X_\mu$ are positive real numbers, so the minimum eigenvalue $z_{\min}$ is also positive.  
This is different from the symmetric distribution of the $X$ eigenvalues, which is invariant under $z\rightarrow -z$. 

In light of the potential importance of symmetry breaking, 
we also consider the quadratic singlets of the subgroups, such as the $\mathrm{O}(D)$ singlet
\begin{equation} 
T_d=X_1 X^1+X_2 X^2+\dots +X_d X^d\,, 
\end{equation}
where $d=1,2,\dots, D-2,D-1$. 
If $d$ is of order $D^0$, we have
\begin{align}
\big\langle  \tr\, (T_d)^1\big\rangle&=\frac{d}{(2 D)^{1/2}}\left(1+\frac 7 {6 D}+O(D^{-2})\right),
\\[6pt]
\big\langle  \tr\, (T_d)^2\big\rangle&= \frac{d}{2D}\left((d+1)+\frac{2d+3}{D}+O(D^{-2})\right),
\\[6pt]
\big\langle  \tr\, (T_d)^3\big\rangle&= \frac{d}{(2D)^{3/2}}
\left(d^2+3d+1+\frac{5d^2+23d+11}{2D}+O(D^{-2})\right),
\\[6pt]
\big\langle  \tr\, (T_d)^4\big\rangle&= \frac{d}{(2D)^{2}}
\left((d+1)(d^2+5d+1)+\frac{2(4d^3+40d^2+55d+13)}{3D}+O(D^{-2})\right)\,,
\end{align} 
together with $\langle  \tr\, (T_d)^0\big\rangle=\langle  \tr\,\mathbb{I}\big\rangle=1$.
These expressions also apply to the case of $d=O(D^1)$, but one needs to be careful about the orders of the leading terms. 
To subleading order in $1/D$, the general expression can be expressed as
\begin{align}
	\langle\tr (T_d)^j\rangle&=\sum_{k=0}^{1} D^{k/2}
	\left( a_{0,k} + \frac{a_{1,k}}{D} \right)
	\bigg[-\frac{4\, \sqrt{z_{\max}\,z_{\min}}}{(\sqrt{z_{\max}}-\sqrt{z_{\min}})^{2}}\,
	\delta_{0,\,k{+}j}+
	\frac{(\sqrt{z_{\max}}+\sqrt{z_{\min}})^{2}}{(\sqrt{z_{\max}}-\sqrt{z_{\min}})^{2}}
	\nonumber\\
	&\quad\times
	\; {}_2F_1\!\left(
	1-k{-}j, -k-j; 2;
	\frac{(\sqrt{z_{\max}}+\sqrt{z_{\min}})^{2}}{(\sqrt{z_{\max}}-\sqrt{z_{\min}})^{2}}
	\right) \left( \frac{\sqrt{z_{\max}}-\sqrt{z_{\min}}}{2} \right)^{2(k{+}j)}
	\bigg], \label{Td-general-expression}
\end{align}
where $z_{\max}, z_{\min}$ are the maximal and minimal eigenvalues of $T_d$, 
\begin{align}
z_{\max}(T_d) &=
\frac{(\sqrt{\text{d}} + 1)^{2}}{\sqrt{2D}}
\left( 1 + \left( \frac{7}{6} - \frac{d - 2}{3 \sqrt{d}} \right)\frac 1 D \right) ,
\\[6pt]
z_{\min}(T_d) &=
\frac{(\sqrt{d} - 1)^{2}}{\sqrt{2D}}
\left( 1 + \left( \frac{7}{6} + \frac{d - 2}{3 \sqrt{d}} \right)\frac 1 D \right),
\end{align}
and the coefficients are
\begin{equation}
	a_{0,0}=1\,,\quad a_{0,1}=0\,,\quad a_{1,0}=\frac{2-d}{3}\,,\quad a_{1,1}=\frac{\sqrt{2}(d-2)}{3d}.
\end{equation}
We verify that the general expression \eqref{Td-general-expression} is consistent with the large $D$ expansion results for length $L\leq 12$. 
For $d=1$, we have $T_d=XX$, so \eqref{Td-general-expression} reduces to $\big\langle \tr  X^{2j}    \big\rangle$,  
and the leading term is associated with the Catalan numbers, 
as mentioned in section \ref{1-length-large-D}. 
For $d=2$, the leading term is instead related to the large Schröder numbers.
In general, the leading term for $d>1$ is related to the number of Schröder paths with $(d-1)$ possible colors. See e.g. \cite{Barry} for more details.
 
Accordingly, the eigenvalue density of $T_d$ can be written as\footnote{The ${}_2F_1$ function in \eqref{Td-general-expression} is first derived for the free singlet moments. 
Then we use the large length expansion, i.e., the $1/j$ expansion,  to deduce the eigenvalue density of the free singlet, 
which is proportional to $\sqrt{(z_\text{max}-z)(z-z_\text{min})}/z$. 
See \cite{Li:2023ewe,Li:2024rod,Huang:2025sua} for earlier applications of the large length expansion in  quantum mechanical systems. }
\begin{equation}\label{rho_Td}
	\rho_{\,T_d,\mathbb{I}}=\frac{ 2}{\pi} 
	\frac{\sqrt{(z_{\max} - z)(z - z_{\min})}}{(\sqrt{z_{\max}} -\sqrt{ z_{\min}})^{2}\, z}
	\left(1 + \frac{a_{1,0}}{D} +\left(a_{0,1}\, z+ \frac{a_{1,1}}{{D}}\right) \, \sqrt{D}\,z \right) .
\end{equation}
We can see that the coefficients $a_{j,k}$ are nothing but the coefficients of the polynomial corrections to the square root form of the Gaussian theory. 
The $\delta_{0, k+j}$ term in \eqref{Td-general-expression} is related to the pole of the resolvent at $z=0$ when $j=0$.
It would be interesting to find the more general eigenvalue density 
that interpolates between the case of $d=O(D^0)$ in \eqref{rho_Td} and that of $d=D$ in \eqref{rho_XmuXmu}.\footnote{There exist some ambiguities in the  perturbative expression of an eigenvalue density, i.e., a different expression may give the same moments to subleading order in  $1/D$. These differences may be crucial to the more general expression. }

\subsection{Analytic trajectory bootstrap for finite $D$ }
Above we used the large $D$ expansion results to derive explicit formulas for the analytic trajectories and eigenvalue distributions. 
They provide good approximations at large $D$, but become less accurate at finite $D$. 
Below we propose some ansatz for the finite $D$ trajectories. 
They resemble the functional forms of the large $D$ formulas, but the coefficients of the polynomial approximations are free parameters.  
We then use symmetries, contraction limit and the loop equations to determine the free parameters, 
and solve  the Yang-Mills matrix model at finite $D$. 
\subsubsection{Leading ansatz} \label{sec:Leading ansatz}
The leading large $D$ behaviors are captured by a Gaussian saddle point. 
The free moments satisfy the freeness property
\begin{equation}
	\Big\langle \tr
	\big[(X^{\mu_1})^{i_1} - \big\langle \tr (X^{\mu_1})^{i_1} \big\rangle\big]
	\big[(X^{\mu_2})^{i_2} - \big\langle \tr (X^{\mu_2})^{i_2} \big\rangle\big]
	\dots
	\big[(X^{\mu_n})^{i_n} - \big\langle \tr (X^{\mu_n})^{i_n} \big\rangle\big]
	\Big\rangle = 0,
	\label{eq:free-independence}
\end{equation}
where 
$X^{\mu_j} \neq X^{\mu_{j+1}}$ for $j=1,\dots,n-1$ and $X^{\mu_n} \neq X^{\mu_{1}}$.
Recursively, all mixed moments 
can be expressed in terms of single-matrix moments $\big\langle \tr (X^{\mu})^{n} \big\rangle$. 
Furthermore, if a mixed moment has duplicate matrices, 
then the power of single-matrix moments can take the sum of its different powers. 
For example, the leading large $D$ expression of the  mixed moment 
$\big\langle \tr X^{n_1} Y^{n_2} X^{n_3} Y^{n_4}\big\rangle$ 
involves
$\big\langle \tr X^{n_1} \big\rangle$, 
$\big\langle \tr X^{n_3} \big\rangle$, 
$\big\langle \tr Y^{n_2} \big\rangle$, 
$\big\langle \tr Y^{n_4} \big\rangle$
and 
$\big\langle \tr X^{n_1+n_3} \big\rangle$, 
$\big\langle \tr Y^{n_2+n_4} \big\rangle$.

To bootstrap the Yang-Mills integrals at finite $D$, 
we make some ansatz for the multi-length trajectories,  
whose construction is inspired by the large $D$ limit discussed above. 
For a general mixed
word,
\begin{equation}
	\big\langle \tr\; \bigl(X^{1}\bigr)^{n_1}\bigl(X^{2}\bigr)^{n_2}\dots \bigl(X^{j_\text{max}}\bigr)^{n_{j_\text{max}}}\big\rangle,
	\qquad X^{j_i}\neq X^{j_{i+1}} ,
\end{equation}
we would like to take all  partitions of  the set $\{n_1,...,n_{j_\text{max}}\}$ such that only the lengths $n_j$ associated with the same matrix can be in the same subset.
In the summation below, \(\pi=\{b_1,b_2,\cdots,b_{m_\pi}\}\) runs over all partitions obtained in this way, while \(b_k\) ranges over all subsets in a given partition \(\pi\).  The number of subsets in $\pi$ is denoted by $m_\pi$.
We will take the leading ansatz as

\begin{equation}
	\big\langle \tr\; \bigl(X^{1}\bigr)^{n_1}\bigl(X^{2}\bigr)^{n_2}\dots \bigl(X^{j_\text{max}}\bigr)^{n_{j_\text{max}}}\big\rangle
	=
	\sum_{\pi} C^{(X^1,X^2,\dots, X^{j_\text{max}})}_{\pi}\prod_{k=1}^{m_\pi}
	P({\textstyle\sum}_{n_i\in b_k} n_i),
	\label{leading ansatz}
\end{equation}
where the building block is taken from the single-matrix moments of the Gaussian theory,
\begin{equation}
	P(n)=\frac{1+(-1)^n}{2}\,
	\frac{\,(1/2)_{\,n/2}}{(2)_{\,n/2}}(z_\text{max})^n.
\end{equation}
Due to $\mathrm{O}(D)$ symmetry, we assume that the maximum eigenvalues of $X^\mu$ are the same. 
Below are some explicit examples of the leading ansatz:
\begin{align}
\big\langle \tr X^{n_1} \big\rangle
&=P(n_1)\,,
\\[6pt]
\big\langle \tr X^{n_1} Y^{n_2} \big\rangle
&=C^{(1,2)}_{1}\,P(n_1)P(n_2)\,,
\\[6pt]
\big\langle \tr X^{n_1} Y^{n_2}Z^{n_3} \big\rangle
&=C^{(1,2,3)}_{1}\,P(n_1)P(n_2)P(n_3)\,,
\\[6pt]
\big\langle \tr X^{n_1} Y^{n_2}Z^{n_3}W^{n_4} \big\rangle
&=C^{(1,2,3,4)}_{1}\,P(n_1)P(n_2)P(n_3)P(n_4)\,,
\end{align}
\begin{align}
\big\langle \tr X^{n_1} Y^{n_2}X^{n_3}Z^{n_4} \big\rangle
&=C^{(1,2,1,3)}_{1}\,P(n_1)P(n_2)P(n_3)P(n_4)
+C^{(1,2,1,3)}_{2}\,P(n_1+n_3)P(n_2)P(n_4)\,,
\\[6pt]
\big\langle \tr X^{n_1} Y^{n_2}X^{n_3}Y^{n_4} \big\rangle
&=C^{(1,2,1,2)}_{1}\,P(n_1)P(n_2)P(n_3)P(n_4)
+C^{(1,2,1,2)}_{2}\,P(n_1+n_3)P(n_2)P(n_4)
\nonumber\\&\quad
+C^{(1,2,1,2)}_{3}\,P(n_1)P(n_2+n_4)P(n_3)
+C^{(1,2,1,2)}_{4}\,P(n_1+n_3)P(n_2+n_4)
\,.
\label{trXYXY-leading}
\end{align}
In the concrete examples, we simplify the subscripts and superscripts of the coefficients $C$.

To determine the unknown parameters, we impose several consistency constraints.

First, we impose discrete symmetries directly at the level of trajectories, which encode infinitely many words. These include the cyclicity of the trace, the reversal symmetry induced by $X_\mu\rightarrow X_\mu^T$, and the $S_D$ permutation symmetry among the matrix species. Here $S_D$ is a discrete subgroup of the original $\mathrm{O}(D)$ symmetry and corresponds to relabelling the $D$ matrices. These discrete symmetries relate equivalent trajectory structures and identify some coefficients in the ansatz. For example, the equality
\begin{equation}
	\big\langle \tr X^{n_1}Y^{n_2}X^{n_3}Y^{n_4}\big\rangle
	=
	\big\langle \tr X^{n_2}Y^{n_1}X^{n_4}Y^{n_3}\big\rangle 
	\label{two-equal}
\end{equation}
 gives \(C^{(1,2,1,2)}_{2}=C^{(1,2,1,2)}_{3}\) in \eqref{trXYXY-leading}.
This equality can be obtained in two steps. First, cyclicity of the trace together with the reversal symmetry gives
\(	\big\langle \tr X^{n_1}Y^{n_2}X^{n_3}Y^{n_4}\big\rangle
	=
	\big\langle \tr Y^{n_2}X^{n_1}Y^{n_4}X^{n_3}\big\rangle .\)
Then, applying the $S_D$ permutation symmetry, which  permutes the  matrices, gives
\(	\big\langle \tr Y^{n_2}X^{n_1}Y^{n_4}X^{n_3}\big\rangle
	=
	\big\langle \tr X^{n_2}Y^{n_1}X^{n_4}Y^{n_3}\big\rangle .\)

Second, the ansatz should satisfy some matching conditions associated with the contraction limits. 
In a contraction limit $n_j=0$, a $J$-length trajectory may reduce to a $(J-1)$-length trajectory or a $(J-2)$-length trajectory. 
For example, 
the $n_2\rightarrow 0$ limit of the two-length trajectory 
$\big\langle \tr X^{n_1} Y^{n_2} \big\rangle$ reduces to the one-length trajectory $\big\langle \tr X^{n_1} \big\rangle$.
For an example of a two-length reduction, we can take 
the $n_2\rightarrow 0$ limit of the four-length trajectory $\big\langle \tr X^{n_1} Y^{n_2}X^{n_3}Z^{n_4} \big\rangle$, 
which gives the two-length trajectory $\big\langle \tr X^{n_1+n_3}Z^{n_4} \big\rangle$.

Third, we require that the moments are consistent with the $O(D)$ singlet decomposition of the covariant moments in \eqref{general-A-decomposition}, up to a finite length cutoff. This step differs from the first step because it uses the full $O(D)$ symmetry and is imposed only at a finite length cutoff.

In this way, we reproduce the freeness solution
\begin{equation}
A_{1,2}=\frac {z_\text{max}^2} 4 \,,\quad
A_{1,2,3,4}=\frac {z_\text{max}^4} {16}\,,\quad
A_{1,3,2,4}=0\,,
\end{equation}
\begin{equation}
C^{(1,2)}_{1}=C^{(1,2,3)}_{1}=C^{(1,2,3,4)}_{1}=C^{(1,2,1,3)}_{2}
=C^{(1,2,1,2)}_{2}=C^{(1,2,1,2)}_{3}=1\,,
\end{equation}
\begin{equation}
C^{(1,2,1,2)}_{1}=-1\,,\quad
C^{(1,2,1,3)}_{1}=C^{(1,2,1,2)}_{4}=0\,.
\end{equation}
Finally, the loop equation \eqref{leqA-1} implies that 
\begin{equation}
z_\text{max}^2={\frac{2\sqrt 2}{\sqrt{D-1}}}\,,
\end{equation}
so we have
\begin{equation}\label{leading-ansatz-sol}
\big\langle \tr XX\big\rangle=\frac{1}{\sqrt{2(D-1)}}\,,
\quad
\big\langle \tr XXXX\big\rangle=\frac{1}{{D-1}}\,,
\end{equation}
which are consistent with the leading large $D$ results in section \ref{Large D expansion}. 
For $D=3,\dots,10$, the estimates for $\big\langle \tr XX\big\rangle$ and $\big\langle \tr XXXX\big\rangle$ are
\begin{align}\label{ansatz-leading-1}
(0.5,\, 0.5)_{D=3}\,,\quad&
(0.408,\, 0.333)_{D=4}\,,\quad
(0.354,\, 0.25)_{D=5}\,,\quad
(0.316,\, 0.2)_{D=6}\,,\\
(0.289,\, 0.167)_{D=7}\,,\quad&
(0.267,\, 0.143)_{D=8}\,,\quad
(0.25,\, 0.125)_{D=9}\,,\quad
(0.236,\, 0.111)_{D=10}\,.
\end{align}
We notice that these simple estimates for $D$ are closer to the Monte Carlo results at $D+1$,  
which suggests that the subleading terms also have important effects.  
To improve the accuracy of the analytic trajectory bootstrap solutions, 
we introduce some subleading terms into the multi-length ansatz.

\subsubsection{Subleading ansatz}

In section \ref{Multi-length trajectories}, we obtained the multi-length
trajectories at subleading order in the large \(D\) expansion. Compared with the
leading terms, the new feature is that a leading factor of the form
\begin{equation}
	P({\textstyle\sum}_{n_i\in b_k} n_i)
\end{equation}
in \eqref{leading ansatz} may acquire an increase $s_{b_k}=1,2$ in its argument at subleading order:
\begin{equation}
	P({\textstyle\sum}_{n_i\in b_k} n_i+1),
	\qquad
P({\textstyle\sum}_{n_i\in b_k}n_i+2).
\end{equation}
We only include terms in which the total increase of the arguments of \(P\) is at most four. Schematically, the subleading ansatz is

	\begin{equation}
	\big\langle \tr\; \bigl(X^{1}\bigr)^{n_1}\bigl(X^{2}\bigr)^{n_2}\dots \bigl(X^{j_\text{max}}\bigr)^{n_{j_\text{max}}}\big\rangle
	=
	\sum_{\pi}   
	\sum_{s_1}	\sum_{s_2}\cdots	\sum_{s_{m_\pi}}
	C^{(X^1,X^2,\dots, X^{j_\text{max}})}_{\pi,\{s_{1},s_2,\cdots,s_{m_\pi}\}}\prod_{k=1}^{m_\pi}
	P({\textstyle\sum}_{n_i\in b_k} n_i+s_{k}),
\end{equation}
where
\begin{equation}
		s_{b_k}\in\{0,1,2\},\; \sum_{k=1}^{m_\pi}s_{b_k}\leq 4.
		\label{subleading-1}
\end{equation}

We consider the multi-length trajectories with at most five lengths:
\begin{equation}
\big\langle \tr X^{n_1} \big\rangle\,,\quad
\big\langle \tr X^{n_1}Y^{n_2}  \big\rangle\,,\quad
\big\langle \tr X^{n_1}Y^{n_2}Z^{n_3}  \big\rangle,
\end{equation}
\begin{equation}
\Big\langle \tr X^{n_1}Y^{n_2}X^{n_3}Y^{n_4}\Big\rangle,\quad
\Big\langle \tr X^{n_1}Y^{n_2}X^{n_3}Z^{n_4}\Big\rangle, \quad
\Big\langle \tr X^{n_1}Y^{n_2}Z^{n_3}W^{n_4}\Big\rangle, 
\end{equation}
\begin{equation}
\Big\langle \tr X^{n_1}Y^{n_2}X^{n_3}Y^{n_4}Z^{n_5}\Big\rangle,\quad
\Big\langle \tr X^{n_1}Y^{n_2}X^{n_3}Z^{n_4}W^{n_5}\Big\rangle,\quad
\Big\langle \tr X^{n_1}Y^{n_2}Z^{n_3}W^{n_4}V^{n_5}\Big\rangle.
\end{equation}
Their exact large $D$ expansion expressions can be found in section \ref{Multi-length trajectories}. 
Some explicit examples of the 
ansatz are 
\begin{align}
\big\langle \tr X^{n_1} \big\rangle
&=C^{(1)}_1 P(n_1)+C^{(1)}_2 P(n_1+2)\,,
\\[6pt]
\big\langle \tr X^{n_1} Y^{n_2} \big\rangle
&=C^{(1,2)}_{1}\,P(n_1)P(n_2)+C^{(1,2)}_{2}\,P(n_1+2)P(n_2)
\nonumber\\
&\quad+C^{(1,2)}_{3}\,P(n_1)P(n_2+2)+C^{(1,2)}_{4}\,P(n_1+2)P(n_2+2)\,,
\\[6pt]
\big\langle \tr X^{n_1} Y^{n_2}Z^{n_3} \big\rangle
&=C^{(1,2,3)}_{1}\,P(n_1)P(n_2)P(n_3)+C^{(1,2,3)}_{2}\,P(n_1+2)P(n_2)P(n_3)
\nonumber\\
&\quad+C^{(1,2,3)}_{3}\,P(n_1)P(n_2+2)P(n_3)
+C^{(1,2,3)}_{4}\,P(n_1)P(n_2)P(n_3+2)
\nonumber\\
&\quad+C^{(1,2,3)}_{5}\,P(n_1+2)P(n_2+2)P(n_3)
+C^{(1,2,3)}_{6}\,P(n_1+2)P(n_2)P(n_3+2)
\nonumber\\
&\quad+C^{(1,2,3)}_{7}\,P(n_1)P(n_2+2)P(n_3+2).
\end{align}
Here we have already erased some vanishing terms. A four-length example with duplicate matrices is given in appendix \ref{Appendix:details}. 

There are significantly more free parameters in the subleading ansatz. 
We again impose that the multi-length trajectories are compatible with the covariant moments \eqref{general-A-decomposition} up to length $10$. Accordingly, we make use of the loop equations \eqref{leqA-1}, \eqref{leqA-2},\eqref{leqA-3} and 
those involving moments of length at most $8$.
Here the matching condition becomes more interesting.
For example, 
the $n_2\rightarrow 0$ limit of the two-length trajectory 
$\big\langle \tr X^{n_1} Y^{n_2} \big\rangle$ reduces to the one-length trajectory $\big\langle \tr X^{n_1} \big\rangle$, 
so we have
\begin{align}
	&\left(C^{(1,2)}_{1}+C^{(1,2)}_{3}P(2)\right)\,P(n_1)
	+\left(C^{(1,2)}_{2}+C^{(1,2)}_{4}\,P(2)\right)\,P(n_1+2)
	\nonumber\\
	=&C^{(1)}_1 P(n_1)+C^{(1)}_2 P(n_1+2)\,,
\end{align}
which implies
\begin{equation}
	C^{(1)}_1=C^{(1,2)}_{1}+C^{(1,2)}_{3}P(2),\quad
	C^{(1)}_2=C^{(1,2)}_{2}+C^{(1,2)}_{4}\,P(2)\,.
\end{equation}
The matching conditions from these contraction limits 
lead to nontrivial constraints on the coefficients of the multi-length ansatz.

\begin{figure}[tbp]
	\centering 
	\includegraphics[width=0.8\textwidth]{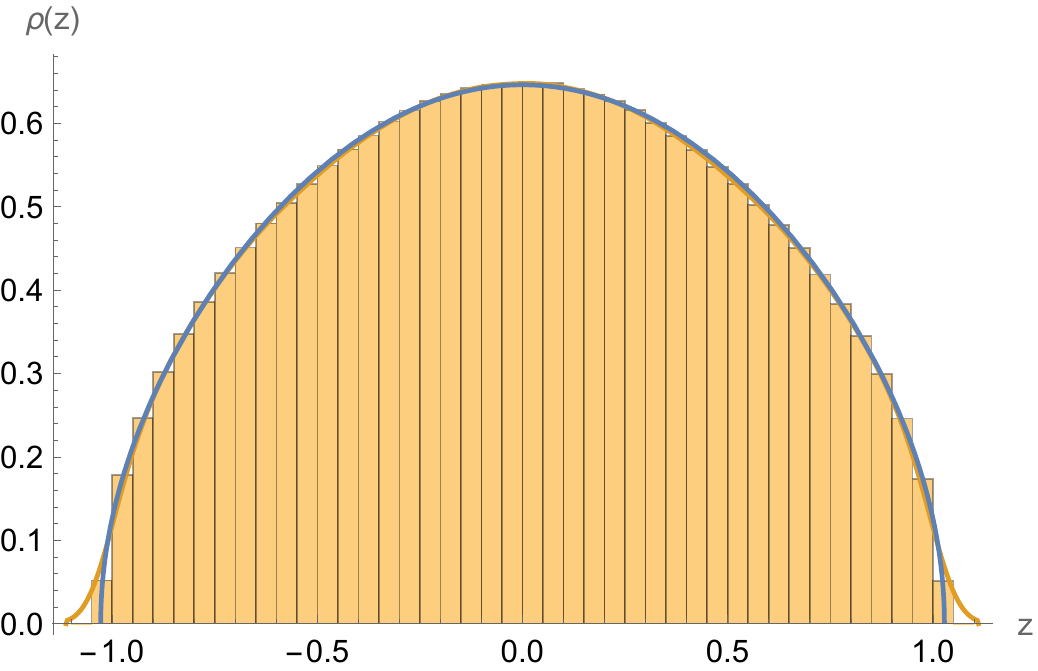}
	\caption{Eigenvalue distribution of  $X$ for $D=10$. The blue curve is the prediction of the analytic trajectory bootstrap in \eqref{rho_D10_ansatz}. The histogram of $X$ eigenvalues is extracted from Monte Carlo simulations with $N=300$ based on the code from \cite{Jha:2021exo}. 
	The orange curve from \texttt{SmoothHistogram} of \texttt{Mathematica} matches well with the blue curve, except near the endpoints where the finite $N$ effects are more significant. 
	}
	\label{rho_D10}
\end{figure}

As the number of constraints is greater than that of free parameters, 
we use the $\eta$ minimization. 
We introduce the error function $\eta$ as a sum of squared loop equations
\begin{equation}
\eta=\sum_j (\text{loop eq})_j^2, 
\label{error-function}
\end{equation}
where $j$ labels the loop equations for $L_\text{max}=8$, 
i.e., the lengths of the words in the loop equations are at most $8$. 
A minimization of the $\eta$ function determines the singlet decomposition coefficients. 
For $D\geq 7$, the accuracy of the analytic trajectory bootstrap results is improved. 
The estimates for $(\langle \tr XX\rangle, \langle \tr XXXX\rangle  )$ are:
\begin{align}\label{ansatz-subleading-1}
(0.317,\, 0.209)_{D=7}\,,
(0.291,\, 0.175)_{D=8}\,,
(0.270,\, 0.150)_{D=9}\,,
(0.253,\, 0.131)_{D=10}\,,
\end{align}
which are inside the $L_\text{max}=8$ bounds and slightly below the Monte Carlo estimates. 
In figure \ref{rho_D10}, we further show that the bootstrap prediction for the $D=10$ eigenvalue distribution 
\begin{equation}\label{rho_D10_ansatz}
\rho_{X,\mathbb{I}}^{D=10}(z)\approx\frac {2\sqrt{z_\text{max}^2-z^2}} {\pi z_\text{max}^2}\left(1.0433-0.1639z^2\right)\,,\quad
z_\text{max}\approx 1.0282\,,
\end{equation}
is in excellent agreement with the histogram of eigenvalues from $N=300$ Monte Carlo simulations, except around the endpoints. 
Using this eigenvalue density \eqref{rho_D10_ansatz}, we can also compute the higher moments
\begin{equation}
\langle \tr X^6\rangle\approx 0.085\,,\quad
\langle \tr X^8\rangle\approx 0.062\,,\quad
\langle \tr X^{10}\rangle\approx 0.049\,,
\end{equation}
which are also compatible with our Monte Carlo estimates 
$0.089(6), 0.066(9), 0.052(5)$.

As the maximum of the total length is increased by 4, 
we may allow the powers of the single-matrix moments to be increased by 3 or 4.
Then there are more possible terms in the multi-length ansatz. Instead of \eqref{subleading-1}, we could have
\begin{equation}
	s_{b_\pi}\in\{0,1,2,3,4\},\; \sum_{k=1}^{m_\pi}s_{b_k}\leq 4.
\end{equation}
The analytic trajectory bootstrap results for $(\langle \tr XX\rangle, \langle \tr XXXX\rangle)$ become 
\begin{align}\label{ansatz-subleading-2}
&(0.355,\, 0.341)_{D=3}\,,\quad
(0.549,\, 0.550)_{D=4}\,,\quad
(0.421,\, 0.340)_{D=5}\,,\quad
(0.357,\, 0.250)_{D=6}\,,
\nonumber\\
&(0.317,\, 0.200)_{D=7}\,,\quad
(0.289,\, 0.167)_{D=8}\,,\quad
(0.268,\, 0.144)_{D=9}\,,\quad
(0.251,\, 0.126)_{D=10}\,.
\end{align}
In comparison with \eqref{ansatz-subleading-1}, 
the bootstrap results become slightly less accurate at relatively large $D$, but more stable at small $D$. 
For $D=3$, the subleading ansatz results are still 
worse than the estimates from the leading ansatz in \eqref{ansatz-leading-1}.  

For comparison, 
the finite $D$ predictions of the large $D$ expansion are
\begin{equation}
\left(\big\langle \tr XX\big\rangle,\big\langle \tr XXXX\big\rangle \right)_{\frac 1 D,\, \text{subleading}}=
\left(\frac{1}{\sqrt{2D}}+\frac{7}{6D^{3/2}}, \frac{1}{D}+\frac{5}{2D^{2}}\right)\,.
\end{equation}
For $D\geq 4$, the corresponding estimates are less accurate than the above  bootstrap results. 
	For instance, the $D=4,5,6$ estimates from the $1/D$ series are
\begin{equation}
	\text{large}\; D\;\text{expansion}:\quad
	(0.457, 0.406)_{D=4}, \quad
	(0.390, 0.300)_{D=5}, \quad
	(0.345,0.236)_{D=6}\,.
\end{equation} 
They exhibit more significant deviations from the Monte Carlo results in table \ref{tab:boot-vs-mc}.  
The analytic trajectory bootstrap results in \eqref{ansatz-subleading-2} provide 
improvements for these relatively small $D$ estimates. 

However, the $D=3$ case is not under good control in our analytic trajectory bootstrap study. 
The reasons are twofold. 
\begin{enumerate}
\item
It may be intrinsically harder to bootstrap the small $D$ Yang-Mills matrix integrals.  

In the introduction, we mentioned that the large $N$ Yang-Mills matrix integrals are well-defined for $D\geq 3$. 
Although the number of degrees of freedom for $D=3$ is smaller than the higher $D$ cases,\footnote{
It is easier to perform both the direct enumeration of the positive semi-definite matrix 
and Monte Carlo simulations for $D=3$, so we can push to higher truncations 
with the same amount of computational resources. } 
the matrix integral may be more dangerous.  
If we view $D$ as a continuous parameter, 
then the small $D$ results are more inclined to be affected by 
the divergences at $D=2$.

As shown in section \ref{Positivity}, the positivity bounds for small $D$ are also weaker. 
For instance, the $D=3$ island appears at the length cutoff $L_\text{max}=12$, which is much larger than  $L_\text{max}=8$ for the $D\geq 4$ islands. 
In the $\mathrm{O}(D)$ basis, this manifests as vanishing irreducible representations, and thus a reduced number of nontrivial positive semi-definite constraints at small $D$. 
(See appendix \ref{Appendix A} for more details.)
On the other hand, $\mathrm{O}(D)$ symmetry is less useful in reducing the number of free parameters at small $D$. 
It is easier but also more necessary to increase the truncation order for $D=3$. 

\item
The ansatz in the analytic trajectory bootstrap  may be less suitable for small $D$. 

As the functional form of the ansatz is inspired by the large $D$ formulas, it is not surprising that this form is less suitable for small $D$. A better choice of ansatz may significantly improve the $D=3$ analytic trajectory bootstrap results, 
but this requires an in-depth understanding of the $D=3$ Yang-Mills matrix model. 

As $D$ increases, the positivity bounds can reproduce the results of the large $D$ expansion, 
so the existence of a large $D$ saddle also simplifies the bootstrap studies. 
The leading large $D$ saddle is associated with decoupled matrices,  
and we may view $1/D$ as a measure of the strength of the matrix interaction. 
Then the small $D$ cases are expected to be harder due to strong coupling effects. 
In terms of ansatz, the eigenvalue distributions of the large $D$ saddle are given by 
Wigner's semi-circle in the square root form, together with subleading corrections in the polynomial form. 
(See \eqref{rho_D10_ansatz} for example.)
A better ansatz for $D=3$ may exhibit more intricate analytic properties.  
\end{enumerate}
One of the basic features in the ansatz construction is the large length factorization, 
i.e., the matrices are asymptotically non-interacting as the lengths are much larger than the interaction strength. 
In this way, we can use a one-variable function to build the ansatz. 
Otherwise, we may need to use more sophisticated building blocks with nontrivial dependence on multiple variables.

\section{Discussion}
\label{sec:discussion}

In this work, we have studied the $D$-matrix Yang--Mills integrals by two complementary bootstrap methods, i.e., the positivity bootstrap and the analytic trajectory bootstrap.

In section \ref{Preliminaries}, we derived the loop equations and used the $\mathrm{O}(D)$ singlet decomposition to reduce the numbers of unknowns and loop equations.
We also extended the large $D$ expansion results in~\cite{Hotta:1998en} to a large number of matrix moments of length $4\leq L\leq 12$.  
The explicit expressions for these moments suggest that they can be organized by lengths and form analytic trajectories in lengths. 

In section~\ref{Positivity}, we used positive semi-definiteness to derive bootstrap bounds for $\big\langle \tr XX\big\rangle$ and $\big\langle \tr XXXX\big\rangle$ up to the  length cutoff \(L_{\max}=12\). 
Naively, the positivity bootstrap is computationally expensive for multi-matrix models because the dimension of the explicit basis grows rapidly with $L_{\max}$.  
Therefore, we used the \(\mathrm{O}(D)\) representation theory to reduce the sizes of positive semi-definite matrices. 
We present this procedure in detail and derive analytic lower bounds for $L_{\max}=4,6$. 
The leading large $D$ prediction lies on the boundary of these allowed regions. 
For $L_{\max}=8,10,12$, we obtain numerical bounds. 
For $D \ge 4$, the allowed region forms an island when $L_{\max}\ge 8$, 
but the $D=3$ island appears only at $L_{\max}=12$.  
These positive regions shrink rapidly with increasing $L_{\max}$. 
Some of the $L_{\max}=12$ islands give highly accurate predictions, whose precision is comparable to that of Monte Carlo simulations. 
We further examined some cases of relatively large $D$ and confirmed the subleading large $D$ behavior from the $1/D$ expansion.

In section \ref{Analytic}, we used the analytic trajectory bootstrap method, which does not rely on positivity assumptions. 
We unified the concrete results of the large $D$ expansion by analytic continuation in lengths, 
from one-length trajectories to the higher generalization of multi-length trajectories. 
They are closely related to resolvents and eigenvalue densities. 
Based on these analytic formulas, we proposed some ansatz for bootstrapping the finite $D$ Yang-Mills integrals, 
and derived accurate results for finite $D> 3$.

Below are several comments on what can be improved in this work.
 \begin{itemize}
 	\item  	 In our positivity bootstrap, the use of irreducible representations leads to positivity constraints whose matrix entries are rational functions of $D$.
 	This makes it possible to analytically continue these constraints to non-integer values of $D$. 
It would be interesting to use this continuation to study the behavior of the bootstrap bounds as $D$ is varied continuously.  In particular, one may decrease $D$ from $3$ toward $2$ and examine whether the bootstrap constraints reflect the known convergence condition of the bosonic Yang-Mills matrix integral in  \cite{Austing:2001bd,Austing:2001pk,Krauth:1998yu}. 
However, it is less clear if positivity conditions should be satisfied at non-integer $D$. 
 	We may also improve some technical treatments to handle the cases of relatively large $D$ at $L_{\text{max}}\geq 12$, which should determine  the higher order terms in the $1/D$ expansion. 
 	
 	\item On the analytic trajectory bootstrap side, the construction of the ansatz is based on the explicit results from large $D$ expansion. 
	To apply this method to other models, we need to develop a more systematic procedure for extracting the analytic structures, which requires a deeper understanding of the matrix models.
 	On the other hand, the eigenvalue densities associated with the singlet trajectories in section \ref{sec:singlet-trajectories} are,  to our knowledge, discussed for the first time. 
	The  ansatz associated with these singlet trajectories may also lead to useful constraints. 
 \end{itemize}

Some comments on future directions are in order.
\begin{itemize}
	\item The positivity bootstrap study can be extended to other matrix models. 
	We may consider other bosonic multi-matrix integrals that admit a controlled expansion in a large parameter.
	In this work, the large parameter is the number of matrices $D$, but similar ideas may apply to other limits. For example, the large coupling limit is particularly interesting.
	It is worthwhile to see if the leading behaviors and their corrections can also be extracted from the positivity bounds.
	One potential goal is to derive the strong coupling expansion of the Hoppe model using the matrix bootstrap, as discussed in \cite{Guerrieri:2025ytx}.
	We may also consider supersymmetric matrix integrals. 
	If the Pfaffian from the fermion path integral is real and positive, 
	we can still impose the positive semidefinite condition on the positivity matrices constructed from bosonic observables,
	such as the $D=4$ super-Yang-Mills matrix integral and its mass deformations
	\cite{Krauth:1998xh,Ambjorn:2000bf,Martina:2025kwc}. 
	In matrix models, we may also derive positivity constraints from non-negative eigenvalue densities, 
	which is closer to the unitarity assumption in field theories. 
	\item As the sign or complex phase problem obstructs a direct use of the positivity bootstrap, an appealing direction is to study supersymmetric matrix models by the analytic trajectory bootstrap. 
	In some cases, the eigenvalue spectra can be computed analytically in certain limits, or inferred from Monte Carlo data. 
	In a super-Yang-Mills matrix integral,  the asymptotic behavior of the
eigenvalue distribution may decay as a power law, and the higher moments may be divergent. 
See \cite{Krauth:1998yu, Krauth1999,Ambjorn:2000dx,Austing:2001pk} for some early discussions. 
Then the  eigenvalue distributions are significantly different from the Wigner semi-circle type, which can be viewed as the starting point of the ansatz in this work. 
As a result, we may need to make some dramatic changes in the functional forms of the ansatz in the analytic trajectory bootstrap. 
Analytic results from the infinite mass limit of the mass-deformed Yang-Mills matrix integrals may be useful. 
	\item 
	Another important direction is to extend the Yang-Mills matrix bootstrap to higher-dimensional spacetime. 
	The case of lattice gauge theories has been studied in \cite{Anderson:2016rcw,Kazakov:2022xuh,Kazakov:2024ool,Li:2024wrd,Guo:2025fii}. 
	In continuum descriptions, one needs to use a different regularization scheme, such as a momentum cutoff or dimensional regularization, and be careful about the nonperturbative formulation \cite{Roberts:1994dr}. 
	Before a full fledged study of higher dimensional Yang-Mills gauge theory, 
	it is also interesting to bootstrap Wilson loops in the zero dimensional context. 
	The use of exponential operators will lead to new observables beyond the simple matrix moments. 
\end{itemize}

\acknowledgments
We thank Zechuan Zheng for discussions and Raghav G. Jha for correspondence. 
This work was partly supported by the Natural Science
Foundation of China (Grants No. 12522504 and No. 12205386), 
and 
college students innovation and entrepreneurship training program, Sun Yat-sen University. 
This research was supported in part by Perimeter Institute for Theoretical Physics. 
Research at Perimeter Institute is supported by the Government of Canada through the Department of Innovation, Science, and Economic Development, and by the Province of Ontario through the Ministry of Colleges and Universities.

\appendix

\section{Construction of $\mathrm{O}(D)$ irreducible representations}
\label{Appendix A}

In this appendix, we present a concrete procedure for constructing irreducible representations of $\mathrm{O}(D)$. See e.g. \cite{Cvitanovic:2008zz,Fulton:1997} for references.

To construct the irreducible representations of the orthogonal group $\mathrm{O}(D)$, it is convenient to introduce Young diagrams and Young tableaux, which are widely used for labelling irreps of various groups due to the fact that standard Young tableaux label the irreps of the general linear group GL($n$). In particular, we also use them to label the irreps of $\mathrm{O}(D)$.

A Young diagram is associated with a partition of an integer $k$. Concretely, we partition $k$ as a sum of positive integers arranged in non-increasing order. Each part specifies the number of boxes in a row, so the corresponding Young diagram $T_{k,r}$ is a left-justified array of boxes whose row lengths are given by the parts of the partition.  Each distinct Young diagram labels an inequivalent irreducible representation of $\mathrm{O}(D)$.
A standard Young tableau $T_{k,r,a}$ is a filling of $T_{k,r}$ with entries from $\{1,\ldots,k\}$ in which each number appears exactly once, and the entries increase strictly along rows and down columns. Different tableaux of the same Young diagram correspond to equivalent irreps.

For the $\mathrm{O}(D)$ group, contractions of the $\mathrm{O}(D)$ indices with  $\delta_{ij}$  yield $\mathrm{O}(D)$-invariant tensors. Therefore, one must first separate the trace and traceless parts 
in the construction of irreducible components. Then, a standard Young tableau indexes a symmetry type of tensor indices and defines the associated Young symmetrizer $F_T$. Acting with $F_T$ on rank-$k$ tensors produces a tensor with the prescribed symmetry type.

This procedure is described in detail below.
\begin{enumerate}
	\item For a given word length $L$ and target rank $k$, list all possible index contractions that yield rank-$k$ words of length $L$.
	
	\smallskip
	\noindent
	Below are some examples. For $L=1,3$ and $k=1$, the possible rank-$1$ words are listed in \eqref{rank-1-decomposition}. 
	For $L=2,4$ and $k=2$, the possible rank-$2$ words are
	\begin{equation}
		\begin{aligned}
			&X^{\mu_1}X^{\mu_2},\;
			\frac{1}{D}\,X^{\mu_1}X^{\mu_2}X^{\mu_3}X^{\mu_3},\;
			\frac{1}{D}\,X^{\mu_1}X^{\mu_3}X^{\mu_2}X^{\mu_3},\;
			\frac{1}{D}\,X^{\mu_1}X^{\mu_3}X^{\mu_3}X^{\mu_2},\;\\
			&
			\frac{1}{D}\,X^{\mu_3}X^{\mu_1}X^{\mu_2}X^{\mu_3},\;
			\frac{1}{D}\,X^{\mu_3}X^{\mu_1}X^{\mu_3}X^{\mu_2},\;
			\frac{1}{D}\,X^{\mu_3}X^{\mu_3}X^{\mu_1}X^{\mu_2}.
		\end{aligned}
	\end{equation}
	For $L=3,5$ and $k=3$, the possible rank-$3$ words include
	\begin{equation}
		\begin{aligned}
			&X^{\mu_1} X^{\mu_2} X^{\mu_3},\;
			\frac{1}{D}\,X^{\mu_1} X^{\mu_2} X^{\mu_3} X^{\mu_4} X^{\mu_4},\;
			\frac{1}{D}\,X^{\mu_1} X^{\mu_2} X^{\mu_4} X^{\mu_3} X^{\mu_4},\;\\
			&\frac{1}{D}\,X^{\mu_1} X^{\mu_2} X^{\mu_4} X^{\mu_4} X^{\mu_3},\;
			\frac{1}{D}\,X^{\mu_1} X^{\mu_4} X^{\mu_2} X^{\mu_3} X^{\mu_4},\;
			\frac{1}{D}\,X^{\mu_1} X^{\mu_4} X^{\mu_2} X^{\mu_4} X^{\mu_3},\;\\
			&\frac{1}{D}\,X^{\mu_1} X^{\mu_4} X^{\mu_4} X^{\mu_2} X^{\mu_3},\;
			\frac{1}{D}\,X^{\mu_4} X^{\mu_1} X^{\mu_2} X^{\mu_3} X^{\mu_4},\;
			\frac{1}{D}\,X^{\mu_4} X^{\mu_1} X^{\mu_2} X^{\mu_4} X^{\mu_3},\;\\
			&\frac{1}{D}\,X^{\mu_4} X^{\mu_1} X^{\mu_4} X^{\mu_2} X^{\mu_3},\;
			\frac{1}{D}\,X^{\mu_4} X^{\mu_4} X^{\mu_1} X^{\mu_2} X^{\mu_3}.
		\end{aligned}
	\end{equation}
	In our normalization convention, we add a factor $1/D$ for each contraction.
	
\item For each rank-$k$ tensor in Step~1, construct its traceless form for $k\ge2$.

	\smallskip
\noindent

First, we construct a traceless projector for a general rank-$k$ tensor $X^{\mu_1}\cdots X^{\mu_k}$. 
The case of rank-$2$ is simple. See \eqref{rank-2-decomposition} for the example of $X^{\mu_1}X^{\mu_2}$, 
where the traceless part is $X^{\mu_1}X^{\mu_2}$ minus the singlet part $S$.  

As an example for higher-$k$, we consider a general rank~3 tensor $\mathcal{T}^{\mu_{1}\mu_{2}\mu_{3}}$. 
Its traceless projection $\Phi^{\mu_{1}\mu_{2}\mu_{3}}$ can be written as
\begin{align}
		\Phi ^{\mu_{1}\mu_{2}\mu_{3}} =
		& \mathcal{T}^{\mu_{1}\mu_{2}\mu_{3}}
		+ \big(\alpha _1\,\delta^{\mu_{2}\mu_{3}}\,\mathcal{T}^{\mu_{1}\mu_{4}\mu_{4}}
		+ \alpha _2\,\delta ^{\mu_{1}\mu_{2}}\,\mathcal{T}^{\mu_{4}\mu_{1}\mu_{4}}
		+ \alpha _3\,\delta ^{\mu_{1}\mu_{2}}\,\mathcal{T}^{\mu_{4}\mu_{4}\mu_{1}}\big)\nonumber\\
		&+ \big(\beta _1\,\delta^{\mu_{1}\mu_{3}}\,\mathcal{T}^{\mu_{2}\mu_{4}\mu_{4}}
		+ \beta_2\,\delta ^{\mu_{1}\mu_{3}}\,\mathcal{T}^{\mu_{4}\mu_{2}\mu_{4}}
		+ \beta _3\,\delta ^{\mu_{1}\mu_{3}}\,\mathcal{T}^{\mu_{4}\mu_{4}\mu_{2}}\big)
		\label{general-traceless-rank-3}\\
		&+ \big(\chi _1\,\delta ^{\mu_{1}\mu_{2}}\,\mathcal{T}^{\mu_{3}\mu_{4}\mu_{4}}
		+ \chi _2\,\delta ^{\mu_{1}\mu_{2}}\,\mathcal{T}^{\mu_{4}\mu_{3}\mu_{4}}
		+ \chi _3\,\delta ^{\mu_{1}\mu_{2}}\,\mathcal{T}^{\mu_{4}\mu_{4}\mu_{3}}\big),\nonumber
\end{align}
where $\alpha$, $\beta$, and $\chi$ are fixed by the traceless conditions
\begin{equation}
	\Phi^{\mu_{1}\mu_{1}\mu_{2}}=0,\qquad
	\Phi^{\mu_{1}\mu_{2}\mu_{1}}=0,\qquad
	\Phi^{\mu_{2}\mu_{1}\mu_{1}}=0.
\end{equation}
The solution is
\begin{align}
	\alpha_1=\beta_2=\chi_3=-\frac{D+1}{(D-1)(D+2)},\\
	\alpha _2=\alpha _3=\beta _1=\beta _3=\chi _1=\chi _2=\frac{1}{(D-1)(D+2)}.
\end{align}
In this way, we obtain the explicit traceless part $\Phi ^{\mu_{1}\mu_{2}\mu_{3}} $ of the general rank-$3$ tensor $\mathcal{T}^{\mu_{1}\mu_{2}\mu_{3}}$. 

Second, we apply the traceless projector to each rank-$k$ tensor from Step~1. 
For example, the action of the rank-$2$ traceless projector on $\frac{1}{D}\,X^{\mu_1}X^{\mu_3}X^{\mu_2}X^{\mu_3}$ gives
\begin{equation}
	\tfrac{1}{D}\Big(X^{\mu_1} X^{\mu_3} X^{\mu_2} X^{\mu_3}
	-\tfrac{1}{D}\,\delta^{\mu_1 \mu_2}\,X^{\mu_4} X^{\mu_3} X^{\mu_4} X^{\mu_3}\Big).
\end{equation}
Applying the rank-$3$ traceless projector on $\frac{1}{D}\,X^{\mu_1} X^{\mu_2} X^{\mu_3} X^{\mu_4} X^{\mu_4}$, we obtain
\begin{equation}
	\begin{aligned}
		&\frac{1}{D(D-1)(D+2)}\Big(
		(D-1)(D+2)\,X^{\mu_1}X^{\mu_2}X^{\mu_3}X^{\mu_4}X^{\mu_4}\\
		&+\delta^{\mu_1\mu_2}\big(X^{\mu_3}X^{\mu_5}X^{\mu_5}X^{\mu_4}X^{\mu_4}
		+X^{\mu_5}X^{\mu_3}X^{\mu_5}X^{\mu_4}X^{\mu_4}
		-(D+1)\,X^{\mu_5}X^{\mu_5}X^{\mu_3}X^{\mu_4}X^{\mu_4}\big)\\
		&+\delta^{\mu_1\mu_3}\big(X^{\mu_2}X^{\mu_5}X^{\mu_5}X^{\mu_4}X^{\mu_4}
		+X^{\mu_5}X^{\mu_5}X^{\mu_2}X^{\mu_4}X^{\mu_4}
		-(D+1)\,X^{\mu_5}X^{\mu_2}X^{\mu_5}X^{\mu_4}X^{\mu_4}\big)\\
		&
		+\delta^{\mu_2\mu_3}\big(X^{\mu_5}X^{\mu_1}X^{\mu_5}X^{\mu_4}X^{\mu_4}
		+X^{\mu_5}X^{\mu_5}X^{\mu_1}X^{\mu_4}X^{\mu_4}
		-(D+1)\,X^{\mu_1}X^{\mu_5}X^{\mu_5}X^{\mu_4}X^{\mu_4}\big)
		\Big).
	\end{aligned}
\end{equation}
	
\item List all standard Young tableaux and read off the Young symmetrizers.
	\smallskip
\noindent

Given a standard Young tableau $T$, its Young symmetrizer is
\begin{equation}
	F_T \;=\; c_{F_T}\sum_{\sigma\in R_T}\ \sum_{\tau\in C_T} \mathrm{sgn}(\tau)\,\tau\, \sigma,
\end{equation}
where $R_T$ is the row group (permutations acting within each row) and $C_T$ is the column group (permutations acting within each column). Here $\mathrm{sgn}(\tau)$ denotes the parity of $\tau$ (it equals \(+1\) for even permutations and \(-1\) for odd permutations). The normalization coefficient $c_{F_T}$ is chosen so that $(F_T)^2=F_T$.

\medskip
\noindent
As a simple example, for the Young tableau
\begin{equation}
	T_{3,2,2} = \young(13,2)\,,
\end{equation}
we have
\begin{equation}
	R_T = \{e, (1\,3)\},\quad C_T = \{e, (1\,2)\},
\end{equation}
where cycle notation is used. 
The corresponding Young symmetrizer is
\begin{equation}
	F_T = c_{F_T}\,(e - (1\,2)) (e + (1\,3))= c_{F_T}\,( e - (1\,2) + (1\,3) - (1\,3\,2)).
	\label{YS-322}
\end{equation}
where $e$ is the identity element. The normalization convention
\begin{equation}
	(F_T)^2=3(c_{F_T})^2( e - (1\,2) + (1\,3) - (1\,3\,2))=F_T
\end{equation}
yields
\begin{equation}
c_{F_T}=\frac{1}{3}\,.
\end{equation}
	\item Apply the Young symmetrizer from Step~3 to the tensors obtained in Step~2 for $k\ge 2$. The action of the Young symmetrizer on tensors has been discussed in footnote~\ref{fn-action}. When $k=0,1$, keep the original tensors from Step~1.
	\smallskip
\noindent

To illustrate, consider the Young tableau
\[
T_{3,2,2} = \begin{ytableau} 1 & 3 \\ 2 \end{ytableau}\,,
\]
whose Young symmetrizer is given in \eqref{YS-322}. 
The group elements act on tensors as follows:

\begin{equation}
	\begin{aligned}
		e &=
		\begin{pmatrix} 1 & 2 & 3 \\ 1 & 2 & 3 \end{pmatrix},
		&
		\begin{array}{l}
			\text{position }1 \to \text{position }1,\quad
			\text{position }2 \to \text{position }2,\\
			\text{position }3 \to \text{position }3,
		\end{array}
		\\[8pt]
		(12) &=
		\begin{pmatrix} 1 & 2 & 3 \\ 2 & 1 & 3 \end{pmatrix},
		&
		\begin{array}{l}
			\text{position }1 \to \text{position }2,\quad
			\text{position }2 \to \text{position }1,\\
			\text{position }3 \to \text{position }3,
		\end{array}
		\\[8pt]
		(13) &=
		\begin{pmatrix} 1 & 2 & 3 \\ 3 & 2 & 1 \end{pmatrix},
		&
		\begin{array}{l}
			\text{position }1 \to \text{position }3,\quad
			\text{position }2 \to \text{position }2,\\
			\text{position }3 \to \text{position }1,
		\end{array}
		\\[8pt]
		(132) &=
		\begin{pmatrix} 1 & 3 & 2 \\ 3 & 2 & 1 \end{pmatrix},
		&
		\begin{array}{l}
			\text{position }1 \to \text{position }3,\quad
			\text{position }3 \to \text{position }2,\\
			\text{position }2 \to \text{position }1.
		\end{array}
	\end{aligned}
\end{equation}

Applying this Young symmetrizer to the general rank-3 traceless tensor 
$\Phi^{\mu_{1}\mu_{2}\mu_{3}}$ in \eqref{general-traceless-rank-3}, we obtain
\begin{align}
	F_{T_{3,2,2}} \Phi^{\mu_{1}\mu_{2}\mu_{3}}
	&=\frac{1}{3} \Big(
	\mathcal{T}^{\mu_1\mu_2\mu_3}
	-\mathcal{T}^{\mu_2\mu_1\mu_3} 
	-\mathcal{T}^{\mu_2\mu_3\mu_1}
	+\mathcal{T}^{\mu_3\mu_2\mu_1}\Big) \quad
	\nonumber\\&+\frac{1}{3(D-1)} \Big(  \delta^{\mu_1\mu_2}(\mathcal{T}^{\mu_4\mu_3\mu_4}- \mathcal{T}^{\mu_3\mu_4\mu_4})
	+\delta^{\mu_2\mu_3}(\mathcal{T}^{\mu_4\mu_1\mu_4}- \mathcal{T}^{\mu_1\mu_4\mu_4}) 
	\nonumber\\
	&\qquad\qquad\qquad+ 2\delta^{\mu_1\mu_3}(\mathcal{T}^{\mu_2\mu_4\mu_4}-\mathcal{T}^{\mu_4\mu_2\mu_4})\Big), 
\end{align}

which is the general form of the traceless rank-3 tensor associated with the Young tableau 
$T_{3,2,2} = \begin{ytableau} 1 & 3 \\ 2 \end{ytableau}.$
This procedure is valid for tensors with or without contracted $\mathrm{O}(D)$ indices. 
For instance, for the Young tableau 
\[
T_{2,1,1} = \begin{ytableau} 1 & 2 \end{ytableau}\,,
\]
the action of the traceless projector and Young symmetrizer on the $L=4,\,k=2$ tensor 
\begin{equation}
	\frac{1}{D} \, X^{\mu_1}X^{\mu_3}X^{\mu_2}X^{\mu_3} 
\end{equation}
gives
\begin{equation}
	\frac{1}{2D}\Big(
	X^{\mu_1}X^{\mu_3}X^{\mu_2}X^{\mu_3}
	- X^{\mu_2}X^{\mu_3}X^{\mu_1}X^{\mu_3}
	-\frac{2}{D}\,\delta^{\mu_1\mu_2}\,X^{\mu_4}X^{\mu_3}X^{\mu_4}X^{\mu_3}
	\Big).
\end{equation}

	Similarly, for  the $L=5,k=3$ tensor $\frac{1}{D}\, X^{\mu_1} X^{\mu_2} X^{\mu_3} X^{\mu_4} X^{\mu_4}$ the action of the traceless projector and the Young symmetrizer associated with the Young tableau
	\(
	T_{3,2,2} = \begin{ytableau} 1 & 3 \\ 2 \end{ytableau}
	\)
	gives
	\begin{equation}
		\begin{aligned}
			&\frac{1}{3(D-1)D}\Big(
			(D-1)\,X^{\mu_1}X^{\mu_2}X^{\mu_3}X^{\mu_4}X^{\mu_4}
			-(D-1)\,X^{\mu_2}X^{\mu_1}X^{\mu_3}X^{\mu_4}X^{\mu_4}\\
			&\quad-(D-1)\,X^{\mu_2}X^{\mu_3}X^{\mu_1}X^{\mu_4}X^{\mu_4}
			+ (D-1)\,X^{\mu_3}X^{\mu_2}X^{\mu_1}X^{\mu_4}X^{\mu_4}\\
			&\quad- X^{\mu_3}X^{\mu_5}X^{\mu_5}X^{\mu_4}X^{\mu_4}\,\delta^{\mu_1\mu_2}
			+ X^{\mu_5}X^{\mu_3}X^{\mu_5}X^{\mu_4}X^{\mu_4}\,\delta^{\mu_1\mu_2}\\
			&\quad+ 2\,X^{\mu_2}X^{\mu_5}X^{\mu_5}X^{\mu_4}X^{\mu_4}\,\delta^{\mu_1\mu_3}
			- 2\,X^{\mu_5}X^{\mu_2}X^{\mu_5}X^{\mu_4}X^{\mu_4}\,\delta^{\mu_1\mu_3}\\
			&\quad- X^{\mu_1}X^{\mu_5}X^{\mu_5}X^{\mu_4}X^{\mu_4}\,\delta^{\mu_2\mu_3}
			+ X^{\mu_5}X^{\mu_1}X^{\mu_5}X^{\mu_4}X^{\mu_4}\,\delta^{\mu_2\mu_3}
			\Big).
		\end{aligned}
	\end{equation}
	
\end{enumerate}

For high ranks, some positive semi-definite matrices are trivially null at small $D$ due to vanishing irreps. 
When the sum of the first two column lengths of a diagram $T_{k,r}$ exceeds $D$, the number of independent components is drastically reduced by the column-antisymmetry. 
The traceless constraints become overdetermined, and thus the traceless tensor and the corresponding matrix $M^{(k,r)}$ vanish. 
For example, the rank-4 Young diagrams
\begin{equation}
	T_{4,1}=\ytableaushort{\, , \, , \, , \,},\qquad
	T_{4,2}=\ytableaushort{\, \, , \, , \,},\qquad
	T_{4,3}=\ytableaushort{\, \, , \, \,}
\end{equation}
 have their first two columns adding up to 4, so they vanish for $D=3$.
Similarly, the rank-5 diagrams
\begin{equation}
	T_{5,1}=\ytableaushort{\, , \, , \, , \, , \,},\qquad
	T_{5,2}=\ytableaushort{\, \, , \, , \, , \,},\qquad
	T_{5,3}=\ytableaushort{\, \, , \, \, , \,}
\end{equation}
vanish for $D=3,4$, while
\begin{equation}
	T_{5,4}=\ytableaushort{\, \, \, , \, , \,},\qquad
	T_{5,5}=\ytableaushort{\, \, \, , \, \,}
\end{equation}
vanish for $D=3$. 
The non-universal features of high-rank tensors at small $D$ were also discussed in \cite{Lin:2025srf}. 

\section{Some technical details}
\label{Appendix:details}
To compute the large $D$ expansion of the sextic singlets, 
we use Wick's theorem to derive 
\begin{equation}
	\begin{aligned}
		&\big\langle X_{a}^{\mu} X_{b}^{\mu} X_{c}^{\nu} X_{d}^{\nu} X_{e}^{\rho} X_{f}^{\rho} \big\rangle
		\\[6pt]
		=\; &\frac{1}{N^{3/2}} \Big\langle 
		D^{3}\,  ({K^{-1}})_{a b} ({K^{-1}})_{c d} ({K^{-1}})_{e f} \\[4pt]
		&\quad\qquad+ D^{2} \Big( ({K^{-1}})_{a f} ({K^{-1}})_{b e} ({K^{-1}})_{c d}
		+ ({K^{-1}})_{a e} ({K^{-1}})_{b f} ({K^{-1}})_{c d} \\[4pt]
		&\quad\qquad\qquad\;\;\; + ({K^{-1}})_{a b} ({K^{-1}})_{c f} ({K^{-1}})_{d e}
		+ ({K^{-1}})_{a b} ({K^{-1}})_{c e} ({K^{-1}})_{d f} \\[4pt]
		&\quad\qquad\qquad\;\;\; + ({K^{-1}})_{a d} ({K^{-1}})_{b c} ({K^{-1}})_{e f}
		+ ({K^{-1}})_{a c} ({K^{-1}})_{b d} ({K^{-1}})_{e f} \Big) \\[4pt]
		&\quad\qquad+ D \Big( ({K^{-1}})_{a f} ({K^{-1}})_{b d} ({K^{-1}})_{c e}
		+ ({K^{-1}})_{a d} ({K^{-1}})_{b f} ({K^{-1}})_{c e} \\[4pt]
		&\quad\qquad\qquad\;\;\; + ({K^{-1}})_{a e} ({K^{-1}})_{b d} ({K^{-1}})_{c f}
		+ ({K^{-1}})_{a d} ({K^{-1}})_{b e} ({K^{-1}})_{c f} \\[4pt]
		&\quad\qquad\qquad\;\;\; + ({K^{-1}})_{a f} ({K^{-1}})_{b c} ({K^{-1}})_{d e}
		+ ({K^{-1}})_{a c} ({K^{-1}})_{b f} ({K^{-1}})_{d e} \\[4pt]
		&\quad\qquad\qquad\;\;\; + ({K^{-1}})_{a e} ({K^{-1}})_{b c} ({K^{-1}})_{d f}
		+ ({K^{-1}})_{a c} ({K^{-1}})_{b e} ({K^{-1}})_{d f} \Big)
		\;\Big\rangle .
	\end{aligned}
\end{equation}
Then the contraction 
with	
\begin{equation}
\operatorname{tr}(t^a t^b t^c t^d t^e t^f), \,
\operatorname{tr}(t^a t^b t^c t^e t^d t^f),\,
\operatorname{tr}(t^a t^b t^c t^e t^f t^d), \,
\operatorname{tr}(t^a t^c t^e t^b t^f t^d), \,
\operatorname{tr}(t^a t^c t^e t^b t^d t^f)
\end{equation} 
gives
\begin{align}
		\big\langle \operatorname{tr}X^\mu X^\mu X^\nu X^\nu X^\rho X^\rho \big\rangle 
		&= \frac{1}{2\sqrt{2}} D^{3/2} + \frac{11}{4\sqrt{2}} D^{1/2} +O(D^{-1/2}) ,\\[6pt]
		\big\langle \operatorname{tr}X^\mu X^\mu X^\nu X^\rho X^\nu X^\rho\big\rangle 
		&= \frac{3}{2\sqrt{2}} D^{1/2} +O(D^{-1/2}) ,\\[6pt]
		\big\langle \operatorname{tr}X^\mu X^\mu X^\nu X^\rho X^\rho X^\nu\big\rangle 
		&= \frac{1}{2\sqrt{2}} D^{3/2} + \frac{29}{12\sqrt{2}} D^{1/2} +O(D^{-1/2}), \\[6pt]
		\big\langle \operatorname{tr}X^\mu X^\nu X^\rho X^\mu X^\rho X^\nu \big\rangle 
		&= \frac{1}{2\sqrt{2}} D^{1/2}+O(D^{-1/2}),  \\[6pt]
		\big\langle \operatorname{tr} X^\mu X^\nu X^\rho X^\mu X^\nu X^\rho \big\rangle 
		&= \frac{3}{2\sqrt{2}} D^{1/2}+O(D^{-1/2}) .
\end{align}

In the analytic trajectory bootstrap, an example for the four-length subleading ansatz associated with duplicate matrices is
\begin{align}
&\quad\big\langle \tr X^{n_1} Y^{n_2}X^{n_3}Z^{n_4} \big\rangle
\nonumber\\[6pt]
&=C^{(1,2,1,3)}_{1}\,P(n_1)P(n_2)P(n_3)P(n_4)
+C^{(1,2,1,3)}_{2}\,P(n_1+n_3)P(n_2)P(n_4)
\nonumber\\&\quad
+C^{(1,2,1,3)}_{3}\,P(n_1+1)P(n_2)P(n_3+1)P(n_4)
+C^{(1,2,1,3)}_{4}\,P(n_1+2)P(n_2)P(n_3)P(n_4)
\nonumber\\&\quad
+C^{(1,2,1,3)}_{5}\,P(n_1)P(n_2+2)P(n_3)P(n_4)
+C^{(1,2,1,3)}_{6}\,P(n_1)P(n_2)P(n_3+2)P(n_4)
\nonumber\\&\quad
+C^{(1,2,1,3)}_{7}\,P(n_1)P(n_2)P(n_3)P(n_4+2)
+C^{(1,2,1,3)}_{8}\,P(n_1+n_3+2)P(n_2)P(n_4)
\nonumber\\&\quad
+C^{(1,2,1,3)}_{9}\,P(n_1+n_3)P(n_2+2)P(n_4)
+C^{(1,2,1,3)}_{10}\,P(n_1+n_3)P(n_2)P(n_4+2)
\nonumber\\&\quad
+C^{(1,2,1,3)}_{11}\,P(n_1+2)P(n_2+2)P(n_3)P(n_4)
+C^{(1,2,1,3)}_{12}\,P(n_1+2)P(n_2)P(n_3+2)P(n_4)
\nonumber\\&\quad
+C^{(1,2,1,3)}_{13}\,P(n_1+2)P(n_2)P(n_3)P(n_4+2)
+C^{(1,2,1,3)}_{14}\,P(n_1)P(n_2+2)P(n_3+2)P(n_4)
\nonumber\\&\quad
+C^{(1,2,1,3)}_{15}\,P(n_1)P(n_2+2)P(n_3)P(n_4+2)
+C^{(1,2,1,3)}_{16}\,P(n_1)P(n_2)P(n_3+2)P(n_4+2)
\nonumber\\&\quad
+C^{(1,2,1,3)}_{17}\,P(n_1+n_3+2)P(n_2)P(n_4)
+C^{(1,2,1,3)}_{18}\,P(n_1+n_3)P(n_2+2)P(n_4)
\nonumber\\&\quad
+C^{(1,2,1,3)}_{19}\,P(n_1+n_3)P(n_2)P(n_4+2)
+C^{(1,2,1,3)}_{20}\,P(n_1+1)P(n_2+2)P(n_3+1)P(n_4)
\nonumber\\&\quad
+C^{(1,2,1,3)}_{21}\,P(n_1+1)P(n_2)P(n_3+1)P(n_4+2)
\,.
\end{align}

\end{document}